\documentclass[12pt,a4paper]{article}

\usepackage[hidelinks]{hyperref}

\usepackage{afterpage, amsfonts, amsmath, amssymb, amsthm, bbm, caption, color, graphicx, epstopdf, enumitem, float, lipsum, longtable, lscape, mathrsfs, multirow, nccmath, nonfloat, pdflscape, rotating, tocloft}

\usepackage[title,titletoc]{appendix}
\usepackage[margin=0.8in]{geometry}

\usepackage[round]{natbib}
\usepackage[doublespacing]{setspace}
\usepackage[table]{xcolor}

\def\be{\begin{equation}}
\def\ee{\end{equation}}
\def\bea{\begin{eqnarray}}
\def\eea{\end{eqnarray}}

\author{}
\title{}

\DeclareMathOperator*{\argmin}{\arg\!\min}
\DeclareMathOperator*{\argmax}{\arg\!\max}

\DeclareMathOperator*{\vect}{\normalfont\textrm{vec}}

\DeclareMathOperator*{\diag}{\normalfont\textrm{diag}}
\DeclareMathOperator*{\tr}{\normalfont\textrm{trace}}
\DeclareMathOperator*{\ic}{\normalfont\textrm{IC}}

\begin{document}
\newcommand\blfootnote[1]{
\begingroup
\renewcommand\thefootnote{}\footnote{#1}
\addtocounter{footnote}{-1}
\endgroup
}

\newtheorem{corollary}{Corollary}
\newtheorem{definition}{Definition}
\newtheorem{lemma}{Lemma}
\newtheorem{proposition}{Proposition}
\newtheorem{remark}{Remark}
\newtheorem{theorem}{Theorem}
\newtheorem{assumption}{Assumption}

\numberwithin{corollary}{section}
\numberwithin{definition}{section}
\numberwithin{equation}{section}
\numberwithin{lemma}{section}
\numberwithin{proposition}{section}
\numberwithin{remark}{section}
\numberwithin{theorem}{section}

\allowdisplaybreaks[4]

\begin{titlepage}

{\small

\begin{center}
{\Large \bf  Binary Response Models for Heterogeneous Panel Data with Interactive Fixed Effects\blfootnote{
\begin{itemize}

\item[] \textit{Correspondence:} Bin Peng, Department of Econometrics and Business Statistics, Monash University, Caulfield East, VIC 3145, Australia. Email: Bin.Peng@monash.edu
\end{itemize}
}

} 
\medskip

{\sc Jiti Gao$^\sharp$ and Fei Liu$^{\ast}$ and Bin Peng$^{\sharp}$ and Yayi Yan$^{\sharp}$}
\medskip

$^\sharp$Monash University, Australia and $^{\ast}$Nankai University, China

\bigskip\bigskip

\today

\bigskip

\begin{abstract}
In this paper, we investigate binary response models for heterogeneous panel data with interactive fixed effects by allowing both the cross-sectional dimension and the temporal dimension to diverge. From a practical point of view, the proposed framework can be applied to predict the probability of corporate failure, conduct credit rating analysis, etc. Theoretically and methodologically, we establish a link between a maximum likelihood estimation and a least squares approach, provide a simple information criterion to detect the number of factors, and achieve the asymptotic distributions accordingly. In addition, we conduct intensive simulations to examine the theoretical findings. In the empirical study, we focus on the sign prediction of stock returns, and then use the results of sign forecast to conduct portfolio analysis. 
\end{abstract}

\end{center}


\medskip


}

\end{titlepage}

\section{Introduction}\label{Section1}

Varieties of binary response panel data models have been proposed and studied over the past a couple of decades, and earlier developments date back at least to \cite{Chamberlain} and the references therein. The challenges in the previous studies often arise due to the identification issues caused by short time periods of data and non-closed form estimators (e.g., \citealp{Manski1987, Chamberlain2010}; among others). With the rise and availability of big and rich datasets, recent studies on binary response panel data models gradually shift focuses to the cases where both the cross-sectional dimension and the temporal dimension are allowed to diverge. An excellent review is given in \cite{FW2018}. Recently, an important strand of the literature is devoted to binary response models with interactive fixed effects (e.g., \citealp{BL2017, Wang2019, Chen2020}). Within these studies, the central questions are a) how to estimate the coefficients together with the factors and the factor loadings? and b) to achieve the optimal efficiency in a),  how to determine the number of factors? 

For linear additive models, the aforementioned questions have been addressed well in the literature by utilizing different techniques. For example, using principal component analysis (PCA) and random matrix theory, \cite{BN2002}, \cite{Onatski}, \cite{LamYao2012} and \cite{AhnHorenstein2013} are able to detect the number of factors for large panel data using information criteria or eigenanalysis; \cite{Pesaran2006} introduces a common correlated effects (CCE) estimator that takes an advantage of a factor structure involving both dependent and independent variables; \cite{Bai2009} and \cite{Moon} develop alternative methods to estimate the coefficients together with the factors and the factor loadings; \cite{LCL2020} and \cite{HJPS}  respectively use the maximum likelihood method and the classifier-Lasso method to consider the cases with heterogeneous coefficients; and so forth. 

However, for non-linear panel data models, especially for binary response panel data models involving interactive fixed effects, there is limited progress, which is mainly due to the fact that a variety of tools adopted for linear additive models may no longer be directly applicable and useful. Below, we comment on the relevant literature. In \cite{BL2017}, the authors extend the CCE approach to a framework with binary responses, in which a key step is to estimate unobservable factors from the regressors. As a consequence, the approach requires an explicit structure of the regressors, and the usual limitation of a CCE type estimator occurs, e.g., the number of unobservable factors cannot be larger than the number of regressors (cf.,  \citealp[eq. 7]{BL2017}). \cite{Wang2019} and \cite{Chen2020} propose similar solutions to binary response panel data models, and the main difference is that the former does not include any regressors in the model. Thereby, we may regard \cite{Wang2019} and \cite{Chen2020} as the binary response counterparts of \cite{BN2002} and \cite{Bai2009}, respectively. Recently, \cite{AB2020}, \cite{AL2020} and \cite{CDG2020} bring attention to additive panel data models with interactive fixed effects, in which closed form estimators are less obvious. Specifically, quantile regressions are investigated in all three papers, of which only \cite{AL2020} include regressors, while \cite{AB2020} propose a Bayesian approach.

In view of the aforementioned literature, we specifically consider a binary response panel data model with interactive fixed effects by incorporating heterogeneous coefficients. From a modelling perspective, it is similar to \cite{BL2017}, but we require less structure on the regressors, which allows us to avoid adding any restriction between the number of regressors and the number of unobservable factors. Our investigation establishes a link between a maximum likelihood estimation and a nonlinear least squares approach. As a consequence, some tools adopted for linear additive models immediately become applicable. For example, the identification restrictions provided in \cite{Bai2009} and \cite{Moon} are readily to be applied to the binary response models with very minor modifications. Meanwhile, we are able to estimate the unknown heterogeneous parameters as well as the unobservable factors and factor loadings, and establish the asymptotic distributions accordingly. Our approach may be considered as the binary response counterpart of that considered in \cite{BN2013}. In addition, we propose a simple information criterion to detect the number of factors. Last but not least, we conduct intensive numerical studies to examine the theoretical findings, and demonstrate the practical relevance. 

From a practical perspective, the proposed framework can be relevant and applicable to the following fields for instance. Predicting the probability of corporate failure has gained its attention since the seminal work of \cite{Altman}. Along this line of research, our paper provides a more generalized framework to extend those panel data driven studies (e.g.,  \citealp{CCL2014}). Similarly, our model and estimation method can  be applied to panel data based credit rating analysis (e.g., \citealp{JJW2015}). In the empirical study of the paper, we pay particular attention to portfolio analysis. More often than not, in order to calculate the optimal weights assigned to each stock, one adopts all stocks to construct a covariance matrix (e.g., \citealp{CLL2019, ELW2019}), which then naturally falls into the category of high dimensional covariance matrix estimation. Thereby, to boost the estimation accuracy, we see the increasing popularity of methods using rank reduction (\citealp{PX2019}), penalization (\citealp{CLL2019}), or both (\citealp{FLM13}) among others. It is worth emphasizing that by default the aforementioned techniques eventually include all stocks in practice, though some of them may have relatively small weights compared to the others. As pointed out in \cite{CD2006} and \cite{Nyberg2011}, the sign of stock market returns may be predictable even if the returns themselves are not predictable. Also, \cite{CD2006} mention that ``\textit{As volatility moves, so too does the probability of a positive return: the higher the volatility, the lower the probability of a positive return}". In connection with the fact that a primary goal of portfolio analysis is to minimize the volatility (\citealp{ELW2019}), a binary response panel data model with interactive fixed effects naturally marries the above studies by modelling the probabilities of positive returns, so we can drop those having low probabilities.

In summary, the main contributions of the paper are as follows. (i). We consider a binary panel data model with both heterogeneous coefficients and interactive fixed effects, and establish a link between the maximum likelihood estimation and the nonlinear least squares estimation. As a consequence, the traditional type of identification conditions (such as those in \citealp{Bai2009} and \citealp{Moon}) for linear panel data models with interactive fixed effects is readily to be applied with very minor modification. (ii). We provide a simply information criterion to select the number of factors. (iii). In addition to extensive simulation studies, we use the newly established model and approach to bridge two strands of studies on stock returns, so that better performance can be achieved for portfolio analysis.

The structure of this paper is as follows. Section \ref{Section2} proposes the model, develops the methodology, and then establishes the asymptotic results. Section \ref{Section3} provides intensive simulations to examine the finite-sample performance of the theoretical findings. Section \ref{Section4} considers an empirical portfolio analysis. Section \ref{Section5} concludes. Appendix \ref{AppA} sketches the outline of the theoretical development, and comments on the bias correction and the average partial effects. For the sake of space, we only provide the proofs of some selected main results in this appendix. The omitted proofs and the preliminary lemmas are given in the online supplementary Appendix \ref{AppB} of the paper.

Before proceeding further, we introduce some mathematical symbols that will be used repeatedly throughout the paper. $\|\cdot\|$ denotes the Euclidean norm of a vector or the Frobenius norm of a matrix; $O(1)$ always stands for a finite positive constant, and may be different at each appearance; $\to_P$ and $\to_D$ stand for convergence in probability and convergence in distribution respectively; $\Pr(A\, | \,B)$ represents the probability of the event $A$ occurring conditional on the event $B$; $E(Y\, | \,X)$ denotes the expectation of the variable $Y$ conditional on the variable $X$; for a matrix $W$ with full column rank, let $M_W = I -P_W$ with $P_W =W(W'W)^{-1}W'$; for a square matrix $W$, $\rho_{\max}(W)$ stands for its largest eigenvalue; $ a\wedge b = \min\{a,b\} $ and $a\vee b= \max\{a,b\} $; for a square matrix $A$, $\rho_{\max}(A)$ returns the maximum eigenvalue.

\section{The Model and Methodology}\label{Section2}

In this section, we present the model and the methodology with an algorithm for numerical implementation in practice, and establish the associated asymptotic results. Specifically, we provide the basic setup in Section \ref{Section2.1}, and present a numerical  estimation procedure for practical implementation; Section \ref{Section2.2} summaries the relevant asymptotic results; and Section \ref{Section2.3} considers the selection of the number of factors.

\subsection{The Setup}\label{Section2.1}

The model we consider is a binary response panel data model with interactive fixed effects of the form:

\begin{eqnarray}\label{EQ21}
y_{it} = \left\{ \begin{array}{cc}
1, & x_{it}'\beta_{0i}+\gamma_{0i}'f_{0t} -\varepsilon_{it} \ge 0 \\
0, & \text{otherwise}
\end{array}\right.,
\end{eqnarray}
where $i=1,\ldots,N$ and $t=1,\ldots,T$. In the model \eqref{EQ21}, we observe the binary dependent variable $y_{it}$ and the $d_\beta\times 1$ explanatory variables $x_{it}$ with $d_\beta$ being finite. For ease of notation, we suppose that $\{\varepsilon_{it}\}$ is an array of identically distributed random errors in $(i,t)$ with the respective probability density function (PDF) and the cumulative distribution function (CDF) being known. Specifically, we denote the PDF and the CDF as $g_\varepsilon(\cdot)$ and $G_\varepsilon(\cdot)$. Both the factor loading $\gamma_{0i}$ and the factor $f_{0t}$ are $d_f\times 1$, where $d_f$ is finite. For the time being, we assume that $d_f$ is known, and we will come back to its estimation with the corresponding asymptotic result in Section \ref{Section2.3} later. In what follows, we are interested in recovering  

\begin{eqnarray}\label{EQBFG}
B_0 =(\beta_{01},\ldots, \beta_{0N})',\quad F_0 =(f_{01},\ldots, f_{0T})',\quad \text{and}\quad \Gamma_0 = (\gamma_{01},\ldots, \gamma_{0N})'.
\end{eqnarray}
For notational simplicity, we let $\theta_{0i}=(\beta_{0i}', \gamma_{0i}')'$, and $\Theta_0=(B_0,\Gamma_0)=(\theta_{01},\ldots,\theta_{0N})'$ throughout the paper.

\begin{remark}\label{RM21}
We now comment on why heteroskedasticity is ruled out in the above setting. When heteroskedasticity occurs, (say, $\varepsilon_{it}\sim N(0 ,\sigma_i^2)$), we can always rewrite the model as follows.

\begin{eqnarray}\label{EQ22}
y_{it} = \left\{ \begin{array}{cc}
1, & x_{it}'\beta_{0i}^*+\gamma_{0i}^{*\prime}f_{0t} -\varepsilon_{it}^*\ge 0, \\
0, & \text{otherwise,}
\end{array}\right.
\end{eqnarray}
where $\beta_{0i}^*  =\frac{\beta_{0i}}{\sigma_i}$, $\gamma_{0i}^{*} =\frac{\gamma_{0i}}{\sigma_i}$, and $\varepsilon_{it}^* =\frac{\varepsilon_{it}}{\sigma_i} $. The transferred model \eqref{EQ22} indicates that we can only estimate the true parameters up to unknown constants $\sigma_i$'s unless some further restrictions are imposed. 
\end{remark}

We now start presenting our estimators for \eqref{EQBFG}. Simple algebra shows that

\begin{eqnarray}\label{EQ23}
\Pr(y_{it} = 1 \, | \, x_{it},\gamma_{0i}, f_{0t}) &=&G_\varepsilon(x_{it}'\beta_{0i}+\gamma_{0i}'f_{0t} ),\nonumber \\
\Pr(y_{it} = 0 \, | \, x_{it},\gamma_{0i}, f_{0t})&=& 1-G_\varepsilon( x_{it}'\beta_{0i}+\gamma_{0i}'f_{0t}),
\end{eqnarray}
which immediately yields $E[y_{it}  \, | \, x_{it},\gamma_{0i}, f_{0t}] = G_\varepsilon(x_{it}'\beta_{0i}+\gamma_{0i}'f_{0t}).$ Thus, the likelihood function is specified as follows:
\begin{eqnarray}\label{EQ25}
L(B, F, \Gamma)=\prod_{i=1}^N\prod_{t=1}^T  \left[1 -G_\varepsilon( x_{it}'\beta_i+\gamma_{i}'f_{t})\right]^{1-y_{it}} \cdot\left[ G_\varepsilon(x_{it}'\beta_i+\gamma_{i}'f_{t}) \right]^{y_{it}} ,
\end{eqnarray}
where $B =(\beta_1,\ldots, \beta_N)'$, $F = (f_1,\ldots, f_T)'$ and $\Gamma = (\gamma_1,\ldots, \gamma_N)'$ are $N\times d_\beta$, $T\times d_{f}$ and $N\times d_{f}$ matrices respectively. Moreover, for the purpose of identification, we require

\begin{eqnarray}\label{EQ26}
F \in   \mathbf{F} = \left\{F\, \big| \, \frac{1}{T}F'F = I_{d_f}\right\}.
\end{eqnarray}
Finally, the log-likelihood function is defined below:
\begin{eqnarray}\label{EQ27}
\log L (B, F, \Gamma) &=&\sum_{i=1}^N\sum_{t=1}^T\Big\{(1-y_{it})\log\left[1-G_\varepsilon(x_{it}'\beta_i+\gamma_{i}'f_{t})\right] \nonumber \\
&&+y_{it} \log G_\varepsilon(x_{it}'\beta_i+\gamma_{i}'f_{t}) \Big\},
\end{eqnarray}
where $F \in   \mathbf{F} $. The estimators are thus given by

\begin{eqnarray}\label{EQ28}
(\widehat{B}, \widehat{F}, \widehat{\Gamma}) =\argmax_{(B,F, \Gamma) } \log L (B, F, \Gamma),
\end{eqnarray}
where $\widehat{B} =(\widehat{\beta}_1,\ldots, \widehat{\beta}_N)'$, $\widehat{F} = (\widehat{f}_1,\ldots,\widehat{f}_T)'$, and $\widehat{\Gamma} = (\widehat{\gamma}_1,\ldots, \widehat{\gamma}_N)'$. 

\medskip

In what follows, the main goals are studying the asymptotic behaviours of the estimators of \eqref{EQ28}. We sketch the strategy of our asymptotic analysis. Before examining any estimator, we first use the Taylor expansion to investigate the log-likelihood function, which allows us to bridge the maximum likelihood estimation of \eqref{EQ28} and a nonlinear least squares approach in Lemma \ref{Lemma2.1} below. As a consequence, the identification restrictions provided in \cite{Bai2009} and \cite{Moon} are readily to be applied with very minor modifications. On this point, some examples are provided in Section \ref{Section2.2} for the purpose of demonstration. Afterwards, the rates of converges and the asymptotic distributions can be established accordingly.

\medskip

Up to this point, it is worth commenting on the practical implementation of \eqref{EQ28}. A variety of algorithms have been proposed for the panel data models with interactive fixed effects, especially for the cases where the nonclosed form estimators are involved, e.g., \cite{Ando}, \cite{Wang2019}, \cite{Chen2020}, \cite{AL2020}, just to name a few. Our numerical implementation largely follows these methods.

\begin{enumerate}
\item[Step 0:] Initial $\widehat{F}^{(0)}$ by using some random number generators, e.g., the standard normal distribution for each element of $\widehat{F}^{(0)}$. Implement the SVD decomposition on $\widehat{F}^{(0)}$, and update $\widehat{F}^{(0)}$ to ensure $\frac{1}{T}\widehat{F}^{(0)'} \widehat{F}^{(0)}=I$.

\item[Step $j$:] For Step $j\ (\ge 1)$, obtain $\widehat{B}^{(j)} =(\widehat{\beta}_{1}^{(j)},\ldots, \widehat{\beta}_{N}^{(j)})'$ and $\widehat{\Gamma}^{(j)} =(\widehat{\gamma}_{1}^{(j)},\ldots, \widehat{\gamma}_{N}^{(j)})'$ by maximizing 

\begin{eqnarray*}
\log L^{(j)} (\beta_i,\gamma_i) &=& \sum_{t=1}^T\Big\{(1-y_{it})\log\left[1-G_\varepsilon(x_{it}'\beta_i+\gamma_{i}'\widehat{f}_{t}^{(j-1)})\right] \\
&&+y_{it} \log G_\varepsilon(x_{it}'\beta_i+\gamma_{i}'\widehat{f}_{t}^{(j-1)}) \Big\}
\end{eqnarray*}
over all $i\ge 1$, where $\widehat{F}^{(j-1)} =(\widehat{f}_1^{(j-1)},\ldots, \widehat{f}_T^{(j-1)})'$ is obtained from the Step $j-1$. Then obtain $\widehat{F}^{(j)} =(\widehat{f}_{1}^{(j)},\ldots, \widehat{f}_{N}^{(j)})'$ by maximizing 

\begin{eqnarray*}
\log L^{(j)} (f_t) &=& \sum_{i=1}^N\Big\{(1-y_{it})\log\left[1-G_\varepsilon(x_{it}'\widehat{\beta}_i^{(j)}+\widehat{\gamma}_{i}^{(j)\prime}f_{t})\right] \\
&&+y_{it} \log G_\varepsilon(x_{it}'\widehat{\beta}_i^{(j)}+\widehat{\gamma}_{i}^{(j)\prime}f_{t}) \Big\}
\end{eqnarray*}
over all $t\ge 1$. Finally, implement the SVD decomposition on $\widehat{F}^{(j)}$, and update $\widehat{F}^{(j)}$ to ensure $\frac{1}{T}\widehat{F}^{(j)'} \widehat{F}^{(j)}=I$.

\item[Stop:] Stop after reaching certain criterion (say, $\frac{1}{\sqrt{N}}\|\widehat{B}^{(j)}-\widehat{B}^{(j-1)}\|\le \epsilon$ in which $\epsilon$ is a sufficiently small number).
\end{enumerate}

Compared to \cite{BL2017}, the above procedure is indeed more time-consuming practically. By utilizing the structure of the regressors, the CCE-type of approach is much appreciated for the computational efficiency. Thus, there is a trade-off between the flexibility of the model and the computational efficiency. In the literature, a few studies aim to justify the aforementioned algorithms theoretically. The early work probably dates back to \cite{Chen2014}, and recently, \cite{Liu2020} and \cite{JYGH2020} further provide theoretical evidences for a semiparametric model and a parametric model respectively. In this paper, we do not pursue any theoretical results along this line of research, as it may lead to a different paper. Finally, in practice, one may follow \cite{Chen2014} to run the algorithm for several initial values and choose the solution that yields the highest value of the log-likelihood.

\subsection{The Asymptotic Results}\label{Section2.2}

The following conditions are necessary before we present the asymptotic results in this subsection.

\begin{assumption}\label{Ass0}
\item

\begin{enumerate}[leftmargin=*]
\item Suppose that $\{\varepsilon_{it}\}$ is an array of identically distributed random variables in $(i,t)$ with known PDF and CDF as $g_\varepsilon(\cdot)$ and $G_\varepsilon(\cdot)$, respectively. 

\item There exists a set $\Xi_{NT}=[\Xi_{NT}^l, \Xi_{NT}^u]$ such that all $z_{it}^0$'s belong to $\Xi_{NT}$ with probability approaching to 1, and $0<G_\varepsilon(\Xi_{NT}^l)<G_\varepsilon(\Xi_{NT}^u)<1$, where $z_{it}^0 =x_{it}'\beta_{0i} +\gamma_{0i}'f_{0t}$.

\item Let $\frac{1}{NT}\sum_{i,j=1}^N\sum_{t,s=1}^T E[ |E[e_{it}e_{js}\, | \, w_{it}^0, w_{js}^0]|  ]=O(1)$, where $w_{it}^0=(x_{it}, \gamma_{0i}, f_{0t})$ and $e_{it} =\frac{1-y_{it}}{1- G_\varepsilon(z_{it}^0)}- \frac{y_{it}}{G_\varepsilon(z_{it}^0)}$.
\end{enumerate}
\end{assumption}

Assumption \ref{Ass0}.1 requires the identical distribution, which is conventional in the literature on nonlinear models (e.g., Assumption 1 of \citealp{Chen2020}). Also, Remark \ref{RM21} partially explains why this condition is reasonable from the perspective of identification.

For Assumption \ref{Ass0}.2, the range of $\Xi_{NT}$ varies respect to the distribution considered. If a distribution is defined on $\mathbb{R}$ (say a normal distribution), we may allow the lower and upper bounds of $\Xi_{NT}$ to diverge to $\pm \infty$ respectively; if an exponential distribution is considered, we may let the lower bound converge to 0 and let the upper bound to diverge; etc.  The design of $\Xi_{NT}$ is not new in the literature. For instance, both \cite{chenxh2015} and \cite{LTG2016} use a similar technique to accommodate some unbounded supports of the regressors.

The current form of Assumption \ref{Ass0}.3 allows for a certain type of weak cross-sectional dependence and time series correlation. Note that  $E[e_{it}\, |\, w_{it}^0] =0$ by construction, so one can regard $e_{it}$ as a newly created residual term with mean 0. Assumption \ref{Ass0}.3 essentially imposes a restriction on the second moment of $e_{it}$, which can be verified for example for the  independent and identically distributed (i.i.d.) case or for the case involving certain mixing conditions (e.g., Assumption \ref{Ass2}.3 below).

\begin{lemma}\label{Lemma2.1} 
Under Assumption \ref{Ass0}, as $(N,T)\to (\infty,\infty)$,   

\begin{eqnarray*}
\frac{1}{NT}\sum_{i=1}^N\sum_{t=1}^T \left[G_\varepsilon(\widehat{z}_{it} )  - G_\varepsilon(z_{it}^0 )\right]^2 =O_P\left(\frac{1}{\sqrt{NT}}\right),
\end{eqnarray*}
where $z_{it}^0$ is defined in Assumption \ref{Ass0}, $\widehat{z}_{it} = x_{it}'\widehat{\beta}_i + \widehat{\gamma}_i'\widehat{f}_t$, and $\widehat{\beta}_i$, $\widehat{\gamma}_i$, and $\widehat{f}_t$ are defined in \eqref{EQ28}.
\end{lemma}

With very limited restrictions, Lemma \ref{Lemma2.1} ensures the overall validity of the approach considered in this paper. It is noteworthy that Lemma \ref{Lemma2.1} still holds even for a model without regressors:

\begin{eqnarray}\label{newmodel}
y_{it} = \left\{ \begin{array}{cc}
1, & \gamma_{0i}'f_{0t} -\varepsilon_{it} \ge 0 \\
0, & \text{otherwise}
\end{array}\right.,
\end{eqnarray}
which is one of the models studied in \cite{Wang2019}. Certainly, the maximum likelihood function should be adjusted in a very obvious manner. We conjecture that Lemma \ref{Lemma2.1} can help simplify the asymptotic development (and maybe assumptions) of \cite{Wang2019}.

In addition, Lemma \ref{Lemma2.1} infers that a maximum likelihood estimation can reduce to a nonlinear least squares estimation more or less. We provide a few examples below.

\medskip

\noindent \textbf{Example 1:} Interestingly, if $\varepsilon_{it}$ follows a uniform distribution, then the expression presented by Lemma \ref{Lemma2.1} completely possesses a form of the least squares approach:

\begin{eqnarray}
\frac{1}{NT}\sum_{i=1}^N\sum_{t=1}^T \left[G_\varepsilon(\widehat{z}_{it} )  - G_\varepsilon(z_{it}^0 )\right]^2 \equiv \frac{1}{NT}\sum_{i=1}^N\sum_{t=1}^T(\widehat{z}_{it}  -z_{it}^0 )^2.
\end{eqnarray}
As a result, most of the arguments made for the term $\widetilde{S}_{NT}(\beta, F)$ on pages 1264-1265 of \cite{Bai2009} will apply. We refer the interested readers to detailed discussions therein.

\medskip

\noindent \textbf{Example 2:} If $\Xi_{NT}$ of Assumption \ref{Ass0} is a compact set with fixed boundaries and $\inf_{z\in \Xi_{NT}} g_\varepsilon (z)\ge c_0 >0$, Lemma \ref{Lemma2.1} immediately yields that

\begin{eqnarray}\label{Example2}
O_P\left(\frac{1}{\sqrt{NT}}\right) =\frac{1}{NT}\sum_{i=1}^N\sum_{t=1}^T \left[G_\varepsilon(\widehat{z}_{it} )  - G_\varepsilon(z_{it}^0 )\right]^2\ge c_0\cdot\frac{1}{NT}\sum_{i=1}^N\sum_{t=1}^T ( \widehat{z}_{it}  - z_{it}^0)^2,
\end{eqnarray}
in which the right hand side again reduces a term identical to  $\widetilde{S}_{NT}(\beta, F)$ of \cite{Bai2009} ignoring the constant $c_0$. One in fact can allow $\Xi_{NT}=[\Xi_{NT}^l, \Xi_{NT}^u]$ to have diverging boundaries, and allow $g_\varepsilon(\cdot)$ to be a density function of normal distribution. If that is the case, we need an extra condition (i.e., Assumption \ref{Ass1}.1 below) to further regulate $\Xi_{NT}$ in order to achieve asymptotic consistency for all estimators of \eqref{EQ28}. Simple algebra shows that under the condition $|\Xi_{NT}^l|+|\Xi_{NT}^u|= o_P(\sqrt{\log (NT)})$, Assumption \ref{Ass1}.1 is fulfilled, so the newly proposed model with the estimation procedure still holds.

\begin{remark}\label{RM2.2}
We now discuss the selection of the number of factors before proceeding further. The topic has been studied in a variety of papers for the parametric linear models (e.g., \citealp{BN2002, Onatski, LamYao2012, AhnHorenstein2013}; and references therein). Among the results and discussions available in the literature, an important one is that many asymptotic results still hold true (but losing some efficiency), if the number of factors is over-specified when conducting the regression (e.g., \citealp{FLM13, Moon}). In this study, we find that such an argument is also valid for the binary response panel data model of \eqref{EQ21} under moderate restrictions. For simplicity, consider the Example 2 above. It is clear that the right hand side of \eqref{Example2} completely reduces to a parametric model, so the arguments made for the parametric linear models apply immediately.
\end{remark}

\medskip

We are now ready to investigate the estimators of \eqref{EQ28}, and emphasize again that we temporarily assume the number of factors is known, and  work on the selection of the number of factors in Section \ref{Section2.3}. The next assumption is essential in order to achieve consistency for each estimator of \eqref{EQ28}.

\begin{assumption}\label{Ass1}
\item 
\begin{enumerate}[leftmargin=*]
\item Suppose that there exists a sequence $\{a_{NT}\}$ satisfying that $\inf_{w\in\Xi_{NT}}g_\varepsilon(w)\ge a_{NT} >0$ and $a_{NT}\sqrt{NT}\to \infty$, in which $a_{NT}$ may converge to 0 or be constant.

\item Let $Z_i =  X_i(\beta_{0i}-\beta_i ) $, and suppose that the following limits exist given $\| B-B_0\|/\sqrt{N}\le C$, where $C$ is a sufficiently large positive constant.

\begin{enumerate}
\item $\sup_{B} \left|\frac{1}{NT}\sum_{i=1}^N (Z_i\otimes Z_i  -E[Z_i\otimes Z_i]) \right| =o_P(1)$;

\item $\sup_{B} \left| \frac{1}{N\sqrt{T}}\sum_{i=1}^N( \gamma_i\otimes  Z_i  - E[\gamma_i\otimes  Z_i]) \right| =o_P(1)$;

\item $\frac{1}{N}\Gamma_{0}'\Gamma_{0}\to_P\Sigma_\gamma$ and $\frac{1}{T}F_0'F_0\to_P\Sigma_f$.
\end{enumerate}

\item Suppose that $0<\inf_{  F\in \mathbf{F} } \Omega (F )$, where $\Omega(F )=\diag\{\frac{1}{T}\Omega_{1T}(F ),\ldots, \frac{1}{T}\Omega_{NT}(F ) \}$, and

\begin{eqnarray*}
\Omega_{iT}(F) &=& E[X_i' M_F X_i\, | \, F ]-E[\gamma_{0i}\otimes (M_F X_i )\, | \, F ]' (\Sigma_\gamma\otimes I_T)^{-1}E[\gamma_{0i}\otimes (M_F X_i )\, | \, F ] .
\end{eqnarray*}
\end{enumerate}
\end{assumption}

Assumption \ref{Ass1}.1 bounds the PDF from below, which is similar to the treatments of \cite{Hansen2008} and \cite{LLL2012} among others. The two examples above should have clearly explained why such a condition is needed. The first two results of Assumption \ref{Ass1}.2 require some uniform consistency, which is due to the fact that the coefficients are indexed by $i$. For the homogeneous coefficient model, such  conditions can be completely removed. These two conditions are similar to Assumption G of \cite{LCL2020}, wherein heterogeneous coefficients are considered as well. Assumption \ref{Ass1}.3 is the identification restriction accounting for the heterogeneous setting of the model which is in the same spirit as Assumption D of \cite{Ando} and the condition imposed on the term $Q_1(F_1)$ of \cite{HJPS}. The reason why Assumption \ref{Ass1}.3 is necessary should be crystal clear in view of the two examples under Lemma \ref{Lemma2.1}.

With Assumption \ref{Ass1} in hand, we summarize the asymptotic consistency in the following lemma.

\begin{lemma} \label{Lemma2.2} 
Under Assumptions \ref{Ass0} and \ref{Ass1}, as $(N,T)\to (\infty,\infty)$,

\begin{enumerate} 
\item $\frac{1}{N}\|\widehat{B}-B_0\|^2 =o_P(1)$,

\item $\frac{1}{NT}  \|\widehat{F}\widehat{\Gamma}'-F_0\Gamma_{0}'\|^2 =o_P(1)$,

\item $\|P_{\widehat{F}} -P_{F_0} \| =  o_P(1)$.
\end{enumerate}
\end{lemma}

In Lemma \ref{Lemma2.2}, the first two results guarantee the consistency of the estimators in \eqref{EQ28}, while the third result infers that the space spanned by the columns of $F_0$ can be recovered consistently.

\medskip

To carry on our analysis further, more structures and notations are needed:

\begin{eqnarray}\label{def_omegas}
l_{it}(w) &=& (1-y_{it})\log\left[1-G_\varepsilon(w)\right] +y_{it} \log G_\varepsilon(w),\nonumber \\
\Sigma_{u,i} &=& \lim_{T\rightarrow\infty}\frac{1}{T}\sum_{t=1}^T  E\left[\frac{[g_\varepsilon(z_{it}^0)]^2 }{[1-G_\varepsilon(z_{it}^0)]G_\varepsilon(z_{it}^0)}  u_{it}^0u_{it}^{0\prime}\right],\nonumber \\
\Sigma_{\gamma,t} &=& \lim_{N\rightarrow\infty}\frac{1}{N}\sum_{i=1}^N E\left[\frac{[g_\varepsilon(z_{it}^0)]^2 }{[1-G_\varepsilon(z_{it}^0)]G_\varepsilon(z_{it}^0)}   \gamma_{0i}\gamma_{0i}'\right],\nonumber \\
\Omega_{u\gamma, i t} &=& E\left[\frac{[g_\varepsilon(z^0_{it})]^2 }{[1-G_\varepsilon(z^0_{it})]G_\varepsilon(z^0_{it})}  u^0_{it}\gamma_{0i}'\right],
\end{eqnarray}
where $u_{it}^0=(x_{it}',f_{0t}')'$. Moreover, let $\Omega_u=\diag\{\Sigma_{u,1},\cdots,\Sigma_{u,N}\}$, $\Omega_\gamma=\diag\{\Sigma_{\gamma,1},\cdots,\Sigma_{\gamma,T}\}$, and $\Omega_{u\gamma} =\{\Omega_{u\gamma, i t}\}_{N(d_\beta+d_f)\times Td_f}$.  For the sake of space, we explain the necessity of these notations in Appendix \ref{Appnot}. 

The following set of conditions are necessary to derive the rates of convergence.

\begin{assumption}\label{Ass2}

\item  
\begin{enumerate}[leftmargin=*]
\item Let $ \max_{i\ge 1,t\ge 1}E\| x_{it}\|^{4+\delta}<\infty$, $\max_{i\ge 1}E\| \gamma_{0i}\|^{4+\delta}<\infty$, and $\max_{t\ge 1}E\| f_{0t}\|^{4+\delta}<\infty$, for some constant $\delta\geq2$. Also, suppose that $\max_{i\ge 1,t\ge 1}\| x_{it}\|=O_P(\log (N T))$, $\max_{i\ge 1}\| \gamma_{0i}\|=O_P(\log N)$ and $\max_{t\ge 1}\| f_{0t}\|=O_P(\log T)$. Let $F_0 \in \mathbf{F}$, and suppose that there exists $\delta^\ast\in(0,\delta)$ such that $\frac{T}{N^{1+\delta^\ast/4}}\rightarrow0$ and $\frac{N}{T^{1+\delta^\ast/4}}\rightarrow0$.

\item  $g_\varepsilon(w)$ is twice differentiable on $\Xi_{NT}$,  and $\sup_{w\in \Xi_{NT}} (|l^{(2)}_{it}(w)|+|l^{(3)}_{it}(w)|)<\infty$ uniformly in $(i,t)$, where $l^{(k)}_{it}(w)$ is the $k^{th}$ derivative of $l_{it}(w)$, Moreover, there exist $\rho_1,\rho_2<0$ such that 
\begin{eqnarray*}
\sup_{w\in \Xi_{NT}} \rho_{\max}\left(\frac{1}{T}\sum_{t=1}^T l^{(2)}_{it}(w)u_{it}^0u_{it}^{0'}\right)\leq \rho_1, \, \sup_{w\in \Xi_{NT}} \rho_{\max}\left(\frac{1}{N}\sum_{i=1}^N l^{(2)}_{it}(w)\gamma_{0i}\gamma_{0i}^\top\right)\leq \rho_2,
\end{eqnarray*} 
with probability one, where $\rho_{\max}(\cdot)$ has been defined in the last paragraph of Section \ref{Section1}.

\item $\{\varepsilon_{it}\}$ is independent of $\{(x_{it}, \gamma_{0i}, f_{0t}): i\geq 1, \, t\geq 1\}$. Let $\{\varepsilon_{it}, x_{it},f_{0t}\}$ be strictly stationary and $\alpha$-mixing across $t$, and let $\alpha_{ij}(|t-s|)$ represent the  $\alpha$-mixing coefficient. Moreover, assume that $\sum_{i,j=1}^N\sum_{t=1}^\infty(\alpha_{ij}(t))^{\delta/(4+\delta)}=O(N)$, $\sum_{i,j=1}^N (\alpha_{ij}(0))^{\delta/(4+\delta)}=O(N)$, and $\max_{i\geq 1} \sum_{t=1}^\infty (\alpha_{ii}(t))^{\delta/(4+\delta)}=O(1)$.

\item Suppose that $\frac{a^2_{NT}\sqrt{NT}}{[\log (NT)]^2}\rightarrow \infty$, $\rho_{\max}(\Omega_{u})<\infty$, $\rho_{\max}(\Omega_{\gamma})<\infty$, $\rho_{\max}\left(\frac{1}{NT}\Omega_{u\gamma}\Omega_{\gamma}^{-1}\Omega_{u\gamma}'\right)<\infty$, and $\rho_{\max}\left(\frac{1}{NT}\Omega_{u\gamma}'\Omega_{u}^{-1}\Omega_{u\gamma}\right)<\infty$.
\end{enumerate}
\end{assumption}

Assumption \ref{Ass2}.1 imposes moments restrictions on $x_{it}$, $\gamma_{0i}$ and $f_{0t}$, which are standard in the literature. In addition, the rates $\log (NT)$, $\log N$ and $\log T$ can be easily fulfilled when measuring the maximum value of a series of random observations. See Assumption A7 of \cite{CHL2012} for example. We restrict the divergence  of $N$ and $T$ by $\frac{T}{N^{1+\delta^\ast/4}}\rightarrow0$ and $\frac{N}{T^{1+\delta^\ast/4}}\rightarrow0$ which can be satisfied in many cases such as $N/T\rightarrow c$ where $c$ is a constant. This condition is used to establish the uniform convergence of the estimators (i.e., the first two results of Lemma \ref{Lemmauni} below). The condition $F_0 \in \mathbf{F}$ is for the purpose of identification only. For instance, given $F_0$, we can always find a rotation matrix $W$ such that

\begin{eqnarray}
\frac{1}{T}W'F_0'F_0W =I_{d_f}.
\end{eqnarray}
As a consequence, we can write

\begin{eqnarray}
\gamma_{0i}'f_{0t} = \gamma_{0i}'W^{-1}\cdot Wf_{0t},
\end{eqnarray}
which infers that instead of having $\gamma_{0i} $  and $f_{0t} $ as true parameters, we in fact use $ \gamma_{0i}'W^{-1}$ and $ Wf_{0t}$ as true parameters under Assumption \ref{Ass2}.1.  For linear models (e.g., \citealp{Bai2009} among others), a rotational matrix usually kicks in through the PCA procedure. However, for nonlinear models without closed-form estimators as in this paper, achieving an analytic form of such a rotation matrix seems to be impossible to the best of our knowledge. Similar issues can also be seen in \cite{AB2020}, and \cite{AL2020}. 

Assumption \ref{Ass2}.2 slightly relaxes Assumption 2 of \cite{Wang2019} and Assumption 1.iv of \cite{Chen2020}. The Probit and Logit models are obviously covered by this assumption.

Assumptions \ref{Ass2}.3 strengthens Assumption \ref{Ass0}.3 by imposing mixing conditions. This idea is consistent with Assumption A.2 of \cite{SC2013} and Assumption \ref{Ass2}.3 of \cite{FPSY2019}. We can then establish the asymptotic distribution of $\widehat{\theta}_i$ from this set of low level conditions. 

Assumption \ref{Ass2}.4 is used for the derivation involving the inverse of the Hessian matrix. It is mild since both $\Omega_\gamma$ and $\Omega_u$ are diagonal matrices. Using this assumption, we can show that the diagonal elements in the  Hessian matrix play a dominating role when studying the asymptotic properties of the estimators.

\medskip

We are now ready to present the rates of convergence in the next lemma.

\begin{lemma}\label{Lemmauni}
Under Assumptions \ref{Ass0}-\ref{Ass2}, as $(N,T)\to (\infty,\infty)$, 

\begin{enumerate}
\item $\max_{i\geq 1}\|\widehat{\theta}_i-\theta_{0i}\|=o_P(1)$,
\item $\max_{t\geq 1}\|\widehat{f}_t-f_{0t}\|=o_P(1)$,
\item $\frac{1}{N} \|\widehat{\Theta} -\Theta_0 \|^2 = O_P\left(\frac{1}{N\wedge T}\right)$,
\item $\frac{1}{T}\|\widehat{F}-F_0 \|^2 =O_P\left(\frac{1}{N\wedge T}\right)$,
\end{enumerate}
where $\widehat{\Theta} = (\widehat{B},\widehat{\Gamma})$ and $\Theta_0 $ is defined under \eqref{EQBFG}.
\end{lemma}

In Lemma \ref{Lemmauni}, the first two results strengthen Lemma \ref{Lemma2.2}, and show the consistency of $\widehat{\theta}_i$ and $\widehat{f}_t$ uniformly in $i$ and $t$. Based on these two results, we are able to establish the rates of convergence in the last two results of Lemma \ref{Lemmauni}, which lead to the derivation of the asymptotic distributions below. 

\medskip

To  present the asymptotic distributions, the following conditions are necessary.

\begin{assumption}\label{Ass3}
Assume that $\Sigma_{u,i}$, $\Sigma_{\theta,i}$, and $\Sigma_{\gamma,t}$ are positive definite matrices for $\forall i$ and $\forall t$, where $\Sigma_{\theta,i}=\lim_{T\rightarrow\infty} \frac{1}{T}\sum_{t=1}^T\sum_{s=1}^T E [g_\varepsilon(z^0_{it}) g_\varepsilon(z^0_{is}) e_{it}e_{is} u^0_{it}u_{is}^{0\prime}]$.
\end{assumption}

Assumption \ref{Ass3} guarantees the positive definiteness of the asymptotic covariances.  Under the mixing conditions of Assumption \ref{Ass2}.3, we can show that $\Sigma_{\theta,i}$ is the asymptotic covariance matrix of $\frac{1}{\sqrt{T}}\frac{\partial \log L(B_0, F_0,\Gamma_0)}{\partial \theta_i}$. Thus, we are able to establish the asymptotic distributions in the following theorem.

\begin{theorem}\label{Theorem2.1}
Let Assumptions \ref{Ass0}-\ref{Ass3} hold, and $(N,T)\to (\infty,\infty)$.

\begin{enumerate}
\item If $T(\log T)^2/N^2\rightarrow0$, then 

\begin{eqnarray*}
\sqrt{T}(\widehat{\theta}_i -\theta_{0i}) \to_D N(0,\Sigma_{u,i}^{-1}\Sigma_{\theta,i}\Sigma_{u,i}^{-1}),
\end{eqnarray*}
where $\widehat{\theta}_i =(\widehat{\beta}_i',\widehat{\gamma}_i')'$ and $\theta_{0i} =(\beta_{0i}',\gamma_{0i}')'$.

\item If $N(\log N)^2/T^2\rightarrow0$ and $\frac{1}{\sqrt{N}}\frac{\partial \log L(B_0, F_0,\Gamma_0)}{\partial f_t}\to_D N(0,\Sigma_{f,t})$ for $\forall t$, then

\begin{eqnarray*}
\sqrt{N}(\widehat{f}_t -f_{0t}) \to_D N(0,\Sigma_{\gamma,t}^{-1}\Sigma_{f,t}\Sigma_{\gamma,t}^{-1}).
\end{eqnarray*}
\end{enumerate}
\end{theorem}

The conditions $T(\log T)^2/N^2\to 0$ and $N(\log N)^2/T^2\to 0$ in the body of the theorem are similar to those in Theorem 1 of \cite{BN2013}. One may regard the above theorem as the binary response counterpart of Theorem 1 of \cite{BN2013}. 

If the error terms $\varepsilon_{it}$ are i.i.d., we have the following consistent estimators for the unknown matrices involved in Theorem \ref{Theorem2.1}:
\begin{eqnarray}\label{EstCov}
\widehat{\Sigma}_{\theta,i} &=&\frac{1}{T}\sum_{t=1}^T  \mathsf{g}_{it} (\widehat{z}_{it})^2  \widehat{u}_{it} \widehat{u}_{it}',\quad \widehat{\Sigma}_{u,i}=\frac{1}{T}\sum_{t=1}^T \mathfrak{g} (\widehat{z}_{it})\widehat{u}_{it} \widehat{u}_{it}^{\prime},\nonumber \\ 
\widehat{\Sigma}_{f,t} &=& \frac{1}{N}\sum_{i=1}^N \mathsf{g}_{it} (\widehat{z}_{it})^2 \widehat{\gamma}_i \widehat{\gamma}_i',\quad\widehat{\Sigma}_{\gamma,t}=\frac{1}{N}\sum_{i=1}^N  \mathfrak{g} (\widehat{z}_{it})\widehat{\gamma}_i \widehat{\gamma}_i',
\end{eqnarray}
where $\widehat{u}_{it}=(x_{it}',\widehat{f}_t')'$, $\mathsf{g}_{it} (w) =\frac{[y_{it}-G_\varepsilon(w )] g_\varepsilon(w)  }{[1-G_\varepsilon(w)]G_\varepsilon(w)}$, and $\mathfrak{g} (w)=   \frac{[g_\varepsilon(w)]^2 }{[1-G_\varepsilon(w)]G_\varepsilon(w )}  $. While there are weak serial correlation or cross-sectional dependence involved in $\varepsilon_{it}$'s, the constructions of $\widehat{\Sigma}_{\theta,i} $ and $\widehat{\Sigma}_{u,i}$ need to be adjusted accordingly, which in fact is a quite complex problem itself even just in the literature of time series analysis, see, Chapter 2 of \cite{FanYao}, for example. Therefore, we do not further purse these estimators when weak correlation is involved over $(i,t)$.

\medskip

Finally, we consider a mean group type of estimator for $\beta_{0i}$'s, which is often studied when heterogeneous coefficients are involved. Traditionally, the models with heterogeneous coefficients always assume that

\begin{eqnarray}\label{eqPS}
\beta_i = \beta_0 +\eta_i,
\end{eqnarray}
where $\eta_i$ is i.i.d. over $i$ and has mean 0. One of the most cited works is \cite{Pesaran2006}. As $\eta_i$ is only indexed by $i$, the estimate of $\beta_0$ is always achieved at a slow rate $\frac{1}{\sqrt{N}}$. 

Interestingly, from the perspective of hypothesis testing, another strand of the literature considers the so-called ``small departure" (e.g., Assumption 3 of \citealp{goncalves_2011} and Eq. (3.5) of \citealp{zhang2012inference}), which can be formalized using the following assumption.

\begin{assumption}\label{Ass4}
There is a vector of unknown parameters-of-interest $\beta_0$ such that $\beta_{0i} = \beta_0 + \frac{1}{N^{\alpha}} \, \eta_i$, where $0\le \alpha\leq \frac{1}{2}$, and $\eta_i$ is a vector of independent and identically distributed (i.i.d.) random errors with $E[\eta_i]=0$ and ${\rm Var}[\eta_i] =\Sigma_{2}>0$. Moreover, $\{\eta_i\}$ is independent of $\{(x_{it}, \varepsilon_{it},\gamma_{0i}, f_{0t}): i\geq 1, \, t\geq 1\}$.
\end{assumption}

Assumption \ref{Ass4} imposes a localized version on $\beta_{0i}$, and it narrows down the usual ``departure" of the form: $\beta_{0i} = \beta_0 + \eta_i$ as commonly assumed in the relevant literature. If we do consider a testing problem such as $H_0: \beta_{0i} ={\beta}_0$, the choice of $\beta_{0i} = {\beta}_0 + \frac{1}{\sqrt{N}} \, \eta_i$ is naturally considered as a sequence of local alternatives with an optimal rate of convergence of an order $N^{-1/2}$ in the conventional parametric setting (see, for example, \cite{dong_gao_2018} for more details about testing small departures). In what follows, we aim to bridge both strands of the literature and explore under what condition an optimal rate $\frac{1}{\sqrt{NT}}$ can be achieved.

\medskip

That said, we now consider a mean group type of estimator for $\beta_{0i}$'s using Assumption \ref{Ass4}. Define ${\widehat{\beta}}= \frac{1}{N} \sum_{i=1}^N \widehat{\beta}_i$ and $\overline{\beta}_{0} = \frac{1}{N} \sum_{i=1}^N \beta_{i0}$. Observe that 
\bea 
\sqrt{T} \, N^{\alpha}  ({\widehat{\beta}} - {\beta}_0 ) &=& \sqrt{T} \, N^{\alpha} ({\widehat{\beta}} - \overline{\beta}_{0} + \overline{\beta}_{0} - {\beta}_0 ) 
\nonumber\\
&=& \frac{\sqrt{T}}{N^{1-\alpha}} \sum_{i=1}^N  (\widehat{\beta}_i-\beta_{i0} ) + \sqrt{\frac{T}{N}} \frac{1}{\sqrt{N}} \sum_{i=1}^N \eta_i.
\label{jiti2.1}
\eea
Equation (\ref{jiti2.1}) indicates that the fast rate of convergence is achievable under Assumption 5. Depending on the limiting behaviour of $\rho_1 \equiv \frac{\sqrt{T}}{N^{1-\alpha}}$ and $\rho_2 \equiv \sqrt{\frac{T}{N}}$, the form of the asymptotic distribution of $\sqrt{T} \, N^{\alpha}  ({\widehat{\beta}} - {\beta}_0 )$ may depend on both terms of (\ref{jiti2.1}). In the case of $\rho_2 \rightarrow 0$, the first term contributes to the asymptotic distribution. When $\rho_1\rightarrow 0$, the second term mainly contributes to the asymptotic distribution. In the case where $\frac{\rho_1}{\rho_2} \rightarrow c\in (0, \infty)$, both terms contribute to the asymptotic distribution. In the relevant literature under the standard setting: $\beta_{i0} = {\beta}_0 + \eta_i$, however, the second term always dominates the asymptotic distribution. We now establish the following results in Theorem \ref{Theorem2.2}.

\begin{theorem}\label{Theorem2.2}

Let Assumptions \ref{Ass0}-\ref{Ass4} hold, and $(N,T)\to (\infty,\infty)$.

\begin{enumerate}

\item
Consider the case of $0\le \alpha <\frac{1}{2}$. If  $\frac{N^{\alpha + \frac{1}{2}}}{T} \rightarrow 0$, then 

\begin{eqnarray*}
N^{\alpha + \frac{1}{2}}  ({\widehat{\beta}} - {\beta}_0 ) \rightarrow_D N(0, \Sigma_{2}).
\end{eqnarray*}

\item
Consider the case of $\alpha = \frac{1}{2}$. If $\frac{T}{N} \rightarrow \infty$, then  

\begin{eqnarray*}
N  ({\widehat{\beta}} - {\beta}_0 - {\rm bias}(N) ) \rightarrow_D N(0, \Sigma_{2}),
\end{eqnarray*}
where ${\rm bias}(N) = O_P\left(\frac{1}{N}\right)$.

\item
Consider the case of $\alpha = \frac{1}{2}$. If $\frac{T}{N} \rightarrow \rho \in (0, \infty)$, then as $(N, T)\rightarrow (\infty, \infty)$
\be
\sqrt{N\, T}  ({\widehat{\beta}} - {\beta}_0  - {\rm bias}(N, T) ) \rightarrow_D N(0, \Sigma_{12}),
\label{jiti2.4}
\ee
where $\Sigma_{12} = \Sigma_{1} + \rho^2 \, \Sigma_{2}>0$,  

\begin{eqnarray*}
\Sigma_1 = \lim_{N, T}   \frac{1}{NT}  \sum_{i,j=1}^N \sum_{t,s=1}^TE[ \mathsf{g}_{it} (z_{it}^0) \mathsf{g}_{it} (z_{js}^0) \Sigma_{u, i}^{(d_\beta)} u_{it}^0 u_{js}^{0\prime} \Sigma_{u, j}^{(d_\beta)\prime}],
\end{eqnarray*}
$\mathsf{g}_{it}(\cdot)$ is defined under \eqref{EstCov}, and $\Sigma^{(d_\beta)}_{u, i}$ includes the first $d_\beta$ rows of $\Sigma^{-1}_{u,i}$.
\end{enumerate}
\end{theorem}

Theorem \ref{Theorem2.2} shows that we can establish the asymptotic distributions for the case of $\frac{T}{N} \rightarrow \rho \in (0, \infty]$. For the case of $\frac{T}{N} \rightarrow 0$, however, we have not been able to establish an asymptotic distribution for ${\widehat{\beta}}$. This is because we cannot improve the higher-order term $O_P\left(\frac{1}{N\wedge T}\right)$ involved in the leading-order approximation:

\begin{eqnarray*}
\widehat{\beta}_i - \beta_{0i} =  \frac{1}{T} \sum_{t=1}^T \frac{[y_{it}-G_\varepsilon(z_{it}^0)]g_\varepsilon(z_{it}^0)\, \Sigma_{u, i}^{(d_{\beta})}}{[1-G_\varepsilon(z_{it}^0)]G_\varepsilon(z_{it}^0)}   u_{it}^0 + O_P\left(\frac{1}{N\wedge T}\right)
\end{eqnarray*}
as shown at the beginning of the proof of Theorem \ref{Theorem2.2}.

The first result of Theorem \ref{Theorem2.2} shows that the rate of convergence can be as fast as $N^{-(\alpha + \frac{1}{2})}$ for the case of $0\leq \alpha <\frac{1}{2}$ and $\frac{N^{\alpha + \frac{1}{2}}}{T} \rightarrow 0$. When $\alpha=0$, it reduces to the standard rate as established in the relevant literature (see, for example, \citealp{Pesaran2006}). The third result of Theorem \ref{Theorem2.2} shows that the conventional parametric rate of $(N T)^{-1/2}$ is achievable, but there is a bias term involved due to the ``trade-off" between the asymptotic bias and the asymptotic variance. In Appendix \ref{BS}, we further discuss how to deal with biases. When $\{\varepsilon_{it}\}$ is i.i.d. over $i$ and $t$, we may be able to estimate $\Sigma_1$ by

\begin{eqnarray*}
\widehat{\Sigma}_1 =\frac{1}{NT}\sum_{i=1}^N  \sum_{t=1}^T \mathsf{g}_{it} (\widehat{z}_{it}) ^2 \widehat{\Sigma}^{(d_\beta)}_{u, i} u_{it}^0  u_{it}^{0\prime}\widehat{\Sigma}^{(d_\beta)\prime}_{u, i}.
\end{eqnarray*}
The discussions made under \eqref{EstCov} still apply here.

We next move on to provide an information criterion to select the number of factors, and present a numerical implementation procedure.

\subsection{Selection of the Number of Factors}\label{Section2.3}

In connection with Lemma \ref{Lemma2.1}, we define the following information criterion:
\begin{eqnarray}\label{EQ211}
\ic(\mathsf{d})  = \frac{1}{NT}\sum_{i=1}^N \sum_{t=1}^T\left[y_{it} - G_\varepsilon (\widehat{z}_{it}^{\mathsf{d}})\right]^2 + \mathsf{d}\cdot \frac{ \xi_{NT}}{\sqrt{NT}},
\end{eqnarray}
where $\widehat{z}_{it}^{\mathsf{d}}$'s are obtained using \eqref{EQ28} by setting the number of factors as $\mathsf{d}$, $\xi_{NT}\to \infty$, and $ \frac{ \xi_{NT}}{\sqrt{NT}}\to 0$. We estimate $d_f$ by minimizing \eqref{EQ211}:
\begin{eqnarray}\label{EQ212}
\widehat{\mathsf{d}} =\argmin_{0\le \mathsf{d}\le d_{\max}} \ic (\mathsf{d}),
\end{eqnarray}
where $d_{\max} (\geq d_f)$ is a user specified large fixed constant. 

 In general, the criterion always has a form as follows.

\begin{eqnarray}\label{CIC}
\text{A measurement of estimation errors}\quad  + \quad \text{A penalty term}
\end{eqnarray}
Such a form has been widely adopted by a wide range of selection procedures, e.g., traditional AIC/BIC, LASSO (\citealp{HHM2008}), and factor analysis (\citealp{BN2002}). The logic is that when under-selection occurs, a bias or a contradictory result will arise. As a consequence, the measurement of estimation errors in \eqref{CIC} will be significantly large. When over-selection occurs, the estimation error will be asymptomatically  the same as the situation of the correct-selection. However, due to the efficiency loss caused by the over-selection, the penalty term comes to play, and will yield a rate of convergence much larger than that associated with the estimator errors. That is why $\xi_{NT}$ needs to satisfy certain conditions specified under \eqref{EQ211}. In the literature, a variety of log forms have been proposed for $\xi_{NT}$. Whether $\xi_{NT}$ has an optimal form remains unclear, but $\log\sqrt{N+T}$ works well for different data generating processes in our simulations below. Similar discussions have been provided to $g(N,T)$ of \cite{BN2002}. 

Last but not least, for the choice of the measurement of estimation errors, two well adopted forms are likelihood type of presentation (Eq. (2.9) of \citealp{ChuZhuWang}) or mean squared errors (Eq. (6) of \citealp{HHM2008}). In view of the development of Lemma \ref{Lemma2.1}, the two forms are almost equivalent in our case. We adopt the latter for simplicity, and summarize the asymptotic property  in the next theorem.
 
\begin{theorem}\label{Theorem2.3}
Under Assumptions \ref{Ass0} and \ref{Ass1}, suppose further that $\xi_{NT}\to \infty$ and $ \frac{ \xi_{NT}}{\sqrt{NT}}\to 0$. As $(N,T)\to (\infty,\infty)$, $\Pr(\widehat{\mathsf{d}}=d_f)\to 1$.
\end{theorem} 

It is worth mentioning that Theorem \ref{Theorem2.3} only requires very limited conditions, i.e., Assumptions \ref{Ass0} and \ref{Ass1}. For each given value of the number of factors, we can implement the estimation procedure of Section \ref{Section2.1}, which then allows us to calculate the information criterion \eqref{EQ211} to select the number of factors.

\section{Simulation}\label{Section3}

In this section, we conduct simulations to examine the theoretical findings of Section \ref{Section2}, and specifically consider the following data generating process.

\begin{eqnarray*}
y_{it} = \left\{ \begin{array}{cc}
1, & x_{it}'\beta_{0i}+\gamma_{0i}'f_{0t} -\varepsilon_{it} \ge 0 \\
0, & \text{otherwise}
\end{array}\right..
\end{eqnarray*}
The factors and loadings are generated by $f_{0t,j}\sim U(-2.5, 2.5)$ and $\gamma_{0i,j} \sim U(0, 6)$, where $\gamma_{0i,j}$ and $f_{0t,j}$ stand for the $j^{th}$ elements of $\gamma_{0i}$ and $f_{0t}$ respectively and $j=1,\ldots, d_f$. In order to introduce correlation between the regressors and the factor structure, we let $x_{it,j} = N(0,1)+0.5( |\gamma_{0i,1}|+|f_{0t,1}|)$, where $x_{it,j}$ stands for the $j^{th}$ element of $x_{it}$, and $j=1,\ldots, d_\beta$. For the coefficients, let $\beta_{0i,j} =i/N$, where $i=1,\ldots, N$, $j=1,\ldots, d_\beta$, and $\beta_{0i,j}$ stands for the $j^{th}$ element of $\beta_{0i}$. We consider two distributions for the error term, and for each distribution we vary the time series correlation and cross-sectional dependence of the error term in order to examine the sensitivity of the method proposed.

\medskip

\textbf{Case 1} -- light tailed distribution

\begin{itemize}
\item[] DGP 1:  $\varepsilon_{it} \sim N(0,1)$, where $N(0,1)$ is the standard normal distribution.

DGP 2: Let $\varepsilon_t =  \rho_\varepsilon \cdot \varepsilon_{t-1} +\Sigma_\nu^{1/2}\cdot\nu_t$, where $\rho_{\varepsilon} =0.3$, $\Sigma_\nu = \{0.3^{|i-j|}\}_{N\times N}$, $\varepsilon_t= (\varepsilon_{1t},\ldots, \varepsilon_{Nt})'$, and $\nu_t= (\nu_{1t},\ldots, \nu_{Nt})'$ with each $\nu_{it}$ being an independent draw from $N(0,1)$.

DGP 3: Let $\rho_\varepsilon$ of DGP 2 be 0.7, and keep the rest settings identical to those of DGP 2.
\end{itemize}

\textbf{Case 2} -- heavy tailed distribution

\begin{itemize}
\item[] We still consider three DGPs as in Case 1, but replace all normal distributions with logistic distributions.
\end{itemize}

\medskip

For each generated dataset, we first select the number of factors using the information criterion defined in \eqref{EQ211}, and then conduct the estimation. We repeat the above procedure $M$ times, and report the following values to evaluate the finite sample performance:

\begin{eqnarray}\label{measure}
&&P_c= \frac{1}{M}\sum_{j=1}^M I(\widehat{\mathsf{d}}_j =d_f), \quad P_u= \frac{1}{M}\sum_{j=1}^M I(\widehat{\mathsf{d}}_j <d_f),\quad P_o= \frac{1}{M}\sum_{j=1}^M I(\widehat{\mathsf{d}}_j >d_f),\nonumber \\
&& \text{RMSE}_{B_0} = \sqrt{\frac{1}{M}\sum_{j=1}^M\frac{1}{N} \|\widehat{B}_j-B_0 \|^2},\quad \text{RMSE}_{F_0} = \sqrt{\frac{1}{M}\sum_{j=1}^M  \| P_{\widehat{F}_j}- P_{F_0} \|^2},\nonumber \\
&& \text{Std}_{\beta_{0}^{(\ell)}} = \frac{1}{N}\sum_{i=1}^N \sqrt{ \frac{1}{M}\sum_{j=1}^M (\widehat{\beta}_{i,j}^{(\ell)} -\beta_{0i}^{(\ell)})^2 } ,
\end{eqnarray}
where $\widehat{\mathsf{d}}_j $, $\widehat{B}_j$ and $\widehat{F}_j$ stand for the estimated number of factors, the estimated value of $B_0=(\beta_{01},\ldots, \beta_{0N})'$, and the estimated value of $F_0=(F_{01},\ldots, F_{0T})'$ at the $j^{th}$ iteration respectively; and $\beta_{0i}^{(\ell)}$  and $\widehat{\beta}_{i,j}^{(\ell)} $ stand for the $\ell^{th}$ element of $\beta_{0i}$ and its estimate at the $j^{th}$ replication.  

We comment on these measures. $P_c$, $P_u$, and $P_o$ measure the probabilities of correctly, under, and over selecting the number of factors respectively. As explained in Section \ref{Section2}, $\widehat{F}_j$ yields a consistent estimation of $F_0$ up to a rotation matrix, so we measure the distance between $P_{\widehat{F}_j}$ and $ P_{F_0}$ in \eqref{measure}.  In stead of looking at $B_0$ as a whole, $\text{Std}_{\beta_{0}^{(\ell)}} $ examines the stability of the estimation procedure. Specifically, the quantity $ \sqrt{ \frac{1}{M}\sum_{j=1}^M (\widehat{\beta}_{i,j}^{(\ell)} -\beta_{0i}^{(\ell)})^2 } $ provides the standard deviation associated with the estimates of $\beta_{0i}^{(\ell)}$. As $i$ runs from 1 to $N$, we further take the average over $i$.

We let\footnote{Due to the restrictions of computational power, we no longer explore larger values of $M$. In practice, heavier tails require longer time to compute, which should be expected.} $d_\beta=2$, $d_f=2$, $N,T\in \{50,100, 150\}$, and $M=500$.  Moreover, let $\xi_{NT}$ of \eqref{EQ211} be $\log\sqrt{N+T}$ throughout the numerical studies without loss of generality. The results are summarized in Table \ref{table1} and Table \ref{table2} below. First, we point out that the newly proposed methodology works well regardless the tail behaviour of the error term $\varepsilon_{it}$, because the values associated with DGPs 1-3 are roughly same across both tables. Second, in Table \ref{table1}, the values of $P_c$ go up to 1, as the sample sizes increase. The pattern is same for DGPs 1-3. It is noteworthy that when the sample sizes are relatively small, the information criterion tends to under select the number of factors, which has a clear impact on the values of $\text{RMSE}_{F_0}$ presented in Table \ref{table2}. Third, in Table \ref{table2}, the values of $\text{RMSE}_{B_0}$, $\text{RMSE}_{F_0}$, and $ \text{Std}_{\beta_{0}^{(1)}} $ converge to 0 in general as the sample sizes go up, which should be expected. The exception is the case with $T=50$, in which the values of $ \text{Std}_{\beta_{0}^{(1)}} $ increase slightly as $N$ increases. Again, the pattern is same for DGPs 1-3. Fourth, compared to the Logit model, the Probit model tends to yield smaller values of Std$_{\beta_0}^{(\ell)}$, which is due to the fact that the Probit model has thin tails.

\begin{table}[h]
\scriptsize
\caption{The percentages of correctly, under and over identifying the number of factors. }\label{table1}
\begin{tabular}{lllrrrrrrrrrrr}
\hline \hline
&  &  & \multicolumn{3}{c}{$P_c$} & \multicolumn{1}{l}{} & \multicolumn{3}{c}{$P_u$} & \multicolumn{1}{l}{} & \multicolumn{3}{c}{$P_o$} \\
& & $N\setminus T$ & 50 & 100 & 150 &  & 50 & 100 & 150 &  & 50 & 100 & 150 \\
Case 1 & DGP 1 & 50 & 0.246 & 0.742 & 0.914 &  & 0.754 & 0.258 & 0.086 &  & 0.000 & 0.000 & 0.000 \\
 &  & 100 & 0.622 & 0.954 & 1.000 &  & 0.378 & 0.046 & 0.000 &  & 0.000 & 0.000 & 0.000 \\
 &  & 150 & 0.806 & 0.980 & 0.992 &  & 0.194 & 0.012 & 0.000 &  & 0.000 & 0.008 & 0.008 \\
 & DGP 2 & 50 & 0.258 & 0.770 & 0.924 &  & 0.742 & 0.230 & 0.074 &  & 0.000 & 0.000 & 0.002 \\
 &  & 100 & 0.634 & 0.964 & 0.998 &  & 0.366 & 0.034 & 0.002 &  & 0.000 & 0.002 & 0.000 \\
 &  & 150 & 0.822 & 0.988 & 1.000 &  & 0.174 & 0.006 & 0.000 &  & 0.004 & 0.006 & 0.000 \\
 & DGP 3 & 50 & 0.194 & 0.676 & 0.882 &  & 0.806 & 0.324 & 0.118 &  & 0.000 & 0.000 & 0.000 \\
 &  & 100 & 0.554 & 0.956 & 0.998 &  & 0.446 & 0.044 & 0.002 &  & 0.000 & 0.000 & 0.000 \\
 &  & 150 & 0.794 & 0.988 & 1.000 &  & 0.204 & 0.012 & 0.000 &  & 0.002 & 0.000 & 0.000 \\
Case 2 & DGP 1 & 50 & 0.236 & 0.748 & 0.934 &  & 0.764 & 0.252 & 0.066 &  & 0.000 & 0.000 & 0.000 \\
 &  & 100 & 0.696 & 0.992 & 1.000 &  & 0.304 & 0.008 & 0.000 &  & 0.000 & 0.000 & 0.000 \\
 &  & 150 & 0.804 & 1.000 & 1.000 &  & 0.000 & 0.000 & 0.000 &  & 0.000 & 0.000 & 0.000 \\
 & DGP 2 & 50 & 0.228 & 0.772 & 0.946 &  & 0.772 & 0.228 & 0.054 &  & 0.000 & 0.000 & 0.000 \\
 &  & 100 & 0.734 & 1.000 & 1.000 &  & 0.266 & 0.000 & 0.000 &  & 0.000 & 0.000 & 0.000 \\
 &  & 150 & 0.904 & 1.000 & 1.000 &  & 0.096 & 0.000 & 0.000 &  & 0.000 & 0.000 & 0.000 \\
 & DGP 3 & 50 & 0.138 & 0.510 & 0.808 &  & 0.862 & 0.490 & 0.192 &  & 0.000 & 0.000 & 0.000 \\
 &  & 100 & 0.486 & 0.930  & 0.998 &  & 0.514 & 0.070 & 0.002 &  & 0.000 & 0.000 & 0.000 \\
 &  & 150 & 0.778 & 0.994 & 1.000 &  & 0.222 & 0.006 & 0.000 &  & 0.000 & 0.000 & 0.000 \\
 \hline \hline
\end{tabular}
\end{table}

\begin{table}[h]
\scriptsize
\caption{Different measurements of the estimation procedure }\label{table2}
\hspace*{-1.4cm}\begin{tabular}{lllrrrrrrrrrrrrrrr}\hline\hline
 &  &  & \multicolumn{3}{c}{$\text{RMSE}_{B_0} $}  & \multicolumn{1}{l}{} &\multicolumn{3}{c}{$\text{RMSE}_{F_0}$} & \multicolumn{1}{l}{} & \multicolumn{3}{c}{$ \text{Std}_{\beta_{0}^{(1)}} $} & \multicolumn{1}{l}{} & \multicolumn{3}{c}{$ \text{Std}_{\beta_{0}^{(2)}} $} \\
 &  & N\textbackslash{}T & 50 & 100 & 150 &  & 50 & 100 & 150 &  & 50 & 100 & 150 &  & 50 & 100 & 150 \\
Probit & DGP 1 & 50 & 0.6456 & 0.5751 & 0.5327 &  & 1.0214 & 0.8483 & 0.7862 &  & 0.2896 & 0.2409 & 0.2140 &  & 0.2883 & 0.2397 & 0.2145 \\
 &  & 100 & 0.6394 & 0.5605 & 0.5174 &  & 0.8814 & 0.7249 & 0.6838 &  & 0.2936 & 0.2387 & 0.2073 &  & 0.2935 & 0.2373 & 0.2081 \\
 &  & 150 & 0.6365 & 0.5569 & 0.5159 &  & 0.8087 & 0.6771 & 0.6560 &  & 0.2947 & 0.2341 & 0.2073 &  & 0.2952 & 0.2335 & 0.2058 \\
 &  &  &  &  &  &  &  &  &  &  &  &  &  &  &  &  &  \\
 & DGP 2 & 50 & 0.6401 & 0.5640 & 0.5147 &  & 1.0135 & 0.8345 & 0.7709 &  & 0.2966 & 0.2429 & 0.2154 &  & 0.2942 & 0.2453 & 0.2173 \\
 &  & 100 & 0.6304 & 0.5458 & 0.5018 &  & 0.8649 & 0.7050 & 0.6722 &  & 0.3004 & 0.2417 & 0.2089 &  & 0.3025 & 0.2433 & 0.2102 \\
 &  & 150 & 0.6281 & 0.5397 & 0.4983 &  & 0.7742 & 0.6595 & 0.6372 &  & 0.3012 & 0.2382 & 0.2064 &  & 0.3021 & 0.2397 & 0.2064 \\
 &  &  &  &  &  &  &  &  &  &  &  &  &  &  &  &  &  \\
 & DGP 3 & 50 & 0.6499 & 0.5775 & 0.5286 &  & 1.0247 & 0.8502 & 0.7649 &  & 0.2885 & 0.2317 & 0.2025 &  & 0.2886 & 0.2332 & 0.2038 \\
 &  & 100 & 0.6402 & 0.5573 & 0.5128 &  & 0.8769 & 0.6764 & 0.6345 &  & 0.2916 & 0.2309 & 0.1980 &  & 0.2939 & 0.2332 & 0.1965 \\
 &  & 150 & 0.6360 & 0.5526 & 0.5104 &  & 0.7641 & 0.6211 & 0.5929 &  & 0.2957 & 0.2280 & 0.1931 &  & 0.2958 & 0.2281 & 0.1933 \\
 &  &  &  &  &  &  &  &  &  &  &  &  &  &  &  &  &  \\
Logit & DGP 1 & 50 & 0.6476 & 0.5490 & 0.4952 &  & 1.0282 & 0.8848 & 0.8233 &  & 0.4452 & 0.3869 & 0.3436 &  & 0.4470 & 0.3851 & 0.3427 \\
 &  & 100 & 0.6447 & 0.5335 & 0.4686 &  & 0.8547 & 0.6963 & 0.6700 &  & 0.4519 & 0.3730 & 0.3257  &  & 0.4521 & 0.3729 & 0.3259  \\
 &  & 150 & 0.6404 & 0.5272 & 0.4559 &  & 0.7695 & 0.6162 & 0.5975 &  & 0.4495 & 0.3680 & 0.3174 &  & 0.4493 & 0.3667 & 0.3173 \\
 &  &  &  &  &  &  &  &  &  &  &  &  &  &  &  &  &  \\
 & DGP 2 & 50 & 0.6405 & 0.5420 & 0.4858 &  & 1.0158 & 0.8501 & 0.7861 &  & 0.4250 & 0.3809 & 0.3330 &  & 0.4209 & 0.3807 & 0.3368 \\
 &  & 100 & 0.6394 & 0.5262 & 0.4608 &  & 0.7916 & 0.6517 & 0.6319 &  & 0.4499 & 0.3696 & 0.3143 &  & 0.4481 & 0.3686 & 0.3156 \\
 &  & 150 & 0.6382 & 0.5203 & 0.4484 &  & 0.6624 & 0.5805 & 0.5570 &  & 0.4476 & 0.3602 & 0.3145 &  & 0.4472 & 0.3621 & 0.3149 \\
 &  &  &  &  &  &  &  &  &  &  &  &  &  &  &  &  &  \\
 & DGP 3 & 50 & 0.6453 & 0.5474 & 0.4823 &  & 1.0384 & 0.9278 & 0.8310 &  & 0.4232 & 0.3659 & 0.3280 &  & 0.4234 & 0.3674 & 0.3281 \\
 &  & 100 & 0.6476 & 0.5327 & 0.4581 &  & 0.8854 & 0.6752 & 0.5989 &  & 0.4389 & 0.3661 & 0.3110 &  & 0.4387 & 0.3660 & 0.3078 \\
 &  & 150 & 0.6503 & 0.5254 & 0.4547 &  & 0.7222 & 0.5463 & 0.5181 &  & 0.4491 & 0.3590 & 0.3016 &  & 0.4499 & 0.3596 & 0.3006  \\
 \hline\hline
\end{tabular}
\end{table}

\section{A Case Study}\label{Section4}

In this section, we apply the model and methodology to portfolio analysis using daily returns data of S\&P 500 stocks. In both economics and finance disciplines, a vast amount of efforts have been devoted to portfolio analysis and management. Among them, a fundamental one is to select the optimal weights associated with each stock when constructing a portfolio. Mathematically, it is realized by the next minimisation problem.

\begin{eqnarray}\label{eq5.1}
\min_w \ w^\prime \Sigma_p w \quad\text{subject to} \quad w^\prime 1_{N} =1,
\end{eqnarray}
where $1_{N}$ is a $N\times 1$ vector of ones, $\Sigma_p $ is a positive definite matrix and is usually decided by the data of stock returns, and $w$ includes the weights assigned to each stock. The analytic solution to the above minimisation problem is

\begin{eqnarray}\label{eq5.2}
w = \frac{\Sigma_p^{-1} 1_{N}}{1_{N}^\prime\Sigma_p^{-1} 1_{N}}.
\end{eqnarray}

More often than not, one adopts all stocks to construct $\Sigma_p$ (e.g., \citealp{CLL2019, ELW2019}), which naturally falls into a category of high dimensional matrix estimation. Thereby, to boost the estimation accuracy, we see the increasing popularity of methods using rank reduction (\citealp{PX2019}), penalization (\citealp{CLL2019}), or both (\citealp{FLM13}) among others. By default, the aforementioned techniques account for all stocks in practice, though some of them may have relatively small weights compared to the others. As pointed out in \cite{CD2006} and \cite{Nyberg2011}, the sign of stock market returns may be predictable even if the returns themselves are not predictable.  Also, \cite{CD2006} mention that ``\textit{As volatility moves, so too does the probability of a positive return: the higher the volatility, the lower the probability of a positive return}". In connection with the fact that a primary goal of portfolio analysis is to minimize the volatility (\citealp{ELW2019}), a binary response panel data model with interactive fixed effects naturally marries the above studies by modelling the probabilities of positive returns, so we can drop those having low probabilities.

\subsection{Data}

The stock prices data are collected from \url{https://www.kaggle.com} over the time period between 2 January 2008 and 31 December 2018. After removing the companies which have missing stock returns during the whole time period, we end up with $319$ stocks ($N= 319$). We adopt the log-normalised CBOE volatility index (VIX) (\url{http://www.cboe.com}) as the regressor, which is widely viewed as a good indicator of market sentiment (e.g.,  \citealp{CD2006, PX2019}). Furthermore, we collect risk-free interest (RFI) data from the U.S. Department of the Treasury (\url{https://www.treasury.gov}) to construct the Sharpe ratio later on in order to evaluate the performance of the proposed method.

\subsection{Empirical Analysis}
Below, we conduct a rolling-window analysis, and focus on the out-of-sample forecast. For each window, we estimate $\Sigma_p$ with the sample information on the most recent 505 trading days (roughly two years) using the next model.

\begin{eqnarray*}
y_{i,t+1} = \left\{ \begin{array}{cc}
1, & x_{it}\beta_{0i}+\gamma_{0i}'f_{0t} -\varepsilon_{it} \ge 0 \\
0, & \text{otherwise}
\end{array}\right.,
\end{eqnarray*}
where 1 and 0 stand for positive and non-positive returns respectively. We always count the first available trading day of each rolling-window as 0, and use $t=0, \ldots, 503$ to implement estimation. $x_{it}$ includes the value of VIX at day $t$, and $y_{i,t+1}$ is the sign of stock $i$ at day $t+1$. For each estimation, we first select the number of factors, and then conduct the estimation. This will provide us the following quantities: $\widehat{\beta}_i$'s, $\widehat{\gamma}_i$'s and $\widehat{f}_t$ with $t=1,\ldots, 503$. Afterwards, we bring the value of $x_{it}$ at $t=504$ in the estimated model to forecast the probabilities of the stocks having positive returns at $t=505$. Since $f_{0t}$ with $t=504$ is still unknown, we replace it by $\widehat{f}_t$ with $t=503$ as an approximation. The reason is that although $f_{0t}$ may vary over $t$, we do not expect any sudden jump at any particular time point. Alternatively, to forecast the next period of the unknown factors, one may follow \cite{BBE} to impose more structure such as a VAR process. By doing so, one cannot select the number of factors in each estimation. Also, the number of optimal lags should be considered before one can conduct a more comprehensive investigation as in \cite{BBE}. In this study, we on longer pursue the empirical results along this line in order not to deviate from our main goal.

We keep the stocks with estimated probabilities  of positive returns greater than or equal to 0.5 to construct $\Sigma_p$ of \eqref{eq5.1}, and then calculate the weight vector $w$ using \eqref{eq5.2}. If all estimated probabilities are less than 0.5, then record $w=0$ (i.e., no transaction made for the day). Finally, we calculate the weighted average of returns for $t=505$. We repeat the above forecasting process from the first available window till the end, and consider the Probit model for the error term. The results of using other distributions (e.g., $t$-distribution and Logit model) for the error terms are quite similar, so we focus on the results of Probit model below. We follow \cite{ELW2019}  to consider the problem of estimating the global minimum variance portfolio, in the absence of short-sales constraints. 

As a comparison, when predicting the sign of  we also consider a model with fixed effects only  (referred to as FE below) and the model of \cite{BL2017} (referred to as BL below). In addition, we also consider two traditional approaches which utilize the entire stocks. Specifically, we consider the equal-weighted portfolio (referred to as EW below), which is a standard benchmark and has been promoted by \cite{DGU2007}, among others. Also, we use correlation matrix for $\Sigma_p$ (referred to as CM below), which is mentioned in \cite{ELW2019}. We acknowledge that many other methods are available when constructing the portfolio, such as the penalization method in \cite{FLM13}, the nonparametric approaches in \cite{CLL2019} and \cite{PX2019}, the dynamic covariance matrix estimation in \cite{ELW2019}, etc. It is extremely hard to exhaust all possible methods in one study, as it may lead to a comprehensive review paper. Thus, we no longer compare with these methods in this article.

In what follows, we report (1). the average of weighted returns across the entire available period (Mean), which is annualized by multiplying 252; (2). the standard deviation of weighted returns across the entire available period (Std), which annualized by multiplying $\sqrt{252}$; (3). the information ratio defined as the ratio of Mean to Std (IR); and (4) the Sharp ratio defined as the mean of returns minus the risk free interest normalised by Std (SR). The annualized Mean and Std are consistent with those defined in Section 6.2 of \cite{ELW2019}. We refer interested readers to their paper for more relevant discussions. Moreover, since a strand of literature on portfolio analysis is interested in estimation from a factor model with a predetermined number of factors (\citealp{PX2019}), we report the results for the cases where the number of factors are fixed as $1,\ldots, 5$ respectively. We also report the results of the case where IC of \eqref{EQ211} is used to select the optimal number of factors for each estimation. The results are summarized in Table \ref{OutofSample}.

\begin{table}[h]
\small
\caption{The results of the out-of-sample forecasts. ``FE" and ``IFE" refer to the results associated with the models with fixed effects and the models with interactive fixed effects respectively. For IFE models, ``Optimal" refers to the results when the number of factors is selected using the information criterion \eqref{EQ211} in each estimation.}\label{OutofSample}
 \begin{tabular}{lrrrrr}
\hline\hline
 & Factor No.  & Mean (\%) & Std & IR & SR \\
IFE & 1 & 14.92 & 11.56 & 1.29 & 1.26 \\
 & 2 & 15.98 & 13.82 & 1.16 & 1.13 \\
 & 3 & 13.38 & 13.40 & 1.00 & 0.97 \\
 & 4 & 13.61 & 12.82 & 1.06 & 1.03 \\
 & 5 & 13.84 & 12.61 & 1.10 & 1.07 \\
 & Optimal & 16.11 & 12.33 & \cellcolor{blue!25}1.31 & \cellcolor{blue!25}1.28 \\ \cline{2-6}
 &  &  &  &  &  \\
BL & & 9.82 & 10.46 & 0.94 & 0.90 \\
FE & & 8.56 & \cellcolor{blue!25}10.27 & 0.83 & 0.80 \\
EW & & 13.35 & 15.36 & 0.87 & 0.84 \\ 
CM & & \cellcolor{blue!25}24.54 & 19.96 & 1.23 & 1.21 \\
 \hline \hline
\end{tabular}
\end{table}

In general, we are looking for a strategy, which generates a small value of Std, but large values of Mean, IR and SR. The CM approach yields the largest Mean, but also has the largest Std, which should be least preferred due to the high risk caused by the high volatility. The FE model yields the smallest Std, which in a sense should be expected due to simplicity of the model. It also produces the minimum values of Mean, IR, and SR respectively. The IFE models (with the ``Optimal" number of factors and one factor respectively) outperform the rest models regarding the criteria IR and SR, which are commonly adopted to measure the performance of portfolios (e.g., \citealp{PX2019, ELW2019}).  Overall, the newly proposed framework has a reasonably good performance. 

Finally, we acknowledge the limit of the current empirical study. For example, one may adopt a VAR structure for the unobservable factors as in \cite{BBE}, and investigate further  the impulse responses and the optimal number of lags. Also, one may further conduct the penalization estimation for the stocks having high probability of getting positive returns, which bridges the literature of binary response models (e.g., \citealp{Chen2020, Wang2019}) and the literature of high-dimensional covariance matrix estimation (e.g., \citealp{FLM13}). In order not to deviate from our main goal, we do not pursue these results in the current study.

\section{Conclusion}\label{Section5}

In this paper, we investigate binary response models for heterogeneous panel data with interactive fixed effects by allowing both the cross-sectional dimension and the time dimension to diverge.  From the modelling perspective, our setting is similar to \cite{BL2017}, but we do not require a specific structure on the regressors, which allows us to avoid putting any restriction between the number of regressors and the number of unobservable factors. Our investigation establishes a link between a maximum likelihood estimation and a least squares approach. As a consequence, the identification restrictions provided in \cite{Bai2009} and \cite{Moon} are readily to be applied to binary response models with very minor modifications. We further establish asymptotic distributions for the unobservable factors and their loadings, which can be considered as the binary response counterpart of those established in \cite{BN2013}. In addition, we provide a simple information criterion to detect the number of factors. Last but not least, we conduct intensive numerical studies to examine the finite sample performance of the newly proposed model and methodology, and demonstrate the practical relevance. 

From a practical perspective, the framework can be applied to predict the probability of corporate failure (\citealp{CCL2014}), conduct credit rating analysis (\citealp{JJW2015}), etc. Following \cite{CD2006} and \cite{Nyberg2011}, in the empirical study, we focus on the sign prediction of stock returns, and then use the results of sign forecast to to conduct portfolio analysis. By implementing rolling-window out of sample forecasts, we demonstrate the practical relevance of the paper.

In the future work, it might be interesting to consider a network model such as those considered in \cite{Yanetal2019} and \cite{Dzemski}. We conjuncture that a result like Lemme \ref{Lemma2.1} might be achievable to simplify the asymptotic development. Moreover, a high dimensional model with sparse coefficients such as that in \cite{ChuZhuWang} is also worth to be investigated.

\section*{Acknowledgements}

Gao acknowledges financial support from the Australian Research Council Discovery Grants Program under Grant Numbers: DP170104421 and DP200102769. Peng acknowledges the Australian Research Council Discovery Grants Program for its financial support under Grant Number DP210100476.

{\footnotesize
\bibliography{Refs}
}

{\small

\begin{appendices}

\newpage

\section{}\label{AppA}

The structure of Appendix A is as follows. In Appendix \ref{Appnotation}, we first  outline the strategy of the theoretical development of the paper. Appendix \ref{Appnot} provides some notations which are repeatedly used throughout the derivation. Appendix \ref{BS} comments on how to deal with bias correction. Appendix \ref{AppAPE} provides a  result on the average partial effects.   Finally, we present the proofs of Theorem \ref{Theorem2.1} in Appendix \ref{AppProof}. Due to the limit of space, we regulate the proofs of the omitted theorems and the preliminary lemmas with their proofs to the online supplementary Appendix \ref{AppB} of this paper.

\subsection{Outline of the Theoretical Development} \label{Appnotation}

We first outline the strategy of the theoretical development of the paper. In Lemma \ref{Lemma2.1}, we first use the Taylor expansion to investigate the log-likelihood function. By doing so, we are able to establish a link between a maximum likelihood estimation and a nonlinear least squares approach. As a consequence, the identification restrictions provided in \cite{Bai2009} and \cite{Moon} (e.g., the conditions for the term $\widetilde{S}_{NT}(\beta, F)$ on \citealp[p. 1264]{Bai2009}) are readily to be applied to the binary response models with very minor modifications. It then immediately yields the consistency of Lemma \ref{Lemma2.2}. After that, we further establish the uniform consistency in the first two results of Lemma \ref{Lemmauni} with some mild restrictions. We then look at the first order conditions of the log-likelihood function, and study the Hessian matrix to further derive the rates associated with different parameters. The results are presented in the third and fourth results of Lemma \ref{Lemmauni}. After investigating the rates of convergence, the leading terms become clear, so we establish the asymptotic distributions  (i.e., Theorem \ref{Theorem2.1}) accordingly. Theorem \ref{Theorem2.2} and Theorem \ref{Theorem2.3} can be regarded as extensions of the above development.

\subsection{Notations} \label{Appnot}

We introduce some notations to facilitate the development. In what follows, $O(1)$ always stands for a constant, and may be different at each appearance. Recall that $B_0 =(\beta_{01},\ldots, \beta_{0N})'$,  $F_0 =(f_{01},\ldots, f_{0T})'$, $\Gamma_0 = (\gamma_{01},\ldots, \gamma_{0N})'$, $\theta_{0i}=(\beta_{0i}', \gamma_{0i}')'$, and $\Theta_0=(B_0,\Gamma_0)=(\theta_{01},\ldots,\theta_{0N})'$. Also, recall that $\Omega_u $, $\Omega_\gamma$ and $\Omega_{u\gamma}$ have been defined under \eqref{def_omegas}. Throughout the derivation, we define further that 

\begin{eqnarray*}
&& \Theta  =(B,\Gamma)=(\theta_{1},\ldots,\theta_{N})',\quad \theta_{i}=(\beta_{i}', \gamma_{i}')',\\
&&\Theta_{0v} = (\theta_{01}', \cdots,\theta_{0N}')', \quad F_{0v} = (f_{01}', \cdots,f_{0T}')',\\
&&\widehat{\Theta}_{v} = (\widehat{\theta}_1', \cdots,\widehat{\theta}_N')',\quad \widehat{F}_{v} = (\widehat{f}_1', \cdots,\widehat{f}_T')',\nonumber \\
&&z_{it}=x_{it}'\beta_i+\gamma_{i}'f_{t},\quad z_{it}^0=x_{it}'\beta_{0i}+\gamma_{0i}'f_{0t},\quad  \widehat{z}_{it}=x_{it}'\widehat{\beta}_i+\widehat{\gamma}_{i}'\widehat{f}_{t}.
\end{eqnarray*}

Simple algebra shows that

\begin{eqnarray*}
\frac{\partial \log L(\Theta, F )}{\partial \Theta_v} &=&\vect \left( \frac{\partial \log L(\Theta, F )}{\partial \theta_1},\cdots, \frac{\partial \log L(\Theta, F )}{\partial \theta_N}\right), \nonumber\\
\frac{\partial \log L(\Theta, F )}{\partial F_v} &=& \vect\left( \frac{\partial \log L(\Theta, F )}{\partial f_1},\cdots, \frac{\partial \log L(\Theta, F )}{\partial f_T}\right),
\end{eqnarray*}
where 

\begin{eqnarray*} 
\frac{\partial \log L(\Theta, F )}{\partial \theta_i}
&=&\sum_{t=1}^T \frac{[y_{it}-G_\varepsilon(z_{it})]g_\varepsilon(z_{it}) }{[1-G_\varepsilon(z_{it})]G_\varepsilon(z_{it})}   u_{it}, \quad
\frac{\partial \log L(\Theta, F)}{\partial f_t} 
=\sum_{i=1}^N \frac{[y_{it}-G_\varepsilon(z_{it})]g_\varepsilon(z_{it}) }{[1-G_\varepsilon(z_{it})]G_\varepsilon(z_{it})}  \gamma_i.
\end{eqnarray*}

The second derivatives of the log-likelihood functions are as follows.

\begin{eqnarray*}
\frac{\partial^2 \log L(\Theta, F )}{\partial \Theta_v \partial\Theta_v'}&=&\diag\left\{ \frac{\partial^2 \log L(\Theta, F)}{\partial \theta_1\partial\theta_1'},\ldots, \frac{\partial^2 \log L(\Theta, F)}{\partial \theta_N\partial\theta_N'}\right\},\nonumber \\
\quad \frac{\partial^2 \log L(\Theta, F )}{\partial F_v \partial F_v'} &=&\diag\left\{\frac{\partial^2 \log L(\Theta, F)}{\partial f_1\partial f_1'},\ldots, \frac{\partial^2 \log L(\Theta, F)}{\partial f_T\partial f_T'} \right\},\nonumber \\
\frac{\partial^2 \log L(\Theta, F_v )}{\partial \Theta \partial F_v'} &=&\left\{\frac{\partial^2 \log L(\Theta, F)}{\partial \theta_i\partial f_t'} \right\}_{N(d_\beta+d_f)\times Td_f},
\end{eqnarray*}
where

\begin{eqnarray*}
&&\frac{\partial^2 \log L(\Theta, F)}{\partial \theta_i\partial\theta_i'}=-\sum_{t=1}^T \left\{\frac{[g_\varepsilon(z_{it})]^2 }{[1-G_\varepsilon(z_{it})]G_\varepsilon(z_{it})}  \right\} u_{it}u_{it}', 
\nonumber\\
&&+\sum_{t=1}^T \left\{\frac{[y_{it}-G_\varepsilon(z_{it})][g_\varepsilon^{(1)}(z_{it})G_\varepsilon(z_{it})(1-G_\varepsilon(z_{it}))+[g_\varepsilon(z_{it})]^2 (1-2G_\varepsilon(z_{it}))] }{[1-G_\varepsilon(z_{it})]^2[G_\varepsilon(z_{it})]^2}  \right\} u_{it}u_{it}', \nonumber\\ \nonumber \\
&&\frac{\partial^2 \log L(\Theta, F)}{\partial f_t\partial f_t'}=-\sum_{i=1}^N \left\{\frac{[g_\varepsilon(z_{it})]^2 }{[1-G_\varepsilon(z_{it})]G_\varepsilon(z_{it})}  \right\} \gamma_{i}\gamma_{i}', \nonumber\\
&&+\sum_{i=1}^N \left\{\frac{[y_{it}-G_\varepsilon(z_{it})][g_\varepsilon^{(1)}(z_{it})G_\varepsilon(z_{it})(1-G_\varepsilon(z_{it}))+[g_\varepsilon(z_{it})]^2 (1-2G_\varepsilon(z_{it}))] }{[1-G_\varepsilon(z_{it})]^2[G_\varepsilon(z_{it})]^2}  \right\}\gamma_{i}\gamma_{i}',  \nonumber\\  \nonumber \\
&&\frac{\partial^2 \log L(\Theta, F)}{\partial \theta_i\partial f_t'}=-\frac{[g_\varepsilon(z_{it})]^2 }{[1-G_\varepsilon(z_{it})]G_\varepsilon(z_{it})}  u_{it}\gamma_{i}',  \nonumber\\
&&+ \frac{[y_{it}-G_\varepsilon(z_{it})][g_\varepsilon^{(1)}(z_{it})G_\varepsilon(z_{it})(1-G_\varepsilon(z_{it}))+[g_\varepsilon(z_{it})]^2 (1-2G_\varepsilon(z_{it}))] }{[1-G_\varepsilon(z_{it})]^2[G_\varepsilon(z_{it})]^2}  u_{it}\gamma_{i}'.
\end{eqnarray*}

\subsection{On Bias Correction}\label{BS}

For simplicity, we let $\beta_{0i}\equiv \beta_0$, and focus on the third result of Theorem \ref{Theorem2.2}. By the proof of Theorem \ref{Theorem2.1}, we can obtain that

\begin{eqnarray*}
\widehat{\beta} - \beta_{0} &=&  \frac{1}{NT}\sum_{i=1}^N \sum_{t=1}^T \frac{[y_{it}-G_\varepsilon(z_{it}^0)]g_\varepsilon(z_{it}^0)\, \Sigma_{u, i}^{(d_{\beta})}}{[1-G_\varepsilon(z_{it}^0)]G_\varepsilon(z_{it}^0)}   u_{it}^0 +  \frac{1}{T}\text{Bias}_1+\frac{1}{N}\text{Bias}_2 \nonumber \\
&:=&  \frac{1}{NT}\sum_{i=1}^N \sum_{t=1}^T w_{it} +  \frac{1}{T}\text{Bias}_1+\frac{1}{N}\text{Bias}_2,
\end{eqnarray*}
where the definition of $w_{it}$ is obvious, $\text{Bias}_1=O_P(1)$, and $\text{Bias}_2=O_P(1)$. We have omitted the detailed formulas of both biased terms for simplicity, and the two terms are also similar to those in \cite{Chen2020}.

We now explain how the half panel jackknife technique can be applied to remove the asymptotic biases. The technique is initially proposed in \cite{DJ2015}, and has been further discussed in \cite{FW2018} and \cite{Chen2020}. The following approach can be considered as a modified version of \cite{Chen2020}, as we need to account for the nature ordering along the time dimension, and also would like to allow for possible smooth transit over time (i.e., certain heteroscedasticity along the time dimension). For example, although we impose mixing conditions on $f_t$, one in fact can relax these conditions by assuming

\begin{eqnarray}
\frac{1}{T}F_0'F_0\to_P\Sigma_f,
\end{eqnarray}
which implicitly allows for heteroscedasticity over time. Thus, we consider the following procedure.

First, Let $N$ be an even number without loss of generality, and randomly divide the individuals into two new sets $S_{1} $ and $S_{2}$ along the cross-sectional dimension such that

\begin{eqnarray*}
S_1  \cap S_2  = \emptyset,\quad S_1  \cup S_2  =\{ 1,\ldots, N\},  \quad \text{and}\quad \sharp S_1 = \sharp S_2 =\frac{N}{2},
\end{eqnarray*}
where $\sharp S_j$ stands for the cardinality of $S_j$ for $j=1,2$. For the time dimension, we also create another two new sets:

\begin{eqnarray*}
S_{\text{odd}} = \{t\in [T]\text{ and $t$ is odd} \} \quad \text{and}\quad S_{\text{even}} =  \{t\in [T]\text{ and $t$ is even} \}.
\end{eqnarray*}
Note that splitting the time points using even and odd indices allows us to preserve the behaviour of the data along the time dimension, so that certain heteroscedasticity can be allowed (e.g., \citealp{gao_linton_peng_2020}). We then define the bias corrected estimator as follows.

\begin{eqnarray}\label{estBC}
\widehat{\beta}_{\text{bc}}  = 3\widehat{\beta}  - (\widehat{\beta}_{S_{1} } +\widehat{\beta}_{S_{2} } +\widehat{\beta}_{S_{\text{odd}}} +\widehat{\beta}_{S_{\text{even}}})/2,
\end{eqnarray}
where $\widehat{\beta}_{S_{\ell} } $ is obtained using $ S_{\ell}\otimes \{1,\ldots, T \}$ for $\ell =1,2$, and $\widehat{\beta}_{S_{\text{odd}}} $ and $\widehat{\beta}_{S_{\text{even}}} $ are obtained using sample from $\{1,\ldots, N \}\otimes S_{\text{odd}} $ and $\{1,\ldots, N \}\otimes S_{\text{even}} $ respectively. 

We now briefly explain why \eqref{estBC} works. Write

\begin{eqnarray*}
\sqrt{NT}(\widehat{\beta}_{\text{bc}}  -\beta_0) &=&3\sqrt{NT}(\widehat{\beta}  -\beta_0) -\frac{1}{\sqrt{2}} \sqrt{(N/2)T}(\widehat{\beta}_{S_{1}} -\beta_0) - \frac{1}{\sqrt{2}}\sqrt{(N/2)T}(\widehat{\beta}_{S_{2}} -\beta_0)\\
&&- \frac{1}{\sqrt{2}}\sqrt{N(T/2)}(\widehat{\beta}_{S_{\text{odd}}} -\beta_0)-\frac{1}{\sqrt{2}} \sqrt{N(T/2)}(\widehat{\beta}_{S_{\text{even}}}-\beta_0).
\end{eqnarray*}
Direct calculation shows that

\begin{eqnarray*}
\sqrt{NT}(\widehat{\beta}_{\text{bc}}   -\beta_0) &=& \Big\{  \frac{3}{\sqrt{NT}}\sum_{i=1}^N\sum_{t=1}^T w_{it} - 3\sqrt{\frac{N}{T}}\text{Bias}_1- 3\sqrt{\frac{T}{N}}\text{Bias}_2 \\
&&-\frac{1}{\sqrt{2}}\left(\frac{1}{\sqrt{(N/2)T}}\sum_{i\in S_{1} }\sum_{t=1}^T w_{it}-  \sqrt{\frac{N/2}{T}}\text{Bias}_1- \sqrt{\frac{T}{N/2}}\text{Bias}_2  \right)\\
&&-\frac{1}{\sqrt{2}}\left(\frac{1}{\sqrt{(N/2)T}}\sum_{i\in S_{2} }\sum_{t=1}^T w_{it} -  \sqrt{\frac{N/2}{T}}\text{Bias}_1- \sqrt{\frac{T}{N/2}}\text{Bias}_2  \right)\\
&&-\frac{1}{\sqrt{2}}\left(\frac{1}{\sqrt{ N(T/2)}}\sum_{i=1}^N\sum_{t\in S_{\text{odd}}}  w_{it} -  \sqrt{\frac{N}{T/2}}\text{Bias}_1- \sqrt{\frac{T/2}{N }}\text{Bias}_2  \right)\\
&&-\frac{1}{\sqrt{2}}\left(\frac{1}{\sqrt{N(T/2)}}\sum_{i=1}^N\sum_{t\in S_{\text{even}}} w_{it} -  \sqrt{\frac{N}{T/2}}\text{Bias}_1- \sqrt{\frac{T/2}{N }}\text{Bias}_2  \right)\Big\} \\
&=& \Big\{\frac{1}{\sqrt{NT}}\sum_{i=1}^N\sum_{t=1}^T w_{it}  -\sqrt{\frac{N}{T}}\left(3\text{Bias}_1-\frac{1}{2}\text{Bias}_1 -\frac{1}{2}\text{Bias}_1-\text{Bias}_1-\text{Bias}_1\right)\nonumber \\
&&-\sqrt{\frac{T}{N}}\left(3\text{Bias}_2 - \text{Bias}_2 - \text{Bias}_2  -\frac{1}{2}\text{Bias}_2 -\frac{1}{2}\text{Bias}_2\right)\Big\}  = \frac{1}{\sqrt{NT}}\sum_{i=1}^N\sum_{t=1}^T w_{it}  .
\end{eqnarray*}
Therefore, the biases vanish.

\subsection{On Average Partial Effects}\label{AppAPE}

We now consider the estimation of average partial effects (APE) based on the binary model \eqref{EQ21}. Let $x_{it,k}$ and $\beta_{0i,k}$ be the $k^{th}$ elements of $x_{it}$ and $\beta_{0i}$ respectively.  The sample version APE of $x_{it,k}$ on the conditional probability of $y_{it}$ can be defined as 

\begin{eqnarray*}
\Delta_{i,k} =\frac{1}{T}\sum_{t=1}^T g(z^0_{it}) \beta_{0i,k}.
\end{eqnarray*}
Using  $(\widehat{B},\widehat{F},\widehat{\Gamma})$ of Section \ref{Section2}, we can estimate  $\Delta_{i}=(\Delta_{i,1},\ldots,\Delta_{i,d_\beta})'$ as follows.

\begin{eqnarray*}
\widehat{\Delta}_{i}=\frac{1}{T}\sum_{t=1}^T g(\widehat{z}_{it})\widehat{\beta}_i.
\end{eqnarray*}
Then the following result holds immediately.

\begin{lemma} \label{LemmaAPE} 
Under Assumptions \ref{Ass0}-\ref{Ass2}, as $(N,T)\to (\infty,\infty)$, $\max_{i \ge 1} \|\widehat{\Delta}_{i}-\Delta_{i}\|=o_P(1)$.
\end{lemma}

In a fashion similar to Theorem \ref{Theorem2.2}, it is possible to establish the asymptotic normality of $\widehat{\Delta}_{i}$ under additional conditions. As it is not the main focus of the paper, we no longer purse it further, and refer interested readers to \cite{Chen2020} for extensive discussions on APE.

\subsection{Proof  of Theorem \ref{Theorem2.1}}\label{AppProof}

\noindent \textbf{Proof of Theorem \ref{Theorem2.1}:}

(1) We have established the consistency of $\widehat{\theta}_i$ and $\widehat{f}_t$ in Lemma \ref{Lemmauni} and its proof is provided in the online supplementary Appendix \ref{AppB} due to page constraint. With additional conditions on the weak cross-sectional dependence and time series correlation on error terms in Assumption \ref{Ass3}, we can establish a $\sqrt{T}$-consistency for $\widehat{\theta}_i$ and $\sqrt{N}$-consistency for $\widehat{f}_t$ in this theorem. Recall that  $\widehat{\Theta}_{v} = (\widehat{\theta}_1', \cdots,\widehat{\theta}_N')'$ and $\widehat{F}_{v} = (\widehat{f}_1', \cdots,\widehat{f}_T')'$.    We can follow analogous arguments in the proof of Lemma \ref{Lemmauni} to show the following rates of convergence for the individual estimators $\widehat{\theta}_i$ and $\widehat{f}_t$:
\begin{eqnarray}\label{rev60}
\|\theta_i-\theta_{0i} \|=O_P \left(\frac{1}{\sqrt{N}\wedge \sqrt{T}}\right),\quad \|\widehat{f}_t-f_{0t} \| = O_P \left(\frac{1}{\sqrt{N}\wedge \sqrt{T}}\right).
\end{eqnarray}
for each $i=1,\ldots,N$ and $t=1,\ldots,T$. Therefore, we need to show further that $\|\theta_i-\theta_{0i} \|=O_P \left( \frac{1}{\sqrt{T}}\right)$ and  $\|\widehat{f}_t-f_{0t} \| = O_P \left(\frac{1}{ \sqrt{N}}\right)$.

To begin with, recall that we have the first derivatives of log-likelihood functions defined in Appendix \ref{Appnot}. We first derive the leading terms in $\widehat{\beta}_i-\beta_{0i}$ from $\frac{\partial \log L(\Theta, F )}{\partial \theta_i}$. For $\frac{\partial \log L(\Theta, F )}{\partial \theta_i}$, the first order condition implies that 

\begin{eqnarray}\label{aa13}
0 &=& \frac{1}{T}\sum_{t=1}^T  \frac{\left[y_{it} -G_\varepsilon(\widehat{z}_{it})\right] g_\varepsilon(\widehat{z}_{it} )}{[1-G_\varepsilon(\widehat{z}_{it})]G_\varepsilon(\widehat{z}_{it})}\widehat{u}_{it}  = \frac{1}{T}\sum_{t=1}^T\frac{\left[y_{it} -G_\varepsilon( z_{it}^0 )\right] g_\varepsilon(z_{it}^0 )}{[1-G_\varepsilon( z_{it}^0)]G_\varepsilon(z_{it}^0)} u_{it}^0
\nonumber\\
&&+ \frac{1}{T}\sum_{t=1}^T \left\{ \frac{\left[y_{it} -G_\varepsilon(\widehat{z}_{it})\right] g_\varepsilon(\widehat{z}_{it} )}{[1-G_\varepsilon(\widehat{z}_{it})]G_\varepsilon(\widehat{z}_{it})} \widehat{u}_{it}- \frac{\left[y_{it} -G_\varepsilon( z_{it}^0 )\right] g_\varepsilon(z_{it}^0 )}{[1-G_\varepsilon( z_{it}^0)]G_\varepsilon(z_{it}^0)} u_{it}^0\right\}\nonumber \\
&:=&A_{1Ti} +A_{2Ti},
\end{eqnarray}
where $A_{1Ti}$ only depends on the statistical behaviours of $z_{it}^0$ and $u_{it}^0$. We now proceed with $A_{2Ti}$. For $A_{2Ti}$, we have
\begin{eqnarray}\label{aa14}
&& A_{2Ti}= \frac{1}{T}\sum_{t=1}^T a_{it}^{-1}a_{2,it}^\ast+\frac{1}{T}\sum_{t=1}^T (a_{it}^{\dagger-1}-a_{it}^{-1})a_{2,it}^\ast,
=A_{3Ti}+A_{4Ti},
\end{eqnarray}
where $a_{it} = [1-G_\varepsilon(z_{it}^0)]^2[G_\varepsilon(z_{it}^0)]^2$, $a_{it}^\dagger = [1-G_\varepsilon(z_{it}^0)]G_\varepsilon(z_{it}^0)[1-G_\varepsilon(\widehat{z}_{it})]G_\varepsilon(\widehat{z}_{it})$, $a_{2,it}^\ast = [y_{it}-G_\varepsilon(\widehat{z}_{it})]g_\varepsilon(\widehat{z}_{it})[1-G_\varepsilon(z_{it}^0)]G_\varepsilon(z_{it}^0)\widehat{u}_{it} -  [y_{it}-G_\varepsilon(z_{it}^0)]g(z_{it}^0) [1-G_\varepsilon(\widehat{z}_{it})] G_\varepsilon(\widehat{z}_{it}) u_{it}^0$.

Among them, $a_{it}$ is a function of real value $z_{it}^0$, therefore we are interested in the convergence of  $a_{2,it}^\ast$ and $a_{it}^\dagger$.  We start our investigation by looking at $a_{2,it}^\ast$, and write
\begin{eqnarray*}
a_{2,it}^\ast&=&-[G_\varepsilon(\widehat{z}_{it})-G_\varepsilon(z^0_{it})]g_\varepsilon(z_{it}^0)[1-G_\varepsilon(z_{it}^0)]G_\varepsilon(z_{it}^0)u_{it}^0
\nonumber\\
&&+[y_{it}-G_\varepsilon(z_{it}^0)][g_\varepsilon(\widehat{z}_{it})-g_\varepsilon(z_{it}^0)][1-G_\varepsilon(z_{it}^0)]G_\varepsilon(z_{it}^0)u_{it}^0
\nonumber\\
&&
+[y_{it}-G_\varepsilon(z_{it}^0)]g_\varepsilon(z_{it}^0)[G_\varepsilon(\widehat{z}_{it})-G_\varepsilon(z^0_{it})]G_\varepsilon(z_{it}^0)u_{it}^0
\nonumber\\
&&-[y_{it}-G_\varepsilon(z_{it}^0)]g_\varepsilon(z_{it}^0)[1-G_\varepsilon(z_{it}^0)][G_\varepsilon(\widehat{z}_{it})-G_\varepsilon(z^0_{it})]u_{it}^0
\nonumber\\
&&+[y_{it}-G_\varepsilon(z_{it}^0)]g_\varepsilon(z_{it}^0)[1-G_\varepsilon(z_{it}^0)]G_\varepsilon(z_{it}^0)(\widehat{u}_{it}-u_{it}^0)
\nonumber\\
&&-[G_\varepsilon(\widehat{z}_{it})-G_\varepsilon(z^0_{it})][g_\varepsilon(\widehat{z}_{it})-g_\varepsilon(z_{it}^0)][1-G_\varepsilon(z_{it}^0)]G_\varepsilon(z_{it}^0)u_{it}^0
\nonumber\\
&&-[G_\varepsilon(\widehat{z}_{it})-G_\varepsilon(z^0_{it})]g_\varepsilon(z_{it}^0)[1-G_\varepsilon(z_{it}^0)]G_\varepsilon(z_{it}^0)(\widehat{u}_{it}-u_{it}^0)
\nonumber\\
&&+[y_{it}-G_\varepsilon(z_{it}^0)][g_\varepsilon(\widehat{z}_{it})-g_\varepsilon(z_{it}^0)][1-G_\varepsilon(z_{it}^0)]G_\varepsilon(z_{it}^0)(\widehat{u}_{it}-u_{it}^0)
\nonumber\\
&&+[y_{it}-G_\varepsilon(z_{it}^0)]g_\varepsilon(z_{it}^0)[G_\varepsilon(\widehat{z}_{it})-G_\varepsilon(z^0_{it})]^2u_{it}^0
\nonumber\\
&&-[G_\varepsilon(\widehat{z}_{it})-G_\varepsilon(z^0_{it})][g_\varepsilon(\widehat{z}_{it})-g_\varepsilon(z_{it}^0)][1-G_\varepsilon(z_{it}^0)]G_\varepsilon(z_{it}^0)(\widehat{u}_{it}-u_{it}^0)
\nonumber\\
&:=&a_{2,it}^{(1)\ast}+\cdots+a_{2,it}^{(10)\ast},
\end{eqnarray*}
where the definitions of $a_{2,it}^{(1)\ast}$ to $a_{2,it}^{(10)\ast}$ are obvious. Below, we examine the terms on the right hand side one by one.

For $a_{2,it}^{(1)\ast}$, by the Taylor expansion, write
\begin{eqnarray*}
&& a_{2,it}^{(1)\ast} = -g^2_\varepsilon(z_{it}^0)[1-G_\varepsilon(z_{it}^0)]G_\varepsilon(z_{it}^0)u_{it}^0(\widehat{z}_{it}-z_{it}^{0})\\
&&-\frac{1}{2}g^{(1)}_\varepsilon(\dot{z}_{it})g_\varepsilon(z_{it}^0)[1-G_\varepsilon(z_{it}^0)]G_\varepsilon(z_{it}^0)u_{it}^0(\widehat{z}_{it}-z_{it}^{0})^2:=a_{21,it}^{(1)\ast}+a_{22,it}^{(1)\ast},
\end{eqnarray*}
where $\dot{z}_{it}$ lies between $\widehat{z}_{it}$ and $z_{it}^0$, and the definitions of $a_{21,it}^{(1)\ast}$ and $a_{22,it}^{(1)\ast}$ are obvious.

Note that for $a_{21,it}^{(1)\ast}$, we have

\begin{eqnarray*}
&&\frac{1}{T}\sum_{t=1}^T a_{it}^{-1}a_{21,it}^{(1)\ast}=-\frac{1}{T}\sum_{t=1}^T \frac{g^2_\varepsilon(z_{it}^0)}{[1-G_\varepsilon(z_{it}^0)]G_\varepsilon(z_{it}^0)}u_{it}^0(\widehat{z}_{it}-z_{it}^{0})
\nonumber\\
&=&-\frac{1}{T}\sum_{t=1}^T \frac{g^2_\varepsilon(z_{it}^0)}{[1-G_\varepsilon(z_{it}^0)]G_\varepsilon(z_{it}^0)}u_{it}^0x_{it}'(\widehat{\beta}_i-\beta_{0i})
-\frac{1}{T}\sum_{t=1}^T \frac{g^2_\varepsilon(z_{it}^0)}{[1-G_\varepsilon(z_{it}^0)]G_\varepsilon(z_{it}^0)}u_{it}^0(\widehat{\gamma}_i^\prime\widehat{f}_t-\gamma_{0i}^\prime f_{0t})
\nonumber\\
&=&-\frac{1}{T}\sum_{t=1}^T \frac{g^2_\varepsilon(z_{it}^0)}{[1-G_\varepsilon(z_{it}^0)]G_\varepsilon(z_{it}^0)}u_{it}^0u^{0\prime}_{it}(\widehat{\theta}_i-\theta_{0i})
-\frac{1}{T}\sum_{t=1}^T \frac{g^2_\varepsilon(z_{it}^0)}{[1-G_\varepsilon(z_{it}^0)]G_\varepsilon(z_{it}^0)}u_{it}^0\gamma_{0i}^{\prime}(\widehat{f}_t-f_{0t})
\nonumber\\
&&-\frac{1}{T}\sum_{t=1}^T \frac{g^2_\varepsilon(z_{it}^0)}{[1-G_\varepsilon(z_{it}^0)]G_\varepsilon(z_{it}^0)}u_{it}^0(\widehat{\gamma}_{i}-\gamma_{0i})^\prime(\widehat{f}_t-f_{0t})
\nonumber\\
&=&-\frac{1}{T}\sum_{t=1}^T \frac{g^2_\varepsilon(z_{it}^0)}{[1-G_\varepsilon(z_{it}^0)]G_\varepsilon(z_{it}^0)}u_{it}^0u^{0\prime}_{it}(\widehat{\theta}_i-\theta_{0i})-\frac{1}{T}\sum_{t=1}^T \frac{g^2_\varepsilon(z_{it}^0)}{[1-G_\varepsilon(z_{it}^0)]G_\varepsilon(z_{it}^0)}u_{it}^0\gamma_{0i}^{\prime}(\widehat{f}_t-f_{0t})
\nonumber\\
&&+O_P\left(\frac{1}{N\wedge T}\right),
\end{eqnarray*}
where the last equality holds, because
\begin{eqnarray*}
&&\left\|\frac{1}{T}\sum_{t=1}^T \frac{g^2_\varepsilon(z_{it}^0)}{[1-G_\varepsilon(z_{it}^0)]G_\varepsilon(z_{it}^0)}u_{it}^0(\widehat{\gamma}_{i}-\gamma_{0i})^\prime(\widehat{f}_t-f_{0t})\right\| \leq \left\|\widehat{\gamma}_{i}-\gamma_{0i}\right\|\\
&&\times \left\{\frac{1}{T}\sum_{t=1}^T \left\|\frac{g^2_\varepsilon(z_{it}^0)}{[1-G_\varepsilon(z_{it}^0)]G_\varepsilon(z_{it}^0)}u_{it}^0\right\|^2 \right\}^{\frac{1}{2}} \cdot \left\{\frac{1}{T}\sum_{t=1}^T \left\|\widehat{f}_t-f_{0t}\right\|^2 \right\}^{\frac{1}{2}}=O_P\left(\frac{1}{N\wedge T}\right),
\end{eqnarray*}
in which we have used the Cauchy-Schwarz inequality and Lemma \ref{Lemmauni}. Furthermore, by Lemma \ref{Lemmauni} and \eqref{rev60}, we have
\begin{eqnarray*}
&&\frac{1}{T}\sum_{t=1}^T \frac{g^2_\varepsilon(z_{it}^0)}{[1-G_\varepsilon(z_{it}^0)]G_\varepsilon(z_{it}^0)}u_{it}^0u^{0\prime}_{it}(\widehat{\theta}_i-\theta_{0i}) =\Sigma_{u,i}(\widehat{\theta}_i-\theta_{0i})+O_P\left(\frac{1}{\sqrt{(N\wedge T)T}}\right).
\end{eqnarray*}

Therefore, we obtain that
\begin{eqnarray}\label{aa1}
\frac{1}{T}\sum_{t=1}^T a_{it}^{-1}a_{21,it}^{(1)\ast}
&=&-\Sigma_{u,i}(\widehat{\theta}_i-\theta_{0i})-\frac{1}{T}\sum_{t=1}^T \frac{g^2_\varepsilon(z_{it}^0)}{[1-G_\varepsilon(z_{it}^0)]G_\varepsilon(z_{it}^0)}u_{it}^0\gamma_{0i}^{\prime}(\widehat{f}_t-f_{0t})
\nonumber\\
&&+O_P\left(\frac{1}{N\wedge T}\right).
\end{eqnarray}

For $a_{22,it}^{(1)\ast}$,
\begin{eqnarray}\label{aa2old}
&& \frac{1}{T}\sum_{t=1}^T a_{it}^{-1}a_{22,it}^{(1)\ast}= -\frac{1}{T}\sum_{t=1}^T\frac{g^{(1)}_\varepsilon(\dot{z}_{it})g_\varepsilon(z_{it}^0)}{2[1-G_\varepsilon(z_{it}^0)]G_\varepsilon(z_{it}^0)}u_{it}^0(\widehat{z}_{it}-z_{it}^{0})^2
\nonumber\\
&&= -\frac{1}{T}\sum_{t=1}^T\frac{g^{(1)}_\varepsilon(\dot{z}_{it})g_\varepsilon(z_{it}^0)}{2[1-G_\varepsilon(z_{it}^0)]G_\varepsilon(z_{it}^0)}u_{it}^0( u_{it}^{0\prime}(\widehat{\theta}_i-\theta_{0i}) + \gamma_{0i}'(\widehat{f}_t-f_{0t})+ ( \widehat{\gamma}_i -\gamma_{0i})'(\widehat{f}_t-f_{0t}) )^2
\nonumber\\
&&=-\frac{1}{T}\sum_{t=1}^T\frac{g^{(1)}_\varepsilon(\dot{z}_{it})g_\varepsilon(z_{it}^0)}{2[1-G_\varepsilon(z_{it}^0)]G_\varepsilon(z_{it}^0)}u_{it}^0( u_{it}^{0\prime}(\widehat{\theta}_i-\theta_{0i}))^2 -\frac{1}{T}\sum_{t=1}^T\frac{g^{(1)}_\varepsilon(\dot{z}_{it})g_\varepsilon(z_{it}^0)}{2[1-G_\varepsilon(z_{it}^0)]G_\varepsilon(z_{it}^0)}u_{it}^0(\gamma_{0i}'(\widehat{f}_t-f_{0t}))^2
\nonumber\\
&&  -\frac{1}{T}\sum_{t=1}^T\frac{g^{(1)}_\varepsilon(\dot{z}_{it})g_\varepsilon(z_{it}^0)}{2[1-G_\varepsilon(z_{it}^0)]G_\varepsilon(z_{it}^0)}u_{it}^0 (( \widehat{\gamma}_i -\gamma_{0i})'(\widehat{f}_t-f_{0t}) )^2
+\text{interaction terms}.
\end{eqnarray}
For the interaction terms on the right-hand side of \eqref{aa2old}, we can show they are bounded in probability by the first three terms by Cauchy-Schwarz inequality. Therefore, the proof for their probability orders is omitted. We now consider the first three terms one by one. For the first term, 

\begin{eqnarray}\label{aa2a}
&&\left\|\frac{1}{T}\sum_{t=1}^T\frac{g^{(1)}_\varepsilon(\dot{z}_{it})g_\varepsilon(z_{it}^0)}{2[1-G_\varepsilon(z_{it}^0)]G_\varepsilon(z_{it}^0)}u_{it}^0( u_{it}^{0\prime}(\widehat{\theta}_i-\theta_{0i}))^2\right\| \nonumber \\
&\leq&\frac{1}{T}\sum_{t=1}^T \left(\left\|\frac{g^{(1)}_\varepsilon(\dot{z}_{it})g_\varepsilon(z_{it}^0)}{2[1-G_\varepsilon(z_{it}^0)]G_\varepsilon(z_{it}^0)}u_{it}^0 \right\|^2\cdot \|u_{it}^0\|^2\right) \cdot \|\widehat{\theta}_i-\theta_{0i} \|^2= O_P\left(\|\widehat{\theta}_i-\theta_{0i} \|^2\right),
\end{eqnarray}
where the equality holds by  Assumption \ref{Ass2}. 

For the second term on the right-hand side of \eqref{aa2old},   

\begin{eqnarray}\label{aa2b}
&&\left\|\frac{1}{T}\sum_{t=1}^T\frac{g^{(1)}_\varepsilon(\dot{z}_{it})g_\varepsilon(z_{it}^0)}{2[1-G_\varepsilon(z_{it}^0)]G_\varepsilon(z_{it}^0)}u_{it}^0(\gamma_{0i}'(\widehat{f}_t-f_{0t}))^2\right\|\nonumber \\
&\leq & O_P(\log(NT)) \cdot \frac{1}{T}\sum_{t=1}^T\|\widehat{f}_t-f_{0t}\|^2 = O_P\left(\frac{\log(NT)}{N\wedge T}\right),
\end{eqnarray}
where the inequality holds by  Assumption \ref{Ass2} and the equality holds by Lemma \ref{Lemmauni}.

For the third term on the right-hand side of \eqref{aa2old},  by   Assumption \ref{Ass2} and  Lemma \ref{Lemmauni},

\begin{eqnarray}\label{aa2c}
&&\left\|\frac{1}{T}\sum_{t=1}^T\frac{g^{(1)}_\varepsilon(\dot{z}_{it})g_\varepsilon(z_{it}^0)}{2[1-G_\varepsilon(z_{it}^0)]G_\varepsilon(z_{it}^0)}u_{it}^0 (( \widehat{\gamma}_i -\gamma_{0i})'(\widehat{f}_t-f_{0t}) )^2\right\| \nonumber \\
&\leq& O_P(\log(NT))\cdot \|\widehat{\gamma}_i-\gamma_{0i}\|^2\cdot \frac{1}{T}\sum_{t=1}^T\|\widehat{f}_t-f_{0t}\|^2=o_P\left(\frac{\log(NT)}{N\wedge T}\right).
\end{eqnarray}

By \eqref{aa2a}, \eqref{aa2b} and \eqref{aa2c}, 

\begin{eqnarray}\label{aa2}
&& \frac{1}{T}\sum_{t=1}^T a_{it}^{-1}a_{22,it}^{(1)\ast}= -\frac{1}{T}\sum_{t=1}^T\frac{g^{(1)}_\varepsilon(\dot{z}_{it})g_\varepsilon(z_{it}^0)}{2[1-G_\varepsilon(z_{it}^0)]G_\varepsilon(z_{it}^0)}u_{it}^0(\widehat{z}_{it}-z_{it}^{0})^2=O_P\left(\frac{\log(NT)}{N\wedge T}\right).
\end{eqnarray}

By (\ref{aa1}) and (\ref{aa2}), we have

\begin{eqnarray}\label{aa11}
\frac{1}{T}\sum_{t=1}^T a_{it}^{-1}a_{2,it}^{(1)\ast}&=&-\Sigma_{u,i}(\widehat{\theta}_i-\theta_{0i})-\frac{1}{T}\sum_{t=1}^T \frac{g^2_\varepsilon(z_{it}^0)}{[1-G_\varepsilon(z_{it}^0)]G_\varepsilon(z_{it}^0)}u_{it}^0\gamma_{0i}^{\prime}(\widehat{f}_t-f_{0t})
\nonumber\\
&&+O_P\left(\frac{\log(NT)}{N\wedge T}\right).
\end{eqnarray}

After obtaining the leading term in $a_{2,it}^{(1)\ast}$, we proceed with $a_{2,it}^{(2)\ast}$. For $a_{2,it}^{(2)\ast}$, by the Taylor expansion, we have
\begin{eqnarray*}
&& a_{2,it}^{(2)\ast} = [y_{it}-G_\varepsilon(z_{it}^0)]g^{(1)}_\varepsilon(z_{it}^0)[1-G_\varepsilon(z_{it}^0)]G_\varepsilon(z_{it}^0)u_{it}^0(\widehat{z}_{it}-z_{it}^0)
\nonumber\\
&&+\frac{1}{2}[y_{it}-G_\varepsilon(z_{it}^0)]g^{(2)}_\varepsilon(\ddot{z}_{it})[1-G_\varepsilon(z_{it}^0)]G_\varepsilon(z_{it}^0)u_{it}^0(\widehat{z}_{it}-z_{it}^0)^2
:= a_{21,it}^{(2)\ast}+a_{22,it}^{(2)\ast}, 
\end{eqnarray*}
where $\ddot{z}_{it}$ lies between $\widehat{z}_{it}$ and $z_{it}^0$, and the definitions of $a_{21,it}^{(2)\ast}$ are $a_{22,it}^{(2)\ast}$ are obvious. 

For $a_{21,it}^{(2)\ast}$, write
\begin{eqnarray}\label{aa3}
\frac{1}{T}\sum_{t=1}^T a_{it}^{-1}a_{21,it}^{(2)\ast}&=&\frac{1}{T}\sum_{t=1}^T \frac{[y_{it}-G_\varepsilon(z_{it}^0)]g^{(1)}_\varepsilon(z_{it}^0)}{[1-G_\varepsilon(z_{it}^0)]G_\varepsilon(z_{it}^0)}u_{it}^0(\widehat{z}_{it}-z_{it}^0)
\nonumber\\
&=&\frac{1}{T}\sum_{t=1}^T \frac{[y_{it}-G_\varepsilon(z_{it}^0)]g^{(1)}_\varepsilon(z_{it}^0)}{[1-G_\varepsilon(z_{it}^0)]G_\varepsilon(z_{it}^0)}u_{it}^0u_{it}^{0\prime}(\widehat{\theta}_i-\theta_{0i})
\nonumber\\
&&+\frac{1}{T}\sum_{t=1}^T \frac{[y_{it}-G_\varepsilon(z_{it}^0)]g^{(1)}_\varepsilon(z_{it}^0)}{[1-G_\varepsilon(z_{it}^0)]G_\varepsilon(z_{it}^0)}u_{it}^0\gamma_{0i}^{\prime}(\widehat{f}_t-f_{0t})
\nonumber\\
&&+\frac{1}{T}\sum_{t=1}^T \frac{[y_{it}-G_\varepsilon(z_{it}^0)]g^{(1)}_\varepsilon(z_{it}^0)}{[1-G_\varepsilon(z_{it}^0)]G_\varepsilon(z_{it}^0)}u_{it}^0(\widehat{\gamma}_i-\gamma_{0i})^{\prime}(\widehat{f}_t-f_{0t}).
\end{eqnarray} 
Recall that $e_{it}=\frac{y_{it}-G_\varepsilon(z^0_{it})}{[1-G_\varepsilon(z^0_{it})]G_\varepsilon(z^0_{it})}$. For the first term in (\ref{aa3}), note that 

\begin{eqnarray*} 
E\left[\frac{1}{T}\sum_{t=1}^T \frac{[y_{it}-G_\varepsilon(z_{it}^0)]g^{(1)}_\varepsilon(z_{it}^0)}{[1-G_\varepsilon(z_{it}^0)]G_\varepsilon(z_{it}^0)}u_{it}^0u_{it}^{0\prime}\right]=0.
\end{eqnarray*}
In addition, we have
\begin{eqnarray}
&&E \left\|\frac{1}{T}\sum_{t=1}^T \frac{[y_{it}-G_\varepsilon(z_{it}^0)]g^{(1)}_\varepsilon(z_{it}^0)}{[1-G_\varepsilon(z_{it}^0)]G_\varepsilon(z_{it}^0)}u_{it}^0u_{it}^{0\prime}\right\|^2\nonumber \\
&&\leq O(1)\frac{1}{T^2} \sum_{t=1}^T\sum_{s=1}^T  E[ \|u_{it}\|^2 \cdot \| u_{is}^0\|^2 \cdot | E[e_{it}e_{is}\, | \, w_{it}^0, w_{is}^0]] 
\nonumber\\
&&\leq\frac{c_{\delta}}{T^2} \sum_{t=1}^T\sum_{s=1}^T  E\left[ \|u_{it}\|^2 \cdot \| u_{is}^0\|^2 \cdot \alpha_{ii}(|t-s|)^{\delta/(4+\delta)} E\left[|e_{it}|^{2+\delta/2} |\, \mathcal{W}\right]^{2/(4+\delta)}E\left[|e_{is}|^{2+\delta/2} |\, \mathcal{W}\right]^{2/(4+\delta)}\right]
\nonumber\\
&&= O\left(\frac{1}{T}\right),
\label{ab2}
\end{eqnarray}
where $c_{\delta}=(4+\delta)/\delta\cdot2^{(4+2\delta)/(4+\delta)}$, the second inequality  holds by the Davydov's inequality for $\alpha$-mixing process (see pages 19-20 in \citealp{Bosq1996}) and the fact that conditional on $\mathcal{W}=\{w_{it}^0, i,t\geq 1\}$, $e_{is}$ is $\alpha$-mixing, because $\varepsilon_{it}$ is $\alpha$-mixing and independent of $\mathcal{W}$ under Assumption \ref{Ass2} and we first invoke the $\alpha$-mixing conditions on $\varepsilon_{it}$ here. The last equality holds by the moment conditions and conditions on $\alpha$-mixing coefficients  in Assumption \ref{Ass2}. By \eqref{ab2}, 
\begin{eqnarray*} 
\frac{1}{T}\sum_{t=1}^T \frac{[y_{it}-G_\varepsilon(z_{it}^0)]g^{(1)}_\varepsilon(z_{it}^0)}{[1-G_\varepsilon(z_{it}^0)]G_\varepsilon(z_{it}^0)}u_{it}^0u_{it}^{0\prime}=O_P\left(\frac{1}{\sqrt{T}}\right),
\end{eqnarray*}
which in connection with \eqref{rev60} yields that

\begin{eqnarray*}
\frac{1}{T}\sum_{t=1}^T \frac{[y_{it}-G_\varepsilon(z_{it}^0)]g^{(1)}_\varepsilon(z_{it}^0)}{[1-G_\varepsilon(z_{it}^0)]G_\varepsilon(z_{it}^0)}u_{it}^0u_{it}^{0\prime}(\widehat{\theta}_i-\theta_{0i})=O_P\left(\frac{1}{\sqrt{(N\wedge T)T}}\right).
\end{eqnarray*}

Therefore, we have shown that the first term in \eqref{aa3} is $o_P(\frac{1}{\sqrt{T}})$. For the third term in \eqref{aa3}, by Lemma \ref{Lemmauni}, \eqref{rev60} and  Cauchy-Schwarz inequality, we have

\begin{eqnarray*}
&&\left\|\frac{1}{T}\sum_{t=1}^T \frac{[y_{it}-G_\varepsilon(z_{it}^0)]g^{(1)}_\varepsilon(z_{it}^0)}{[1-G_\varepsilon(z_{it}^0)]G_\varepsilon(z_{it}^0)}u_{it}^0(\widehat{\gamma}_i-\gamma_{0i})^{\prime}(\widehat{f}_t-f_{0t})\right\| \\
&\leq& \left\{\frac{1}{T}\sum_{t=1}^T \left\|\frac{[y_{it}-G_\varepsilon(z_{it}^0)]g^{(1)}_\varepsilon(z_{it}^0)}{[1-G_\varepsilon(z_{it}^0)]G_\varepsilon(z_{it}^0)}u_{it}^0\right\|^2\right\}^{\frac{1}{2}} \cdot\left\|\widehat{\gamma}_i-\gamma_{0i}\right\| \cdot\left\{\frac{1}{T}\sum_{t=1}^T \|\widehat{f}_t-f_{0t}\|^2\right\}^{\frac{1}{2}}=O_P\left(\frac{1}{N\wedge T}\right).
\end{eqnarray*}
Therefore, we have

\begin{eqnarray}
\frac{1}{T}\sum_{t=1}^T a_{it}^{-1}a_{21,it}^{(2)\ast}&=&\frac{1}{T}\sum_{t=1}^T \frac{[y_{it}-G_\varepsilon(z_{it}^0)]g^{(1)}_\varepsilon(z_{it}^0)}{[1-G_\varepsilon(z_{it}^0)]G_\varepsilon(z_{it}^0)}u_{it}^0\gamma_{0i}^{\prime}(\widehat{f}_t-f_{0t})+O_P\left(\frac{1}{N\wedge T}\right).\label{aa4}
\end{eqnarray}

For $a_{22,it}^{(2)\ast}$, 

\begin{eqnarray}\label{aa5old}
&&\frac{1}{T}\sum_{t=1}^T a_{it}^{-1}a_{22,it}^{(2)\ast}=\frac{1}{2T}\sum_{t=1}^T \frac{[y_{it}-G_\varepsilon(z_{it}^0)]g^{(2)}_\varepsilon(\ddot{z}_{it})}{[1-G_\varepsilon(z_{it}^0)]G_\varepsilon(z_{it}^0)}u_{it}^0(\widehat{z}_{it}-z_{it}^0)^2
\nonumber\\
&=&\frac{1}{2T}\sum_{t=1}^T \frac{[y_{it}-G_\varepsilon(z_{it}^0)]g^{(2)}_\varepsilon(\ddot{z}_{it})}{[1-G_\varepsilon(z_{it}^0)]G_\varepsilon(z_{it}^0)}u_{it}^0(u_{it}^{0\prime}(\widehat{\theta}_i-\theta_{0i}))^2
+\frac{1}{2T}\sum_{t=1}^T \frac{[y_{it}-G_\varepsilon(z_{it}^0)]g^{(2)}_\varepsilon(\ddot{z}_{it})}{[1-G_\varepsilon(z_{it}^0)]G_\varepsilon(z_{it}^0)}u_{it}^0(\gamma_{0i}^{\prime}(\widehat{f}_t-f_{0t}))^2
\nonumber\\
&&+\frac{1}{2T}\sum_{t=1}^T \frac{[y_{it}-G_\varepsilon(z_{it}^0)]g^{(2)}_\varepsilon(\ddot{z}_{it})}{[1-G_\varepsilon(z_{it}^0)]G_\varepsilon(z_{it}^0)}u_{it}^0((\widehat{\gamma}_i-\gamma_{0i})^{\prime}(\widehat{f}_t-f_{0t}))^2+\text{interaction terms}.
\end{eqnarray} 

Using Cauchy-Schwarz inequality, we can show the interaction terms are bounded in probability by the first three terms on the right-hand side of  \eqref{aa5old}. Therefore, we only consider the first three terms.  Recall that $e_{it}=-\frac{[y_{it}-G_\varepsilon(z_{it}^0)]}{[1-G_\varepsilon(z_{it}^0)]G_\varepsilon(z_{it}^0)}$. For the first term, 

\begin{eqnarray}\label{ag0}
&&\left\|\frac{1}{2T}\sum_{t=1}^T \frac{[y_{it}-G_\varepsilon(z_{it}^0)]g^{(2)}_\varepsilon(\ddot{z}_{it})}{[1-G_\varepsilon(z_{it}^0)]G_\varepsilon(z_{it}^0)}u_{it}^0(u_{it}^{0\prime}(\widehat{\theta}_i-\theta_{0i}))^2\right\|
\nonumber\\
&&\leq \frac{1}{2T}\sum_{t=1}^T \|e_{it}\| \cdot \|g^{(2)}_\varepsilon(\ddot{z}_{it})\|\cdot \|u_{it}^0\|^2 \cdot \|\widehat{\theta}_i-\theta_{0i}\|^2
\nonumber\\
&&\leq  \frac{1}{2T} \left(\sum_{t=1}^T e_{it}^2\right)^{\frac{1}{2}}
\left(\sum_{t=1}^T g^{(2)}_\varepsilon(\ddot{z}_{it})^2\cdot\| u_{it}^0\|^4 \right)^{\frac{1}{2}}\cdot \|\widehat{\theta}_i-\theta_{0i}\|^2=O_P(\|\widehat{\theta}_i-\theta_{0i}\|^2).
\end{eqnarray}
where the second inequality holds by   Assumption \ref{Ass2} and Cauchy-Schwarz inequality. Analogously to \eqref{ag0}, we can compute the  probability orders for the second and third terms on the right-hand side of \eqref{aa5old}, which are $O_P\left(\frac{\log (NT)}{N\wedge T}\right)$ and $o_P\left(\frac{\log (NT)}{N\wedge T}\right)$, respectively.  Therefore, we can obtain the following result for $a_{22,it}^{(2)\ast}$. 

\begin{eqnarray}
\frac{1}{T}\sum_{t=1}^T a_{it}^{-1}a_{22,it}^{(2)\ast}=O_P(\|\widehat{\theta}_i-\theta_{0i}\|^2)+O_P\left(\frac{\log (NT)}{N\wedge T}\right).\label{aa5}
\end{eqnarray}

By \eqref{aa4} and \eqref{aa5},

\begin{eqnarray}\label{aa6}
\frac{1}{T}\sum_{t=1}^T a_{it}^{-1}a_{2,it}^{(2)\ast}&=&\frac{1}{T}\sum_{t=1}^T \frac{[y_{it}-G_\varepsilon(z_{it}^0)]g^{(1)}_\varepsilon(z_{it}^0)}{[1-G_\varepsilon(z_{it}^0)]G_\varepsilon(z_{it}^0)}u_{it}^0\gamma_{0i}^{\prime}(\widehat{f}_t-f_{0t})+O_P(\|\widehat{\theta}_i-\theta_{0i}\|^2)+O_P\left(\frac{\log (NT)}{N\wedge T}\right).
\nonumber\\
\end{eqnarray}

At the current stage, we have obtained the leading terms in $a_{2,it}^{(1)\ast}$ and $a_{2,it}^{(2)\ast}$. Following the argument analogously to that for these two terms, we can derive the leading terms for $a_{2,it}^{(3)\ast}$ and $a_{2,it}^{(4)\ast}$. Therefore, we omit the proofs and provide the results directly: 
\begin{eqnarray}\label{aa7}
&& \frac{1}{T}\sum_{t=1}^T a_{it}^{-1}a_{2,it}^{(3)\ast} = \frac{1}{T}\sum_{t=1}^T \frac{[y_{it}-G_\varepsilon(z_{it}^0)][g_\varepsilon(z_{it}^0)]^2}{[1-G_\varepsilon(z_{it}^0)]^2G_\varepsilon(z_{it}^0)}u_{it}^0\gamma_{0i}^{\prime}(\widehat{f}_t-f_{0t})+O_P(\|\widehat{\theta}_i-\theta_{0i}\|^2) +O_P\left(\frac{\log (NT)}{N\wedge T}\right),\nonumber\\
&& \frac{1}{T}\sum_{t=1}^T a_{it}^{-1}a_{2,it}^{(4)\ast} = -\frac{1}{T}\sum_{t=1}^T \frac{[y_{it}-G_\varepsilon(z_{it}^0)][g_\varepsilon(z_{it}^0)]^2}{[1-G_\varepsilon(z_{it}^0)][G_\varepsilon(z_{it}^0)]^2}u_{it}^0\gamma_{0i}^{\prime}(\widehat{f}_t-f_{0t})+O_P(\|\widehat{\theta}_i-\theta_{0i}\|^2) +O_P\left(\frac{\log (NT)}{N\wedge T}\right).
\nonumber\\
\end{eqnarray}

For $a_{2,it}^{(5)\ast}$, recall that we have $u_{it}^0=(x_{it}',f_{0t}')'$ and $\widehat{u}_{it}=(x_{it}',\widehat{f}_t')'$, we obtain that  
\begin{eqnarray}\label{aa8}
\frac{1}{T}\sum_{t=1}^T a_{it}^{-1}a_{2,it}^{(5)\ast}&=&\frac{1}{T}\sum_{t=1}^T \frac{[y_{it}-G_\varepsilon(z_{it}^0)]g_\varepsilon(z_{it}^0)}{[1-G_\varepsilon(z_{it}^0)]G_\varepsilon(z_{it}^0)}(\widehat{u}_{it}-u^0_{it})
\nonumber\\
&=&\frac{1}{T}\sum_{t=1}^T \frac{[y_{it}-G_\varepsilon(z_{it}^0)]g_\varepsilon(z_{it}^0)}{[1-G_\varepsilon(z_{it}^0)]G_\varepsilon(z_{it}^0)}(0_{d_\beta}^\prime,(\widehat{f}_t-f_{0t})^\prime)^\prime.
\end{eqnarray}

For $a_{2,it}^{(6)\ast}$, by the Taylor expansion, write
\begin{eqnarray*}
a_{2,it}^{(6)\ast}&=&-[G_\varepsilon(\widehat{z}_{it})-G_\varepsilon(z^0_{it})][g_\varepsilon(\widehat{z}_{it})-g_\varepsilon(z_{it}^0)][1-G_\varepsilon(z_{it}^0)]G_\varepsilon(z_{it}^0)u_{it}^0
\nonumber\\
&=&-g_\varepsilon(z^\dagger_{it})g^{(1)}_\varepsilon(z^\ddagger_{it})[1-G_\varepsilon(z_{it}^0)]G_\varepsilon(z_{it}^0)u_{it}^0(\widehat{z}_{it}-z_{it}^0)^2,
\end{eqnarray*}
where $z^\dagger_{it}$ and $z^\ddagger_{it}$ lie between $\widehat{z}_{it}$ and $z_{it}^0$. Then by Lemma \ref{Lemmauni} and \eqref{rev60}, we obtain that
\begin{eqnarray}\label{aa9}
&& \frac{1}{T}\sum_{t=1}^T a_{it}^{-1}a_{2,it}^{(6)\ast} = -\frac{1}{T}\sum_{t=1}^T \frac{g_\varepsilon(z^\dagger_{it})g^{(1)}_\varepsilon(z^\ddagger_{it})}{[1-G_\varepsilon(z_{it}^0)]G_\varepsilon(z_{it}^0)}u_{it}^0(\widehat{z}_{it}-z_{it}^0)^2
\nonumber\\
&& = O_P (\|\widehat{\theta}_i-\theta_{0i}\|^2)+O_P\left(\frac{1}{T}\|\widehat{F}-F_{0} \|^2\right)
= O_P\left(\frac{\log (NT)}{N\wedge T}\right).
\end{eqnarray}

The derivations for the terms with $a_{2,it}^{(7)\ast}$, $a_{2,it}^{(8)\ast}$, $a_{2,it}^{(9)\ast}$ and $a_{2,it}^{(10)\ast}$ are analogously and one can easily shows it by the Taylor expansion, Lemma \ref{Lemmauni} and \eqref{rev60}. Therefore, the detailed proofs for these terms are omitted  and we list the results directly here:
\begin{eqnarray}\label{aa10}
\frac{1}{T}\sum_{t=1}^T a_{it}^{-1}a_{2,it}^{(j)\ast}=O_P (\|\widehat{\theta}_i-\theta_{0i}\|^2)+O_P\left(\frac{\log (NT)}{N\wedge T}\right),
\end{eqnarray}
for $j=7,8,9,10$.

We have finished all the derivations for these ten terms in $a_{2,it}^{\ast}$ and we are ready to combine the leading terms in them. 
By \eqref{aa11}, \eqref{aa6}, \eqref{aa7}, \eqref{aa8}, \eqref{aa9} and \eqref{aa10}, we have
\begin{eqnarray*}
&& A_{3Ti}=\frac{1}{T}\sum_{t=1}^T a_{it}^{-1}a_{2,it}^{\ast} = -\Sigma_{u,i}(\widehat{\theta}_i-\theta_{0i})-\frac{1}{T}\sum_{t=1}^T \frac{g^2_\varepsilon(z_{it}^0)}{[1-G_\varepsilon(z_{it}^0)]G_\varepsilon(z_{it}^0)}u_{it}^0\gamma_{0i}^{\prime}(\widehat{f}_t-f_{0t})
\nonumber\\
&&+\frac{1}{T}\sum_{t=1}^T \frac{[y_{it}-G_\varepsilon(z_{it}^0)]g^{(1)}_\varepsilon(z_{it}^0)}{[1-G_\varepsilon(z_{it}^0)]G_\varepsilon(z_{it}^0)}u_{it}^0\gamma_{0i}^{\prime}(\widehat{f}_t-f_{0t})\nonumber\\
&&+\frac{1}{T}\sum_{t=1}^T \frac{[y_{it}-G_\varepsilon(z_{it}^0)][g_\varepsilon(z_{it}^0)]^2}{[1-G_\varepsilon(z_{it}^0)]^2G_\varepsilon(z_{it}^0)}u_{it}^0\gamma_{0i}^{\prime}(\widehat{f}_t-f_{0t})\nonumber\\
&&-\frac{1}{T}\sum_{t=1}^T \frac{[y_{it}-G_\varepsilon(z_{it}^0)][g_\varepsilon(z_{it}^0)]^2}{[1-G_\varepsilon(z_{it}^0)][G_\varepsilon(z_{it}^0)]^2}u_{it}^0\gamma_{0i}^{\prime}(\widehat{f}_t-f_{0t})\nonumber \\
&&+\frac{1}{T}\sum_{t=1}^T \frac{[y_{it}-G_\varepsilon(z_{it}^0)]g_\varepsilon(z_{it}^0)}{[1-G_\varepsilon(z_{it}^0)]G_\varepsilon(z_{it}^0)}(0_{d_\beta}^\prime,(\widehat{f}_t-f_{0t})^\prime)^\prime +O_P\left(\frac{\log (NT)}{N\wedge T}\right). 
\end{eqnarray*}

We then proceed with  $A_{4Ti}$. Note that
\begin{eqnarray*}
 a_{it}^\dagger-a_{it}&=&[1-G_\varepsilon(z_{it}^0)]G_\varepsilon(z_{it}^0)\left\{[1-G_\varepsilon(\widehat{z}_{it})]G_\varepsilon(\widehat{z}_{it})-[1-G_\varepsilon(z_{it}^0)][G_\varepsilon(z_{it}^0)]\right\}
 \nonumber\\
&=&-[1-G_\varepsilon(z_{it}^0)][G_\varepsilon(z_{it}^0)]^2[G(\widehat{z}_{it})-G(z^0_{it})] \\
&&+[1-G_\varepsilon(z_{it}^0)]^2G_\varepsilon(z_{it}^0)[G(\widehat{z}_{it})-G(z^0_{it})]\\
&&-[1-G_\varepsilon(z_{it}^0)]G_\varepsilon(z_{it}^0)[G(\widehat{z}_{it})-G(z^0_{it})]^2.
\end{eqnarray*}

Then by Taylor expansion and Lemma \ref{Lemmauni}, 
\begin{eqnarray*}
&&\frac{1}{T}\sum_{t=1}^T (a_{it}^{\dagger-1}-a_{it}^{-1})a_{2,it}^{(1)\ast} = -\frac{1}{T}\sum_{t=1}^T a_{it}^{\dagger-1}a_{it}^{-1}(a_{it}^{\dagger}-a_{it})a_{2,it}^{(1)\ast}
\nonumber\\
&=&-\frac{1}{T}\sum_{t=1}^T a_{it}^{\dagger-1}[G_\varepsilon(\widehat{z}_{it})-G_\varepsilon(z^0_{it})]^2G_\varepsilon(z_{it}^0)g_\varepsilon(z_{it}^0)u_{it}^0
\nonumber\\
&&+\frac{1}{T}\sum_{t=1}^T a_{it}^{\dagger-1}[G_\varepsilon(\widehat{z}_{it})-G_\varepsilon(z^0_{it})]^2[1-G_\varepsilon(z_{it}^0)]g_\varepsilon(z_{it}^0)u_{it}^0
\nonumber\\
&&-\frac{1}{T}\sum_{t=1}^T a_{it}^{\dagger-1}[G_\varepsilon(\widehat{z}_{it})-G_\varepsilon(z^0_{it})]^3g_\varepsilon(z_{it}^0)u_{it}^0
+O_P\left(\frac{1}{N\wedge T}\right)
\nonumber\\
&=&-\frac{1}{T}\sum_{t=1}^T a_{it}^{\dagger-1}G_\varepsilon(z_{it}^0)[g_\varepsilon(z^\dagger_{it})]^2g_\varepsilon(z_{it}^0)u_{it}^0(\widehat{z}_{it}-z^0_{it})^2
\nonumber\\
&&+\frac{1}{T}\sum_{t=1}^T a_{it}^{\dagger-1}[1-G_\varepsilon(z_{it}^0)][g_\varepsilon(z^\dagger_{it})]^2g_\varepsilon(z_{it}^0)u_{it}^0(\widehat{z}_{it}-z^0_{it})^2
\nonumber\\
&&-\frac{1}{T}\sum_{t=1}^T a_{it}^{\dagger-1}[g_\varepsilon(z^\dagger_{it})]^3g_\varepsilon(z_{it}^0)u_{it}^0(\widehat{z}_{it}-z^0_{it})^3
+O_P\left(\frac{1}{N\wedge T}\right) = O_P\left(\frac{1}{N\wedge T}\right).
\end{eqnarray*}

Following analogous arguments, we can show that the rest terms in $T^{-1}\sum_{t=1}^T (a_{it}^{\dagger-1}-a_{it}^{-1})a_{2,it}^{\ast}$ are bounded by probability of the order $O_P\left(\frac{1}{N\wedge T}\right)$. Therefore, 
\begin{eqnarray}
A_{4Ti}=O_P\left(\frac{1}{N\wedge T}\right).\label{aa15}
\end{eqnarray}

Finishing the discussions on $A_{3Ti}$ and $A_{4Ti}$,  we have derived the leading terms in $A_{2Ti}$, all of which depend on the convergence of $\widehat{f}_t-f_{0t}$. Since $A_{1Ti}$ only  contains real values which can contribute to the CLT, we leave it for further discussions.  By \eqref{aa13}, \eqref{aa14}, \eqref{aa11} and \eqref{aa15}, we have
\begin{eqnarray*} 
\widehat{\theta}_i-\theta_{0i}&=&\Sigma_{u,i}^{-1}A_{1Ti}-\frac{1}{T}\Sigma_{u,i}^{-1}\sum_{t=1}^T \frac{g^2_\varepsilon(z_{it}^0)}{[1-G_\varepsilon(z_{it}^0)]G_\varepsilon(z_{it}^0)}u_{it}^0\gamma_{0i}^{\prime}(\widehat{f}_t-f_{0t})
\nonumber\\
&&+\frac{1}{T}\Sigma_{u,i}^{-1}\sum_{t=1}^T \frac{[y_{it}-G_\varepsilon(z_{it}^0)]g^{(1)}_\varepsilon(z_{it}^0)}{[1-G_\varepsilon(z_{it}^0)]G_\varepsilon(z_{it}^0)}u_{it}^0\gamma_{0i}^{\prime}(\widehat{f}_t-f_{0t})
\nonumber\\
&&+\frac{1}{T}\Sigma_{u,i}^{-1}\sum_{t=1}^T \frac{[y_{it}-G_\varepsilon(z_{it}^0)][g_\varepsilon(z_{it}^0)]^2}{[1-G_\varepsilon(z_{it}^0)]^2G_\varepsilon(z_{it}^0)}u_{it}^0\gamma_{0i}^{\prime}(\widehat{f}_t-f_{0t})
\nonumber\\
&&-\frac{1}{T}\Sigma_{u,i}^{-1}\sum_{t=1}^T \frac{[y_{it}-G_\varepsilon(z_{it}^0)][g_\varepsilon(z_{it}^0)]^2}{[1-G_\varepsilon(z_{it}^0)][G_\varepsilon(z_{it}^0)]^2}u_{it}^0\gamma_{0i}^{\prime}(\widehat{f}_t-f_{0t})
\nonumber\\
&&+\frac{1}{T}\Sigma_{u,i}^{-1}\sum_{t=1}^T \frac{[y_{it}-G_\varepsilon(z_{it}^0)]g_\varepsilon(z_{it}^0)}{[1-G_\varepsilon(z_{it}^0)]G_\varepsilon(z_{it}^0)}(0_{d_\beta}^\prime,(\widehat{f}_t-f_{0t})^\prime)^\prime +O_P\left(\frac{\log (NT)}{N\wedge T}\right)\nonumber\\
&:=&A_{5Ti}+\cdots+A_{10Ti}+O_P\left(\frac{\log (NT)}{N\wedge T}\right).
\end{eqnarray*}

We can use the results in the proof of Lemma \ref{Lemmauni} to show the convergence of these terms. Since the proofs are analogous to those in the proof of Lemma \ref{Lemmauni}, we omit some details which are repetitive to save pages for the main context.  We proceed with the derivation of $A_{5Ti}$. Recall that we have the following notation: $\Omega_u=\text{diag}(\Sigma_{u,1}, \cdots, \Sigma_{u,N})$, $\Omega_{u\gamma, it} =E\left[\frac{g^2_\varepsilon(z_{it}^0)}{[1-G_\varepsilon(z_{it}^0)]G_\varepsilon(z_{it}^0)}u_{it}^0\gamma_{0i}^{\prime}\right]$, and $\Omega_{u\gamma}$ is the  matrix with its $(i,t)$-th block being $\Omega_{u\gamma, it}$. Let further $\zeta_{it}=\frac{g^2_\varepsilon(z_{it}^0)}{[1-G_\varepsilon(z_{it}^0)]G_\varepsilon(z_{it}^0)}u_{it}^0\gamma_{0i}^{\prime}- \Omega_{u\gamma, it}$ and $C_{4NT}$ be the matrix with its $(i,t)$-th block being $\zeta_{it}$. Let $\mathcal{I}_i=(0_{d_\beta+d_f},\cdots, I_{d_\beta+d_f}', \cdots,0_{d_\beta+d_f})$ to be a $N\times 1$ block matrix with its $i$-th block being the $(d_\beta+d_f)\times (d_\beta+d_f)$ identity matrix and other blocks being $(d_\beta+d_f)\times (d_\beta+d_f)$ zero matrices. With this notation, we have

\begin{eqnarray*} 
A_{6Ti}&=&-\frac{1}{T}\mathcal{I}_i' \Omega_{u}^{-1}\Omega_{u\gamma}(\widehat{F}_v-F_{0v})-\frac{1}{T}\mathcal{I}_i' \Omega_{u}^{-1}C_{4NT}(\widehat{F}_v-F_{0v})
\nonumber\\
&=&A_{11Ti}+A_{12Ti}.
\end{eqnarray*}
For $A_{11Ti}$, by \eqref{rev42} in online supplementary Appendix \ref{AppB},

\begin{eqnarray}\label{rev62}
\frac{1}{T}\mathcal{I}_i' \Omega_{u}^{-1}\Omega_{u\gamma}(\widehat{F}_v-F_{0v}) = \frac{1}{T}\mathcal{I}_i' \Omega_{u}^{-1}\Omega_{u\gamma}\left(\mathcal{P}_{NT,1}+\cdots+\mathcal{P}_{NT,6}\right),
\end{eqnarray}
where $\mathcal{P}_{NT,1}$, $\ldots$, $\mathcal{P}_{NT,6}$ are defined in \eqref{rev42}.

For the first term in the product on the right-hand side of \eqref{rev62}, 

\begin{eqnarray*}
\frac{1}{T}\mathcal{I}_i' \Omega_{u}^{-1}\Omega_{u\gamma}\mathcal{P}_{NT,1}&=&\frac{1}{NT}\mathcal{I}_i' \Omega_{u}^{-1}\Omega_{u\gamma}\Omega_\gamma^{-1}\cdot\frac{\partial \log L(\Theta_0, F_0)}{\partial F_v}
\nonumber\\
&=&\frac{1}{NT}\Sigma_{u,i}^{-1}  \sum_{j=1}^N\sum_{t=1}^T  \frac{[y_{jt}-G_\varepsilon(z_{jt}^0)]g_\varepsilon(z_{jt}^0) }{[1-G_\varepsilon(z_{jt}^0)]G_\varepsilon(z_{jt}^0)} \Omega_{u\gamma,it}\Omega_{\gamma}^{-1} \gamma_{0j}
\nonumber\\
&=&-\frac{1}{NT}\Sigma_{u,i}^{-1}  \sum_{j=1}^N\sum_{t=1}^T  g_\varepsilon(z_{jt}^0)\Omega_{u\gamma,it}\Sigma_{\gamma,t}^{-1} \gamma_{0j} e_{jt}. 
\end{eqnarray*}
For the first moment, we can easily show that $E\left[\frac{1}{T}\mathcal{I}_i' \Omega_{u}^{-1}\Omega_{u\gamma}\mathcal{P}_{NT,1}\right]=0$. For the second moment,

\begin{eqnarray*}
&&E\left\|\frac{1}{T}\mathcal{I}_i' \Omega_{u}^{-1}\Omega_{u\gamma}\mathcal{P}_{NT,1}\right\|^2=O\left(\frac{1}{N^2T^2}\right) E\left\| \sum_{j=1}^N\sum_{t=1}^T  g_\varepsilon(z_{jt}^0)\Omega_{u\gamma,it}\Sigma_{\gamma,t}^{-1} \gamma_{0j} e_{jt}\right\|^2
\nonumber\\
&=&O\left(\frac{1}{N^2T^2}\right) \sum_{j_1=1}^N \sum_{j_2=1}^N\sum_{t=1}^T\sum_{s=1}^TE[\|\gamma_{0j_1}\| \cdot \|\gamma_{0j_2}\| \cdot | E[ e_{j_1t}e_{j_2s}|\,\mathcal{W}]|]   
\nonumber\\
&\leq & O\left(\frac{c_{\delta}}{N^2T^2}\right) \sum_{j_1=1}^N \sum_{j_2=1}^N\sum_{t=1}^T\sum_{s=1}^T
\alpha_{j_1j_2}(|t-s|)^{\delta/(4+\delta)}E\left[\|\gamma_{0j_1}\| \cdot \|\gamma_{0j_2}\|\cdot  E\left[|e_{it}|^{2+\delta/2} |\, \mathcal{W}\right]^{2/(4+\delta)} \right.
\nonumber\\
&&\quad \left.
\cdot E\left[|e_{is}|^{2+\delta/2} |\, \mathcal{W}\right]^{2/(4+\delta)}\right]
\nonumber\\
&=&O\left(\frac{1}{NT}\right),
\end{eqnarray*}
where $c_{\delta}=(4+\delta)/\delta\cdot2^{(4+2\delta)/(4+\delta)}$; the second inequality  holds by Davydov's inequality for $\alpha$-mixing process  and the last equality holds by the $\alpha$-mixing and moment conditions in  Assumption \ref{Ass2}. It immediately yields that

\begin{eqnarray}\label{rev63}
\left\|\frac{1}{T}\mathcal{I}_i' \Omega_{u}^{-1}\Omega_{u\gamma}\mathcal{P}_{NT,1}\right\|=O_P\left(\frac{1}{\sqrt{NT}}\right).
\end{eqnarray}
Analogously, we can show that

\begin{eqnarray}\label{rev64}
\left\|\frac{1}{T}\mathcal{I}_i' \Omega_{u}^{-1}\Omega_{u\gamma}\mathcal{P}_{NT,2}\right\|=O_P\left(\frac{1}{\sqrt{NT}}\right),\quad \left\|\frac{1}{T}\mathcal{I}_i' \Omega_{u}^{-1}\Omega_{u\gamma}\mathcal{P}_{NT,3}\right\|=O_P\left(\frac{1}{\sqrt{NT}}\right).
\end{eqnarray}

We can use analogous arguments in the proofs of  \eqref{rev48}, \eqref{rev49}  and  \eqref{rev56}  in online supplementary Appendix \ref{AppB} to obtain the following results:

\begin{eqnarray}\label{rev65}
\left\|\frac{1}{T}\mathcal{I}_i' \Omega_{u}^{-1}\Omega_{u\gamma}\mathcal{P}_{NT,4}\right\|&=&O_P\left(\frac{1}{\sqrt{NT}} \|\widehat{\Theta}_v-\Theta_{0v}\|\right)+O_P\left(\frac{1}{T}\|\widehat{F}-F_{0v}\|^2\right)+O_P\left(\frac{1}{N}\|\widehat{\Theta}_v-\Theta_{0v} \|^2\right)
\nonumber\\
&=&O_P\left(\frac{1}{N\wedge T}\right),
\end{eqnarray}

\begin{eqnarray}\label{rev66}
\left\|\frac{1}{T}\mathcal{I}_i' \Omega_{u}^{-1}\Omega_{u\gamma}\mathcal{P}_{NT,5}\right\|&=&O_P\left(\frac{1}{T} \|\widehat{F}_v-F_{0v}\|\right)+O_P\left(\frac{1}{T}\|\widehat{F}-F_{0v}\|^2\right)+O_P\left(\frac{1}{N}\|\widehat{\Theta}_v-\Theta_{0v} \|^2\right)
\nonumber\\
&&+O_P\left( \frac{1}{\sqrt{N}}\|\widehat{\Theta}_v-\Theta_{0v}\| \cdot \frac{1}{\sqrt{T}}\|\widehat{F}_v-F_{0v} \|\right)
\nonumber\\
&=&O_P\left(\frac{1}{N\wedge T}\right),
\end{eqnarray}
and

\begin{eqnarray}\label{rev67}
\left\|\frac{1}{T}\mathcal{I}_i' \Omega_{u}^{-1}\Omega_{u\gamma}\mathcal{P}_{NT,6}\right\|&=& O_P\left(\frac{1}{\sqrt{NT}}\right)
+O_P\left( \frac{1}{N}\|\widehat{\Theta}_v-\Theta_{0v} \|^2\right)+O_P\left(  \frac{1}{T}\|\widehat{F}_v-F_{0v} \|^2\right)
\nonumber\\
&&
+O_P\left( \frac{1}{\sqrt{N}}\|\widehat{\Theta}_v-\Theta_{0v}\| \cdot \frac{1}{\sqrt{T}}\|\widehat{F}_v-F_{0v} \|\right)
\nonumber\\
&=&O_P\left(\frac{1}{N\wedge T}\right).
\end{eqnarray}
By \eqref{rev62}, \eqref{rev63}, \eqref{rev64}, \eqref{rev65}, \eqref{rev66} and \eqref{rev67}, 

\begin{eqnarray}\label{rev68}
A_{11Ti} = O_P\left(\frac{1}{N\wedge T}\right).
\end{eqnarray}

For $A_{12Ti}$, since we have discussed the probability order of $\Omega_{u}^{-1}C_{4NT}(\widehat{F}_v-F_{0v})$ in the proof of Lemma \ref{Lemmauni},  we can then follow analogous arguments in the proof of \eqref{rev54}  in online supplementary Appendix \ref{AppB} to show that

\begin{eqnarray}\label{rev69}
A_{12Ti} 
&=& O_P\left(\frac{1}{\sqrt{NT}}\right)
+O_P\left( \frac{1}{N}\|\widehat{\Theta}_v-\Theta_{0v} \|^2\right)+O_P\left(  \frac{1}{T}\|\widehat{F}_v-F_{0v} \|^2\right)
\nonumber\\
&&
+O_P\left( \frac{1}{\sqrt{N}}\|\widehat{\Theta}_v-\Theta_{0v}\| \cdot \frac{1}{\sqrt{T}}\|\widehat{F}_v-F_{0v} \|\right)
\nonumber\\
&=& O_P\left(\frac{1}{N\wedge T}\right).
\end{eqnarray}
By \eqref{rev68} and \eqref{rev69}, 

\begin{eqnarray}\label{rev70}
A_{6Ti} = O_P\left(\frac{1}{N\wedge T}\right).
\end{eqnarray}

For $A_{7Ti}$, $\ldots$, $A_{10Ti}$,  we observe that they have the same probability orders with $\frac{1}{T}\mathcal{I}_i' \Omega_{u}^{-1}C_{5NT}(\widehat{F}_v-F_{0v})$ and $\frac{1}{T}\mathcal{I}_i' \Omega_{u}^{-1}C_{6NT}(\widehat{F}_v-F_{0v})$. Analogously to \eqref{rev54} in the proof of Lemma \ref{Lemmauni}  in online supplementary Appendix \ref{AppB}, we can show that they are also bounded in probability by $O_P\left(\frac{1}{N\wedge T}\right)$. Therefore, we are ready to conclude that 

\begin{eqnarray}\label{subs4}
\widehat{\theta}_i-\theta_{0i}&=&\Sigma_{u,i}^{-1}A_{1Ti}+O_P\left(\frac{\log (NT)}{N\wedge T}\right).
\end{eqnarray}
By  \eqref{subs4}, Lemma \ref{LemmaT4} and the conditions in the body of this theorem, the proof of Theorem \ref{Theorem2.1}.(1) is complete.

(2) The proof of  Theorem \ref{Theorem2.1}.(2) is analogous to that for Theorem \ref{Theorem2.1}.(1). Thus, it is omitted here. 
 \hspace*{\fill}{$\blacksquare$}

\end{appendices}

\newpage
\setcounter{page}{1}

\begin{appendices}
\begin{center}
{\Large \bf Supplementary Appendix B to \\``Binary Response Models for Heterogeneous Panel Data
with Interactive Fixed Effects"}

\medskip

{\sc Jiti Gao$^\sharp$, Fei Liu$^{\ast}$, and Bin Peng$^{\sharp}$ and Yayi Yan$^{\sharp}$}
\medskip

$^\sharp$Monash University and $^{\ast}$Nankai University
\end{center}

In this supplementary file, Appendix \ref{Theo23} presents the proofs of Theorem \ref{Theorem2.2} and Theorem \ref{Theorem2.3}. The proofs of  Lemmas \ref{Lemma2.1}-\ref{Theorem2.3} and Lemma \ref{LemmaAPE} are given in Appendix \ref{ProofLem}. Some secondary results are finally summarized in Appendix \ref{TechLem}.

\section{}\label{AppB}

%
%
%

\subsection{Proofs of Theorem \ref{Theorem2.2} and Theorem \ref{Theorem2.3}}\label{Theo23}

\noindent {\bf Proof of Theorem \ref{Theorem2.2}:}

It follows from equation (\ref{subs4}) that we have 
\be
\widehat{\beta}_i - \beta_{0i} =  \frac{1}{T} \sum_{t=1}^T \frac{[y_{it}-G_\varepsilon(z_{it}^0)]g_\varepsilon(z_{it}^0)\, \Sigma_{u, i}^{(d_\beta)}}{[1-G_\varepsilon(z_{it}^0)]G_\varepsilon(z_{it}^0)}   u_{it}^0 + O_P\left(\frac{1}{N\wedge T}\right),
\label{jitib1}
\ee
where $\Sigma^{(d_\beta)}_{u, i}$ corresponds to the first $d_\beta$ rows of $\Sigma^{-1}_{u,i}$.

Thus, we have
\bea
&& \frac{1}{N} \sum_{i=1}^N \left(\widehat{\beta}_i - \beta_{0i}\right) =  \frac{1}{NT} \sum_{t=1}^T \sum_{i=1}^N \frac{[y_{it}-G_\varepsilon(z_{it}^0)]g_\varepsilon(z_{it}^0)}{[1-G_\varepsilon(z_{it}^0)]G_\varepsilon(z_{it}^0)} \Sigma^{(d_\beta)}_{u, i} u_{it}^0  + O_P\left(\frac{1}{N\wedge T}\right)
\nonumber\\
&& =  \frac{1}{NT} \sum_{t=1}^T \sum_{i=1}^N \frac{[y_{it}-G_\varepsilon(z_{it}^{0})]g_\varepsilon(z_{it}^{0})}{[1-G_\varepsilon(z_{it}^{0})]G_\varepsilon(z_{it}^{0})}   \Sigma^{(d_\beta)}_{u, i} u_{it}^0 + O_P\left(\frac{1}{N\wedge T}\right),
\label{jitib2}
\eea
where $z_{it}^{0} = x_{it}^{\prime} \beta_{0i} + \lambda_{0i}^{\prime} f_t$. Note further that $\beta_{0i}=\beta_0 + O_P\left(\frac{1}{N^{\alpha}}\right)$ as $N\rightarrow \infty$.

In order to establish an asymptotic distribution for each of the cases in Theorem \ref{Theorem2.2}, we will need to deal with the following bias term:
\be
\frac{\sqrt{T} \, N^{\alpha}}{N\wedge T} = \max\left(\frac{\sqrt{T}}{N^{1-\alpha}}, \frac{N^{\alpha}}{\sqrt{T}}\right).
 \label{jitib3}
 \ee

We now complete the proof of Theorem \ref{Theorem2.2}. Let us start with the proof of Theorem \ref{Theorem2.2}(1).

(1). Observe that for $0\le \alpha<\frac{1}{2}$
\be
\sqrt{T} \, N^{\alpha}  ({\widehat{\beta}} - {\beta}_0 ) = \sqrt{T} \, N^{\alpha}  ({\widehat{\beta}} - \overline{\beta}_{0} + \overline{\beta}_{0} - {\beta}_0 ) 
 = \frac{\sqrt{T}}{N^{1-\alpha}} \sum_{i=1}^N (\widehat{\beta}_i-\beta_{i0} ) + \sqrt{\frac{T}{N}} \frac{1}{\sqrt{N}} \sum_{i=1}^N \eta_i.
\label{jitib4}
\ee

By equations (\ref{jitib2})-(\ref{jitib4}), we then have
\bea
&& N^{\alpha + \frac{1}{2}}  ({\widehat{\beta}} - {\beta}_0 ) = N^{\alpha+ \frac{1}{2}} \,  ({\widehat{\beta}} - \overline{\beta}_{0} + \overline{\beta}_{0} - {\beta}_0 ) 
\nonumber\\
&& = \frac{1}{N^{\frac{1}{2}-\alpha}} \, \frac{1}{T\sqrt{ N}} \sum_{t=1}^T \sum_{i=1}^N \frac{[y_{it}-G_\varepsilon(z_{it}^{0})]g_\varepsilon(z_{it}^{0})}{[1-G_\varepsilon(z_{it}^{0})]G_\varepsilon(z_{it}^{0})}  \Sigma^{(d_\beta)}_{u, i} u_{it}^0 + \frac{1}{\sqrt{N}} \sum_{i=1}^N \eta_i 
\nonumber\\
&& + O_P\left( \frac{1}{N^{\frac{1}{2}-\alpha}}\vee \frac{N^{\alpha + \frac{1}{2}}}{T} \right),
\label{jiti5}
\eea
which completes the proof of Theorem \ref{Theorem2.2}(i), in view of $N^{\frac{1}{2}-\alpha} \rightarrow \infty$ and $\frac{N^{\alpha+\frac{1}{2}}}{T}\rightarrow 0$ as $(N, T)\rightarrow (\infty, \infty)$.

\medskip

(2)-(3). For the case of $\alpha = \frac{1}{2}$, we have
 \bea
&& \sqrt{N T} ({\widehat{\beta}} - {\beta}_0 - {\rm bias}(N,T) ) = \sqrt{N T} ({\widehat{\beta}} - \overline{\beta}_{0} - {\rm bias}(N,T) + \overline{\beta}_0 - \beta_0 ) 
\nonumber\\
&& = \sqrt{N T}\cdot\frac{1}{N} \sum_{i=1}^N \left(\widehat{\beta}_i-\beta_{i0} - {\rm bias}(N,T)\right)  + \sqrt{\frac{T}{N}} \frac{1}{\sqrt{N}} \sum_{i=1}^N \eta_i
\nonumber\\
&& =  \frac{1}{\sqrt{NT}} \sum_{t=1}^T \sum_{i=1}^N  \frac{[y_{it}-G_\varepsilon(z_{it}^{0})]g_\varepsilon(z_{it}^{0})}{[1-G_\varepsilon(z_{it}^{0})]G_\varepsilon(z_{it}^{0})} \Sigma^{(d_\beta)}_{u, i}  u_{it}^0  + \sqrt{\frac{T}{N}} \frac{1}{\sqrt{N}} \sum_{i=1}^N \eta_i,
\label{jitib6}
\eea
where ${\rm bias}(N, T) = O_P\left(\frac{1}{N\wedge T}\right)$. Equation (\ref{jitib6}) indicates that we have an asymptotic distribution for each of the cases of $\sqrt{\frac{T}{N}} \rightarrow \rho \in (0, \infty]$.  \hspace*{\fill}{$\blacksquare$}

\bigskip

\noindent \textbf{Proof of Theorem \ref{Theorem2.3}:}

Without loss of generality, we consider the next two cases: Case 1 (over selection) with $\mathsf{d} = d_f+1$; and Case 2 (under selection) with $\mathsf{d} = d_f-1$.

Start with Case 1. Note that

\begin{eqnarray*}
&&\frac{1}{NT}\sum_{i=1}^N \sum_{t=1}^T\left[y_{it}-G_\varepsilon(z_{it}^0)+G_\varepsilon(z_{it}^0) - G_\varepsilon (\widehat{z}_{it}^{\mathsf{d}})\right]^2 \nonumber \\
&= &\frac{1}{NT}\sum_{i=1}^N \sum_{t=1}^T\left[y_{it}-G_\varepsilon(z_{it}^0) \right]^2+\frac{1}{NT}\sum_{i=1}^N \sum_{t=1}^T\left[ G_\varepsilon(z_{it}^0) - G_\varepsilon (\widehat{z}_{it}^{\mathsf{d}})\right]^2\nonumber \\
&&+\frac{2}{NT}\sum_{i=1}^N \sum_{t=1}^T\left[y_{it}-G_\varepsilon(z_{it}^0)\right]\cdot \left[G_\varepsilon(z_{it}^0) - G_\varepsilon (\widehat{z}_{it}^{\mathsf{d}})\right] .
\end{eqnarray*}
By Lemma \ref{Lemma2.1}, we have

\begin{eqnarray*}
\frac{1}{NT}\sum_{i=1}^N \sum_{t=1}^T\left[ G_\varepsilon(z_{it}^0) - G_\varepsilon (\widehat{z}_{it}^{\mathsf{d}})\right]^2 =O_P\left( \frac{1}{\sqrt{NT}}\right)
\end{eqnarray*}
and similar to \eqref{consistency1}, we can show that uniformly in $z_{it}^{\mathsf{d}}$'s

\begin{eqnarray*}
&&\left| \frac{1}{NT}\sum_{i=1}^N \sum_{t=1}^T\left[y_{it}-G_\varepsilon(z_{it}^0)\right]\cdot \left[G_\varepsilon(z_{it}^0) - G_\varepsilon (z_{it}^{\mathsf{d}})\right]\right|=O_P\left( \frac{1}{\sqrt{NT}}\right).
\end{eqnarray*}

Thus, simple algebra shows that

\begin{eqnarray*}
\ic(\mathsf{d}) -\ic(d_f)  &=& O_P\left( \frac{1}{\sqrt{NT}}\right)+ (\mathsf{d}-d_f)\cdot   \frac{\xi_{NT}}{\sqrt{NT}}  >0
\end{eqnarray*}
with probability approaching 1, given $\xi_{NT}\to \infty$ and $\frac{\xi_{NT}}{\sqrt{NT}} \to 0$.

\medskip

Next, consider Case 2. If we under-specify the number of factors, $\frac{1}{NT}\sum_{i=1}^N \gamma_{0i}F_0'M_{\widehat{F}}F_0\gamma_{0i}=o_P(1)$ in the proof of Lemma 2.2 is no longer achievable. By \eqref{consistency2} in the proof of Lemma \ref{Lemma2.2}, it will then yield a non-negligible bias for the term $\frac{1}{NT}\sum_{i=1}^N \sum_{t=1}^T\left[ G_\varepsilon(z_{it}^0) - G_\varepsilon (\widehat{z}_{it}^{\mathsf{d}})\right]^2$.

Based on the above development, the result follows. \hspace*{\fill}{$\blacksquare$}

\subsection{Proofs of the Lemmas}\label{ProofLem}

\noindent \textbf{Proof of Lemma \ref{Lemma2.1}:}

In the following proof, we provide a robust version which assumes that $F$ and $\Gamma$ are $T\times d_{\max}$ and $N\times d_{\max}$ respectively where $d_{\max}\geq d_f$ is a user-specified large fixed constant.

(1). For notational simplicity, let $z_{it} =x_{it}'\beta_i+\gamma_i'f_t$ and $\Delta z_{it} = z_{it} -z_{it}^0$, where $z_{it}^0$ is defined in the beginning of Section \ref{Section2.2}. Note that under Assumption \ref{Ass0}, we need to  investigate the minimization of \eqref{EQ28} under the constraint that $z_{it}$'s $\in  \Xi_{NT}$. Moreover, note that provided $0<x, x_0<1$, we have the following two expressions by the Taylor expansion:

\begin{eqnarray}
\log x &=&\log x_0 + (x -x_0)\frac{1}{x_0} - (x-x_0)^2\frac{1}{2(x^*)^2},\label{MVT1} \\
\log (1-x) &=&\log (1-x_0) - (x -x_0)\frac{1}{1-x_0} -(x -x_0)^2\frac{1}{2(1-x^\dagger)^2} , \label{MVT2}
\end{eqnarray}
where both $x^*$ and $x^\dagger$ lie between $x$ and $x_0$.

We are now ready to start our investigation. By \eqref{MVT1} and \eqref{MVT2}, write

\begin{eqnarray}\label{consistency0}
&&\log L(\beta_0, F_0, \Gamma_0) -\log L(\beta, F, \Gamma)\nonumber \\
&=& -\frac{1}{NT}\sum_{i=1}^N\sum_{t=1}^T(1-y_{it}) \left\{ \log\left[1 - G_\varepsilon(z_{it} ) \right]-\log\left[1 -G_\varepsilon(z_{it}^0) \right]\right\}\nonumber \\
&&-\frac{1}{NT}\sum_{i=1}^N\sum_{t=1}^T y_{it} \left\{ \log G_\varepsilon(z_{it} ) -\log G_\varepsilon(z_{it}^0 ) \right\} \nonumber \\
&=& \frac{1}{NT}\sum_{i=1}^N\sum_{t=1}^T[G_\varepsilon(z_{it} )  - G_\varepsilon(z_{it}^0) ] \cdot  \frac{1-y_{it}}{1- G_\varepsilon(z_{it}^0)}\nonumber \\
&&+\frac{1}{NT}\sum_{i=1}^N\sum_{t=1}^T[G_\varepsilon(z_{it} )  - G_\varepsilon(z_{it}^0 ) ]^2 \cdot  \frac{1-y_{it}}{2(1-G_{it}^\dagger)^2} \nonumber \\
&&- \frac{1}{NT}\sum_{i=1}^N\sum_{t=1}^T[G_\varepsilon(z_{it} )  - G_\varepsilon(z_{it}^0 ) ] \cdot  \frac{y_{it}}{G_\varepsilon(z_{it}^0 )}\nonumber \\
&&+\frac{1}{NT}\sum_{i=1}^N\sum_{t=1}^T[G_\varepsilon(z_{it} )  - G_\varepsilon(z_{it}^0 ) ]^2 \cdot  \frac{y_{it}}{2(G_{it}^*)^2}  \nonumber \\
&=& \frac{1}{NT}\sum_{i=1}^N\sum_{t=1}^T[G_\varepsilon(z_{it} )  - G_\varepsilon(z_{it}^0 ) ] \cdot  \left[\frac{1-y_{it}}{1- G_\varepsilon(z_{it}^0 )}- \frac{y_{it}}{G_\varepsilon(z_{it}^0 )}\right]\nonumber \\
&&+\frac{1}{NT}\sum_{i=1}^N\sum_{t=1}^T[G_\varepsilon(z_{it} )  - G_\varepsilon(z_{it}^0 ) ]^2 \cdot  \left[ \frac{1-y_{it}}{2(1-G_{it}^\dagger)^2} + \frac{y_{it}}{2(G_{it}^*)^2} \right]\nonumber \\
&:= &\mathbb{L}_{1NT} +\mathbb{L}_{2NT},
\end{eqnarray}
where both $G_{it}^*$ and $G_{it}^\dagger$ lie between $G_\varepsilon(z_{it} ) $ and $G_\varepsilon(z_{it}^0 )$, and the definitions of $\mathbb{L}_{1NT}$ and $\mathbb{L}_{2NT}$ are obvious. 

\bigskip

We then consider $\mathbb{L}_{1NT}$ and $\mathbb{L}_{2NT}$ respectively, and start with $\mathbb{L}_{1NT}$. Recall that we have defined $e_{it}$ right above Assumption \ref{Ass1}. Then consider

\begin{eqnarray*}
&&E\left[\max_{z_{it}\text{'s} } \left| \frac{1}{NT}\sum_{i=1}^N\sum_{t=1}^TG_\varepsilon(z_{it} )e_{it}\right|\right]^2 \nonumber \\
&\le &E\left[\max_{\ z_{it}\text{'s}, \ z_{js}\text{'s} } \frac{1}{N^2T^2}\sum_{i,j=1}^N\sum_{t,s=1}^T G_\varepsilon(z_{it} )G_\varepsilon(z_{js} )e_{it}e_{js}\right] \nonumber \\
&\le &E\left[\max_{\ z_{it}\text{'s}, \ z_{js}\text{'s} } \frac{1}{N^2T^2}\sum_{i,j=1}^N\sum_{t,s=1}^T G_\varepsilon(z_{it} )G_\varepsilon(z_{js} )\cdot  |E[e_{it}e_{js}\, | \, z_{it}^0, z_{js}^0]| \right]\nonumber \\
&\le &  \frac{1}{N^2T^2}\sum_{i,j=1}^N\sum_{t,s=1}^T  E[ |E[e_{it}e_{js}\, | \, z_{it}^0, z_{js}^0]| ] =O\left(\frac{1}{NT} \right),
\end{eqnarray*}
in which the third inequality follows from the fact that $ G_\varepsilon(\cdot)\le 1$ uniformly, and the last equality follows from Assumption \ref{Ass0}. Thus, it is easy to know that

\begin{eqnarray}\label{consistency1}
|\mathbb{L}_{1NT}| =O_P\left(\frac{1}{\sqrt{NT}} \right).
\end{eqnarray}

\medskip

We next investigate $\mathbb{L}_{2NT}$. Write

\begin{eqnarray*}
\mathbb{L}_2&=&\frac{1}{NT}\sum_{i=1}^N\sum_{t=1}^T[G_\varepsilon(z_{it} )  - G_\varepsilon(z_{it}^0 ) ]^2 \cdot  \left[ \frac{1-y_{it}}{2(1-G_{it}^\dagger)^2} + \frac{y_{it}}{2(G_{it}^*)^2} \right]\nonumber \\
&\ge &\frac{1}{NT}\sum_{i=1}^N\sum_{t=1}^T[G_\varepsilon(z_{it} )  - G_\varepsilon(z_{it}^0 ) ]^2 \cdot  \left\{ \frac{1-y_{it}}{4[1+(G_{it}^\dagger)^2]} + \frac{ y_{it}}{2(G_{it}^*)^2} \right\}  \nonumber \\
&\ge &\frac{1}{NT}\sum_{i=1}^N\sum_{t=1}^T[G_\varepsilon(z_{it} )  - G_\varepsilon(z_{it}^0 ) ]^2 \cdot  \left\{ \frac{1-y_{it}}{4\cdot 2} + \frac{y_{it}}{2} \right\} \nonumber \\
&\ge &\frac{1}{8}\cdot \frac{1}{NT}\sum_{i=1}^N\sum_{t=1}^T[G_\varepsilon(z_{it} )  - G_\varepsilon(z_{it}^0 ) ]^2  ,
\end{eqnarray*}
where the first inequality follows from $\frac{1}{(a+b)^2} \ge \frac{1}{2a^2 + 2b^2}$ because of $(a+b)^2\le 2a^2 + 2b^2$, the second inequality follows from the fact that $G_{it}^*$ and $G_{it}^\dagger$ lie between $G_\varepsilon(z_{it} ) $ and $G_\varepsilon(z_{it}^0 )$, and the third inequality follows from that $ \frac{1-y_{it}}{4\cdot 2} + \frac{y_{it}}{2} \ge \frac{1}{8}$ because of $y_{it}$ taking the value of 1 or 0 only.

By the fact that $0\ge \log L(B_0, F_0, \Gamma_0) -\log L (\widehat{B}, \widehat{F}, \widehat{\Gamma})$, and \eqref{consistency0} and \eqref{consistency1}, we now can conclude that

\begin{eqnarray*}
\frac{1}{NT}\sum_{i=1}^N\sum_{t=1}^T[G_\varepsilon(\widehat{z}_{it} )  - G_\varepsilon(z_{it}^0 ) ]^2 =O_P\left(\frac{1}{\sqrt{NT}} \right),
\end{eqnarray*}
which completes the proof for the first result of this lemma. \hspace*{\fill}{$\blacksquare$}

\bigskip

\noindent \textbf{Proof of Lemma \ref{Lemma2.2}:}

Again, in the following proof, we provide a robust version which assumes that $F$ and $\Gamma$ are $T\times d_{\max}$ and $N\times d_{\max}$ respectively where $d_{\max}\geq d_f$ is a user specified large fixed constant.

(1). First, note that using Assumption \ref{Ass1}.2, we can write

\begin{eqnarray*}
\frac{1}{NT}\sum_{i=1}^N\vect (Z_i' M_FZ_i) &=& \frac{1}{NT}\sum_{i=1}^N (Z_i'\otimes Z_i') \vect (M_F)=\frac{1}{NT}\sum_{i=1}^N E[Z_i'\otimes Z_i'] \vect (M_F)\cdot (1+o_P(1)),
\end{eqnarray*}
and 

\begin{eqnarray*}
\frac{1}{NT}\sum_{i=1}^N (\beta_{0i} -\beta_i)' X_iM_F F_0\gamma_i &=& \frac{1}{NT}\sum_{i=1}^N \gamma_i'\otimes  [(\beta_{0i} -\beta_i)' X_i']\vect(M_F F_0)\\
&=& \frac{1}{NT}\sum_{i=1}^NE[ \gamma_i'\otimes  ((\beta_{0i} -\beta_i)' X_i')]\vect(M_F F_0)\cdot(1+o_P(1)).
\end{eqnarray*}
Then we again let $z_{it} =x_{it}'\beta_i+\gamma_i'f_t$, and write

\begin{eqnarray}\label{consistency2}
&&\frac{1}{NT}\sum_{i=1}^N \sum_{t=1}^T [G_\varepsilon(z_{it} )  - G_\varepsilon(z_{it}^0 )]^2=\frac{1}{NT}\sum_{i=1}^N \sum_{t=1}^T  [g_\varepsilon(z_{it}^\dagger)\Delta z_{it}]^2\nonumber \\
&=&\frac{1}{NT}\sum_{i=1}^N \left[\mathscr{G}_i  X_i(\beta_i-\beta_{0i} )+  \mathscr{G}_i (F\gamma_i-F_0\gamma_{0i}) \right]'\left[\mathscr{G}_i  X_i(\beta_i-\beta_{0i} )+  \mathscr{G}_i (F\gamma_i-F_0\gamma_{0i}) \right]\nonumber \\
&\ge &\frac{a_{NT}^2}{NT}\sum_{i=1}^N \left[  X_i(\beta_i-\beta_{0i} )+  (F\gamma_i-F_0\gamma_{0i}) \right]'\left[ X_i(\beta_i-\beta_{0i})+ (F\gamma_i-F_0\gamma_{0i}) \right]\nonumber \\
&\ge &\frac{a_{NT}^2}{NT}\sum_{i=1}^N \left[  X_i(\beta_{0i}-\beta_i )+  F_0\gamma_{0i} \right]'M_F\left[ X_i(\beta_{0i}-\beta_i )+  F_0\gamma_{0i} \right]\nonumber \\
&= &\frac{a_{NT}^2}{NT}\sum_{i=1}^N \left[ (\beta_{0i} -\beta_i)'A_i (\beta_{0i} -\beta_i) +\eta 'B_i\eta + (\beta_{0i} -\beta_i)'C_i' \eta\right] +o_P(a_{NT}^2),
\end{eqnarray}
where $\mathscr{G}_i  =  \diag\big\{g_\varepsilon(z_{i1}^\dagger),\ldots, g_\varepsilon(z_{iT}^\dagger )\big\}$ with $z_{it}^\dagger$ lying between $z_{it}$ and $z_{it}^0$ for each $(i,t)$, $A_i = E[X_i' M_F X_i\, | \, F]$, $ B_i =  E[\gamma_{0i}\gamma_{0i}']\otimes I_T $, $C_i = E[\gamma_{0i}\otimes (M_FX_i)\, | \, F]$, $\eta =\vect(M_FF_0)$, the first inequality follows from Assumption \ref{Ass1}.1, and the third equality follows these two equations pointed out in the beginning of the proof.

Note that expect the extra term $a_{NT}^2$, the right hand side of \eqref{consistency2} has the identical form as in \citet[pp. 1265-1266]{Bai2009} for each $i$. In connection with Lemma \ref{Lemma2.1} and Assumption \ref{Ass1}, we are readily to conclude that $\frac{1}{N}\sum_{i=1}^N\|\widehat{\beta}_i -\beta_{0i} \|^2 =o_P (1 )$.  

\medskip

(2). After establishing the second result, we rewrite \eqref{consistency2} as follows.

\begin{eqnarray}\label{consistencyF}
&&\frac{1}{NT}\sum_{i=1}^N \sum_{t=1}^T [G_\varepsilon(\widehat{z}_{it} )  - G_\varepsilon(z_{it}^0 )]^2 \nonumber \\
&\ge &\frac{a_{NT}^2}{NT}\sum_{i=1}^N \left[  X_i(\widehat{\beta}_i-\beta_{0i} )+  (\widehat{F}\widehat{\gamma}_i-F_0\gamma_{0i}) \right]'\left[ X_i(\widehat{\beta}_i-\beta_{0i} )+ (\widehat{F}\widehat{\gamma}_i-F_0\gamma_{0i}) \right]\nonumber \\
&=&\frac{a_{NT}^2}{NT}\sum_{i=1}^N  (\widehat{\beta}_i-\beta_{0i} )'X_i'X_i(\widehat{\beta}_i-\beta_{0i} )+\frac{a_{NT}^2}{NT}\sum_{i=1}^N(\widehat{F}\widehat{\gamma}_i-F_0\gamma_{0i}) ' (\widehat{F}\widehat{\gamma}_i-F_0\gamma_{0i}) \nonumber \\
&&+\frac{2a_{NT}^2}{NT}\sum_{i=1}^N (\widehat{\beta}_i-\beta_{0i} )'X_i'(\widehat{F}\widehat{\gamma}_i-F_0\gamma_{0i}) ,
\end{eqnarray}
which immediately yields that 

\begin{eqnarray*}
\frac{1}{NT}\sum_{i=1}^N \|\widehat{F}\widehat{\gamma}_i-F_0\gamma_{0i}\|^2 =o_P(1).
\end{eqnarray*}
The second result then follows.

\medskip

(3). By \eqref{consistency2} and the second result of this lemma, we obtain that

\begin{eqnarray*}
o_P(1)=\frac{1}{NT}\sum_{i=1}^N \gamma_{0i}F_0'M_{\widehat{F}}F_0\gamma_{0i}=\tr\left\{ \frac{F_0' M_{\widehat{F}}F_0 }{T} \cdot \frac{\Gamma_0'\Gamma_0}{N}\right\},
\end{eqnarray*}
which in connection with $\frac{\Gamma_0'\Gamma_0}{N}\to_P\Sigma_\gamma$ of Assumption \ref{Ass1} yields that

\begin{eqnarray*}
o_P(1)=\tr\left\{ \frac{F_0' M_{\widehat{F}}F_0 }{T} \right\} =\tr\left\{ \frac{F_0'F_0}{T} - \frac{F_0'\widehat{F}}{T} \cdot \frac{\widehat{F}'F_0}{T}\right\}.
\end{eqnarray*}
By $\frac{F_0'F_0}{T}\to_P\Sigma_f$ of Assumption \ref{Ass1}, we can further write

\begin{eqnarray}\label{consistency3}
o_P(1)&=&\tr\left\{ I_{d_f} - \frac{F_0'\widehat{F}}{T} \cdot \frac{\widehat{F}'F_0}{T}\left(\frac{F_0'F_0}{T} \right)^{-1}\right\} \nonumber\\
&=&  \tr\left\{ I_{d_f} - \frac{\widehat{F}' P_{F_0} \widehat{F}}{T}\right\}.
\end{eqnarray}
Note that it is easy to show that

\begin{eqnarray}\label{consistency4}
\left\| P_{\widehat{F}} -P_{F_0}\right\|^2 &=& \text{tr}\left[ (P_{\widehat{F}}-P_{F_0})^2\right] = \text{tr}\left[ P_{\widehat{F}} - P_{\widehat{F}} P_{F_0}- P_{F_0}P_{\widehat{F}}+P_{F_0}\right]\nonumber \\
&=& \text{tr}\left[ I_{d_{\max}}\right] - 2\cdot\text{tr}\left[P_{\widehat{F}} P_{F_0}\right]+\text{tr}\left[I_{d_f}\right]\nonumber \\
&=&(d_{\max} -d_f) + 2 \cdot\text{tr}[ I_{d_f}-\widehat{F}'P_{F_0} \widehat{F}/T],
\end{eqnarray}
which in connection with \eqref{consistency3} yields the third result. The third result follows by letting $d_{\max}=d_f$. The proof is now complete. \hspace*{\fill}{$\blacksquare$}

\bigskip

\noindent \textbf{Proof of Lemma \ref{Lemmauni}:}

(1).  Recall that the log-likelihood function is defined as

\begin{eqnarray*}
\log L (\Theta,F) &=&\sum_{i=1}^N\sum_{t=1}^T\Big\{(1-y_{it})\log\left[1-G_\varepsilon(z_{it})\right] +y_{it} \log G_\varepsilon(z_{it}) \Big\},
\end{eqnarray*}
where $z_{it}=x_{it}'\beta_i+\gamma_{i}'f_{t}$. To study the uniform consistency of $\widehat{\theta}_i$, we introduce the following objective functions for $i=1, \ldots, N$,

\begin{eqnarray*}
L_i(\theta_i, F) = \sum_{t=1}^T\Big\{(1-y_{it})\log\left[1-G_\varepsilon(z_{it})\right] +y_{it} \log G_\varepsilon(z_{it}) \Big\}.
\end{eqnarray*}
For simplicity of notation, we denote $l_{it}(z_{it})=(1-y_{it})\log\left[1-G_\varepsilon(z_{it})\right] +y_{it} \log G_\varepsilon(z_{it})$. By the definition of $l_{it}(z_{it})$ and $L_i(\theta_i, F)$,  we can observe that $\log L (\Theta,F) = \sum_{i=1}^N L_i(\theta_i, F)=\sum_{i=1}^N\sum_{t=1}^Tl_{it}(z_{it})$.

Before we proceed to the proof of uniform consistency, we first show that $\|\widehat{\theta}_i - \theta_{i0}\|=o_P(1)$ for each $i$.  To establish the consistency of $\widehat{\theta}_i$, it suffices to show that

\begin{eqnarray}\label{rrr1}
L_i(\theta_{0i}, F_0)-L_i(\widehat{\theta}_i, \widehat{F})\leq 0,
\end{eqnarray}
with probability approaching 1 and then we can use the arguments analogously to those in the proof of Lemma \ref{Lemma2.2} to prove that  $\|\widehat{\theta}_i - \theta_{i0}\|=o_P(1)$.  We first rewrite the first order conditions (FOCs) for $\widehat{\Theta}$ and $\widehat{F}$. Recall that we have  the following conditions for $\widehat{\Theta}$ and $\widehat{F}$:

\begin{eqnarray*}
\frac{\partial \log L (\widehat{\Theta},\widehat{F})}{\partial \theta_i}=0,\quad \frac{\partial \log L (\widehat{\Theta},\widehat{F})}{\partial f_t}=0,
\end{eqnarray*}
for $t=1,\ldots,T$.

It is equivalent to have
\begin{eqnarray*}
&&\sum_{t=1}^T l_{it}^{(1)}(\widehat{z}_{it})x_{it}=0,\quad \sum_{t=1}^T l_{it}^{(1)}(\widehat{z}_{it})\widehat{f}_t=0,\quad \sum_{i=1}^N l_{it}^{(1)}(\widehat{z}_{it})\widehat{\gamma}_i=0,
\end{eqnarray*} 
where  $l_{it}^{(1)}(z_{it})=\frac{[y_{it}-G_\varepsilon(z_{it})]g_\varepsilon(z_{it}) }{[1-G_\varepsilon(z_{it})]G_\varepsilon(z_{it})}$ is the first derivative of $l_{it}(z_{it})$. 

By Taylor expansions, we have

\begin{eqnarray}\label{rrr2}
L_i(\theta_{0i}, F_0)-L_i(\widehat{\theta}_i, \widehat{F}) &=&\sum_{t=1}^T l_{it}(z_{it}^0)-\sum_{t=1}^T l_{it}(\widehat{z}_{it})
\nonumber\\
&=& \sum_{t=1}^T l^{(1)}_{it}(\widehat{z}_{it})(z_{it}^0-\widehat{z}_{it})+\frac{1}{2}\sum_{t=1}^T l^{(2)}_{it}(\dot{z}_{it})(z_{it}^0-\widehat{z}_{it})^2, 
\end{eqnarray}
where $l^{(2)}_{it}(z_{it})$ is the second derivative of $l_{it}(z_{it})$ and $\dot{z}_{it}$ lies between $z_{it}^0$ and $\widehat{z}_{it}$. For the first term on the right-hand side of \eqref{rrr2},

\begin{eqnarray}\label{rrr3}
\sum_{t=1}^T l^{(1)}_{it}(\widehat{z}_{it})(z_{it}^0-\widehat{z}_{it}) &=& \sum_{t=1}^T l^{(1)}_{it}(\widehat{z}_{it})x_{it}'(\beta_{0i}-\widehat{\beta}_i)+\sum_{t=1}^T l^{(1)}_{it}(\widehat{z}_{it}) \gamma_{0i}' f_{0t}-\sum_{t=1}^T l^{(1)}_{it}(\widehat{z}_{it}) \widehat{\gamma}_{i}' \widehat{f}_{t}
\nonumber\\
&=&\sum_{t=1}^T l^{(1)}_{it}(\widehat{z}_{it}) \gamma_{0i}' f_{0t}
\nonumber\\
&=&\sum_{t=1}^T l^{(1)}_{it}(\widehat{z}_{it}) \gamma_{0i}' (f_{0t}-\widehat{f}_t)
\nonumber\\
&\leq & \left(\sum_{t=1}^T\|l^{(1)}_{it}(\widehat{z}_{it}) \gamma_{0i}\|^2\right)^{\frac{1}{2}}\cdot \left(\sum_{t=1}^T \|f_{0t}-\widehat{f}_t\|^2\right)^{\frac{1}{2}}
\nonumber\\
&=&o_P\left(T\right),
\end{eqnarray}
where the second and third equalities hold by the FOCs; the inequality holds by Cauchy-Schwarz inequality and the last equality holds by the fact $\left\|\widehat{F}-F_0\right\|^2=o_P(T)$ which is implied by  Lemma \ref{Lemma2.2} and Assumption \ref{Ass2}.

For the second term on the right-hand side of \eqref{rrr2}, 

\begin{eqnarray}\label{rrr4}
&&\sum_{t=1}^T l^{(2)}_{it}(\dot{z}_{it})(z_{it}^0-\widehat{z}_{it})^2
\nonumber\\
&=&\sum_{t=1}^T l^{(2)}_{it}(\dot{z}_{it}) ( x_{it}'(\beta_{0i}-\widehat{\beta}_i)+  f_{0t}'(\gamma_{0i}- \widehat{\gamma}_i) + \gamma_{0i}'(f_{0t}-\widehat{f}_t)+ (\gamma_{0i}- \widehat{\gamma}_i)'(f_{0t}-\widehat{f}_t) )^2
\nonumber\\
&=&\sum_{t=1}^T l^{(2)}_{it}(\dot{z}_{it})[x_{it}'(\widehat{\beta}_{i}-\beta_{i0})+f_{t0}'(\widehat{\gamma}_{i}-\gamma_{i0})]^2+o_P\left(T\right)
\nonumber\\
&=&\sum_{t=1}^T l^{(2)}_{it}(\dot{z}_{it})[u_{it}^{0\prime}(\widehat{\theta}_{i}-\theta_{0i})]^2+o_P\left(T\right)
,
\end{eqnarray}
where the second equality holds by the fact:   $\left\|\widehat{F}-F_0\right\|^2=o_P(T)$.

By \eqref{rrr2}, \eqref{rrr3}, \eqref{rrr4} and the condition of uniformly bounded second derivatives in  Assumption \ref{Ass2}, we have  \eqref{rrr1} holds. With \eqref{rrr1}, we can  follow the proofs of Lemma \ref{Lemma2.1} and Lemma \ref{Lemma2.2} to show that $\|\widehat{\theta}_i-\theta_{0i}\|=o_P(1)$ for each $i$. Since the arguments are analogous but tedious, we omit its proof here. 

We now proceed to prove the uniform consistency : $\max_{1\leq i\leq N}\|\widehat{\theta}_i-\theta_{0i}\|=o_P(1)$. 
Recall that we establish the  consistency of the proposed estimators in Lemma \ref{Lemma2.2}. From the derivations of \eqref{consistency2} and \eqref{consistencyF} and the condition $a^2_{NT}\sqrt{NT}/\log (NT)^2\rightarrow \infty$ in Assumption \ref{Ass2}, we can actually have 
\begin{eqnarray*}
\frac{1}{N}\left\|\widehat{\Theta}-\Theta_0\right\|^2=o_P\left(\frac{1}{\log (NT)^2}\right), \,\,\frac{1}{T}\left\|\widehat{F}-F_0\right\|^2=o_P\left(\frac{1}{\log (NT)^2}\right).
\end{eqnarray*}

Therefore, without loss of generality, we only need to discuss the properties of $(\theta_i, F)$ where  $\left\|F-F_{0}\right\|\leq c\sqrt{T}/\log(NT)$ with $c$ as an arbitrary small positive number. 

Define the following parameter sets:
\begin{eqnarray*}
\mathcal{B}_{\theta,i}=\{\theta_i:\left\|\theta_i-\theta_{0i}\right\|\leq \delta_\theta\},\quad \mathcal{B}_{F}=\{F:\left\|F-F_{0}\right\|\leq c\sqrt{T}/\log(NT)\}.
\end{eqnarray*}
Without loss of generality, we define $\mathcal{B}^c_{\theta,i}=\{\theta_i: \delta_\theta< \left\|\theta_i-\theta_{0i}\right\|\leq C_\delta \}$, where $C_\delta$ is a positive and sufficiently large but finite number.  We want to show the probability of the following set goes to zero for any given $\delta_\theta>0$:

\begin{eqnarray*}
&&\left\{\max_i \|\widehat{\theta}_i-\theta_{0i}\|> \delta_\theta, \quad \widehat{F}\in \mathcal{B}_{F}\right\}
\nonumber\\
&=&\left\{\exists i, \, \widehat{\theta}_i\in \mathcal{B}^c_{\theta,i}, \quad \widehat{F}\in \mathcal{B}_{F}\right\}
\nonumber\\
&\subseteq&\left\{\exists i \text{~and~}\theta_i\in \mathcal{B}^c_{\theta,i}, \quad F\in \mathcal{B}_{F},\text{~s.t.~} \frac{1}{T}L_i(\theta_i,F)\geq \frac{1}{T}L_i(\theta_{0i},F_0)\right\}.
\end{eqnarray*}
By Taylor expansions, we have 

\begin{eqnarray}\label{rvs3}
\frac{1}{T}L_i(\theta_i,F)- \frac{1}{T}L_i(\theta_{0i},F_0)&=&
\frac{1}{T}\sum_{t=1}^T l_{it}(z_{it})-\frac{1}{T}\sum_{t=1}^T l_{it}(z_{it}^0)
\nonumber\\
&=& \frac{1}{T}\sum_{t=1}^T l^{(1)}_{it}(z_{it}^0)(z_{it}-z_{it}^0)+\frac{1}{2T}\sum_{t=1}^T l^{(2)}_{it}(\ddot{z}_{it})(z_{it}-z_{it}^0)^2, 
\end{eqnarray}
where $\ddot{z}_{it}$ lies between $z_{it}^0$ and $z_{it}$. 

For the first term on the right-hand side of \eqref{rvs3}, 

\begin{eqnarray}\label{rrr5}
\frac{1}{T}\sum_{t=1}^T l^{(1)}_{it}(z_{it}^0)(z_{it}-z_{it}^0)&=& \frac{1}{T}\sum_{t=1}^T l^{(1)}_{it}(z_{it}^0)(x_{it}'(\beta_i-\beta_{0i})+\gamma_i'f_t-\gamma_{0i}'f_{0t})
\nonumber\\
&=& \frac{1}{T}\sum_{t=1}^T l^{(1)}_{it}(z_{it}^0)(u_{it}^{0\prime}(\theta_i-\theta_{0i})+\gamma_{0i}'(f_t-f_{0t})+(\gamma_i-\gamma_{0i})'(f_t-f_{0t}))
\nonumber\\
&=& \frac{1}{T}\sum_{t=1}^T l^{(1)}_{it}(z_{it}^0)u_{it}^{0\prime}(\theta_i-\theta_{0i})+\frac{1}{T}\sum_{t=1}^T l^{(1)}_{it}(z_{it}^0)u_{it}^{0\prime}\gamma_{0i}'(f_t-f_{0t})
\nonumber\\
&&+ \frac{1}{T}\sum_{t=1}^T l^{(1)}_{it}(z_{it}^0) (f_t-f_{0t})' (\gamma_i-\gamma_{0i})
\nonumber\\
&:=& \widetilde{L}_{1i}+\widetilde{L}_{2i}+\widetilde{L}_{3i}.
\end{eqnarray}

For the second term on the right-hand side of \eqref{rvs3},

\begin{eqnarray}\label{rrr6}
&&\frac{1}{2T}\sum_{t=1}^T l^{(2)}_{it}(\ddot{z}_{it})(z_{it}-z_{it}^0)^2
\nonumber\\
&=&\frac{1}{2T}\sum_{t=1}^T l^{(2)}_{it}(\ddot{z}_{it}) (u_{it}^{0\prime}(\theta_i-\theta_{0i})+\gamma_{0i}'(f_t-f_{0t})+(\gamma_i-\gamma_{0i})'(f_t-f_{0t}))^2
\nonumber\\
&=&\frac{1}{2T}\sum_{t=1}^T l^{(2)}_{it}(\ddot{z}_{it}) (u_{it}^{0\prime}(\theta_i-\theta_{0i}))^2+\frac{1}{2T}\sum_{t=1}^T l^{(2)}_{it}(\ddot{z}_{it})(\gamma_{0i}'(f_t-f_{0t}))^2
\nonumber\\
&&+ \frac{1}{2T}\sum_{t=1}^T l^{(2)}_{it}(\ddot{z}_{it}) ((\gamma_i-\gamma_{0i})'(f_t-f_{0t}))^2
+ \frac{1}{T}\sum_{t=1}^T l^{(2)}_{it}(\ddot{z}_{it})u_{it}^{0\prime}(\theta_i-\theta_{0i})\gamma_{0i}'(f_t-f_{0t})
\nonumber\\
&& +\frac{1}{T}\sum_{t=1}^T l^{(2)}_{it} (\ddot{z}_{it}) u_{it}^{0\prime}(\theta_i-\theta_{0i})(\gamma_i-\gamma_{0i})'(f_t-f_{0t})+ \frac{1}{T}\sum_{t=1}^T l^{(2)}_{it}(\ddot{z}_{it})\gamma_{0i}'(f_t-f_{0t})(\gamma_i-\gamma_{0i})'(f_t-f_{0t})
\nonumber\\
&&=\widetilde{L}_{4i}+\cdots+\widetilde{L}_{9i}.
\end{eqnarray}
For $\widetilde{L}_{4i}$, by  Assumption \ref{Ass2}, we can find a positive finite number $M$ uniformly for $i$ and $\theta_i\in \mathcal{B}^c_{\theta,i}$ such that,

\begin{eqnarray*}
P\left(\max_{1\leq i\leq N } \widetilde{L}_{4i} \leq -M\delta_\theta^2\right)\rightarrow 1,
\end{eqnarray*}
as $N, T\rightarrow\infty$.

For $\widetilde{L}_{5i}, \ldots, \widetilde{L}_{9i}$, since $c$ can be chosen to be arbitrarily small and by Assumption \ref{Ass2}, we can always find a value  $c>0$ such that for $F\in  \mathcal{B}_{F}$,  the following inequalities hold with probability approaching 1: $\max_{1\leq i\leq N } |\widetilde{L}_{5i}|<\frac{1}{6}M\delta_\theta^2$, $\max_{1\leq i\leq N } |\widetilde{L}_{6i}|<\frac{1}{6}M\delta_\theta^2$, $\max_{1\leq i\leq N } |\widetilde{L}_{7i}|<\frac{1}{6}M\delta_\theta^2$, $\max_{1\leq i\leq N } |\widetilde{L}_{8i}|<\frac{1}{6}M\delta_\theta^2$ and $\max_{1\leq i\leq N } |\widetilde{L}_{9i}|<\frac{1}{6}M\delta_\theta^2$. In summary, we can find $c>0$ and $0<M<\infty$ such that  for $\theta_i\in \mathcal{B}^c_{\theta,i}$ and $F\in  \mathcal{B}_{F}$
 
 \begin{eqnarray}\label{rrr7}
P\left(\max_{1\leq i\leq N } (\widetilde{L}_{4i}+\cdots+\widetilde{L}_{9i} )< -\frac{1}{6}M\delta_\theta^2\right)\rightarrow 1,
\end{eqnarray}
as $N, T\rightarrow\infty$. 

Therefore, by \eqref{rrr5}, \eqref{rrr6} and \eqref{rrr7}, we can have

\begin{eqnarray}\label{rrr11}
&&P\left(\exists i \text{~and~}\theta_i\in \mathcal{B}^c_{\theta,i}, \quad F\in \mathcal{B}_{F},\text{~s.t.~} \frac{1}{T}L_i(\theta_i,F)\geq \frac{1}{T}L_i(\theta_{i0},F_0)\right)
\nonumber\\
&\leq&  P\left(\exists i \text{~and~}\theta_i\in \mathcal{B}^c_{\theta,i}, \quad F\in \mathcal{B}_{F},\text{~s.t.~} \widetilde{L}_{1i}+\cdots+\widetilde{L}_{9i}\geq 0  \right)
\nonumber\\
&\leq&  P\left(\exists i \text{~and~}\theta_i\in \mathcal{B}^c_{\theta,i}, \quad F\in \mathcal{B}_{F},\text{~s.t.~} \widetilde{L}_{1i}+\widetilde{L}_{2i}+\widetilde{L}_{3i}\geq \frac{1}{6}M\delta_\theta^2  \right)
\nonumber\\
&&+P\left(\max_{1\leq i\leq N, \, \theta_i\notin \mathcal{B}_{\theta,i}, \, F\in  \mathcal{B}_{F} } (\widetilde{L}_{4i}+\cdots+\widetilde{L}_{9i} )\geq -\frac{1}{6}M\delta_\theta^2\right)
\nonumber\\
&\leq &\sum_{i=1}^N P\left( \widetilde{L}_{1i}+\widetilde{L}_{2i}+\widetilde{L}_{3i}\geq \frac{1}{6}M\delta_\theta^2 \,|\, \theta_i\in \mathcal{B}^c_{\theta,i}, \quad F\in \mathcal{B}_{F}\right)+o(1)
\nonumber\\
&\leq &\sum_{i=1}^N P\left( |\widetilde{L}_{1i}| \geq \frac{1}{18}M\delta_\theta^2 \,|\, \theta_i\in \mathcal{B}^c_{\theta,i}, \quad F\in \mathcal{B}_{F}\right)+ \sum_{i=1}^N P\left( |\widetilde{L}_{2i}| \geq \frac{1}{18}M\delta_\theta^2 \,|\, \theta_i\in \mathcal{B}^c_{\theta,i}, \quad F\in \mathcal{B}_{F}\right)
\nonumber\\
&&+\sum_{i=1}^N P\left( |\widetilde{L}_{3i}| \geq \frac{1}{18}M\delta_\theta^2 \,|\, \theta_i\in \mathcal{B}^c_{\theta,i}, \quad F\in \mathcal{B}_{F}\right)+o(1).
\end{eqnarray}
For $\widetilde{L}_{1i}$, recall that $l_{it}^{(1)}(z^0_{it})=\frac{[y_{it}-G_\varepsilon(z^0_{it})]g_\varepsilon(z^0_{it}) }{[1-G_\varepsilon(z^0_{it})]G_\varepsilon(z^0_{it})}= - g_\varepsilon(z^0_{it})e_{it}$, where $e_{it}=-\frac{[y_{it}-G_\varepsilon(z^0_{it})] }{[1-G_\varepsilon(z^0_{it})]G_\varepsilon(z^0_{it})}$. We can observe that $E[\widetilde{L}_{1i}|\mathcal{W}]=0$ and  for $\theta_i\in \mathcal{B}^c_{\theta,i}$ and $F\in \mathcal{B}_{F}$,

\begin{eqnarray*}
\sum_{i=1}^N P\left( |\widetilde{L}_{1i}| \geq \frac{1}{18}M\delta_\theta^2 \right)&=& 
\sum_{i=1}^N P\left( \left\|\frac{1}{T}\sum_{t=1}^T g_\varepsilon(z^0_{it})e_{it}u^0_{it}\right\| \geq \frac{M\delta_\theta^2}{18 C_\delta}\right)
\nonumber\\
&\leq& \sum_{i=1}^N \frac{E[\|T^{-1}\sum_{t=1}^T g_\varepsilon(z^0_{it})e_{it}u^0_{it} \|^{2+\delta^\ast}]}{M^{ 2+\delta^\ast }_\ast}
\nonumber\\
&=&O\left(\frac{N}{T^{1+\delta^\ast/4}}\right),
\end{eqnarray*}
where $ M_\ast = \frac{M\delta_\theta^2}{18 C_\delta}$, the inequality holds by by Chebyshev's inequality and the second equality holds by Lemma \ref{LemmaT1}. Under the assumption $\frac{N}{T^{1+\delta^\ast/4}}\rightarrow 0$, we have

\begin{eqnarray}\label{rrr8}
\sum_{i=1}^N P\left( |\widetilde{L}_{1i}| \geq \frac{1}{18}M\delta_\theta^2 \,|\, \theta_i\in \mathcal{B}^c_{\theta,i}, \quad F\in \mathcal{B}_{F}\right)&=& o(1).
\end{eqnarray}

Following analogous arguments in the proof of \eqref{rrr8}, we can show that 
\begin{eqnarray}\label{rrr9}
\sum_{i=1}^N P\left( |\widetilde{L}_{2i}| \geq \frac{1}{18}M\delta_\theta^2 \,|\, \theta_i\in \mathcal{B}^c_{\theta,i}, \quad F\in \mathcal{B}_{F}\right)=o(1),
\end{eqnarray}
and 
\begin{eqnarray}\label{rrr10}
\sum_{i=1}^N P\left( |\widetilde{L}_{3i}| \geq \frac{1}{18}M\delta_\theta^2 \,|\, \theta_i\in \mathcal{B}^c_{\theta,i}, \quad F\in \mathcal{B}_{F}\right)=o(1).
\end{eqnarray}

By \eqref{rrr11},  \eqref{rrr9} and \eqref{rrr10}, 
\begin{eqnarray*}
P\left(\exists i \text{~and~}\theta_i\in \mathcal{B}^c_{\theta,i}, \quad F\in \mathcal{B}_{F},\text{~s.t.~} \frac{1}{T}L_i(\theta_i,F)\geq \frac{1}{T}L_i(\theta_{i0},F_0)\right)=o(1), 
\end{eqnarray*}
and therefore it immediately yields $\max_{1\leq i\leq N}\|\widehat{\theta}_i-\theta_{0i}\|=o_P(1)$ as discussed at the beginning of this section. 

\medskip

(2). For the uniform consistency $\max_{1\leq t\leq T}\|\widehat{f}_t-f_{0t}\|=o_P(1)$, we can follow analogous arguments to show this result. Therefore, the proof of Lemma \ref{Lemmauni}.(2) is omitted. 

\medskip

(3)-(4). In the following proofs, we first find expressions for $\widehat{\Theta}_{v}-\Theta_{0v}$ and $\widehat{F}_{v}-F_{0v}$ from the FOCs. After deriving their leading terms, we can then compute the rates of convergence for 
$\frac{1}{N}\sum_{i=1}^N\|\widehat{\theta}_i -\theta_{0i} \|^2$ and $\frac{1}{T}\sum_{t=1}^T\|\widehat{f}_t -f_{0t}\|^2$.

With the notation of derivatives defined in Appendix \ref{Appnot}, we are ready to proceed with the derivations of $\widehat{\Theta}_{v}-\Theta_{0v}$ and $\widehat{F}_{v}-F_{0v}$.  From the FOCs for $\widehat{\Theta}$, we take the Taylor expansion of $\log L(\Theta,F)$ at $(\Theta_0, F_0)$,

\begin{eqnarray}\label{rev7}
0&=&\frac{1}{T}\frac{\partial \log L(\widehat{\Theta}, \widehat{F})}{\partial \Theta_v}
\nonumber\\
&=&\frac{1}{T}\frac{\partial \log L(\Theta_0, F_0)}{\partial \Theta_v}+\frac{1}{T}\frac{\partial^2 \log L(\Theta_0, F_0)}{\partial \Theta_v\partial\Theta_v'}(\widehat{\Theta}_{v}-\Theta_{0v})+\frac{1}{T}\frac{\partial^2 \log L(\Theta_0, F_0)}{\partial \Theta_v\partial F_v'}(\widehat{F}_{v}-F_{0v}) +\mathcal{Q}_{NT},
\nonumber\\
\end{eqnarray}
where $\mathcal{Q}_{NT}$ contains the Taylor expansion residuals: $\mathcal{Q}_{NT}=(\mathcal{Q}_{T,1}',\cdots, \mathcal{Q}_{T, N}')$ with

\begin{eqnarray}\label{rev25}
\mathcal{Q}_{Ti}&:=&\frac{1}{T}\sum_{l=1}^{d_{\beta}+d_f}\frac{\partial^3 \log L(\dot{\Theta}, \dot{F})}{\partial \theta_i\partial\theta_i'\partial \theta_{il}}(\widehat{\theta}_{i}-\theta_{0i})(\widehat{\theta}_{il}-\theta_{0il})+\frac{2}{T}\sum_{t=1}^T\sum_{r=1}^{d_f}\frac{\partial^3 \log L(\dot{\Theta}, \dot{F})}{\partial \theta_i\partial\theta_i'\partial f_{tr}}(\widehat{\theta}_{i}-\theta_{0i})(\widehat{f}_{tr}-f_{0tr})
\nonumber\\
&&+\frac{1}{T}\sum_{t=1}^T\sum_{r=1}^{d_f}\frac{\partial^3 \log L(\dot{\Theta}, \dot{F})}{\partial \theta_i\partial f_t'\partial f_{tr}}(\widehat{f}_{t}-f_{0t})(\widehat{f}_{tr}-f_{0tr}),
\end{eqnarray}
and $(\dot{\Theta}, \dot{F})$ lies between $(\widehat{\Theta}, \widehat{F})$ and $(\Theta_0, F_0)$.

We then proceed to find the leading term in $\frac{1}{T}\frac{\partial^2 \log L(\Theta_0, F_0)}{\partial \Theta_v\partial\Theta_v'}$. Recall that $\frac{1}{T}\frac{\partial^2 \log L(\Theta_0, F_0)}{\partial \Theta_v\partial\Theta_v'}$ is a block-diagonal matrix with its $i$-th diagonal block being

\begin{eqnarray*}
&&\frac{1}{T}\frac{\partial^2 \log L(\Theta_0, F_0)}{\partial \theta_i\partial\theta_i'}= -\Sigma_{u,i}-\left(\frac{1}{T}\sum_{t=1}^T \left\{\frac{[g_\varepsilon(z_{it}^0)]^2 }{[1-G_\varepsilon(z_{it}^0)]G_\varepsilon(z_{it}^0)}  \right\} u_{it}^0u_{it}^{0\prime}-\Sigma_{u,i}\right)
\nonumber\\
&&+\frac{1}{T}\sum_{t=1}^T \frac{[y_{it}-G_\varepsilon(z_{it}^0)]g_\varepsilon^{(1)}(z_{it}^0) }{[1-G_\varepsilon(z_{it}^0)]^2[G_\varepsilon(z_{it}^0)]^2}  \cdot u_{it}^0 u_{it}^{0\prime}
+\frac{1}{T}\sum_{t=1}^T  \frac{[y_{it}-G_\varepsilon(z_{it}^0)][g_\varepsilon(z_{it}^0)]^2 [1-2G_\varepsilon(z_{it}^0)]}{[1-G_\varepsilon(z_{it}^0)]^2[G_\varepsilon(z_{it}^0)]^2}  \cdot u_{it}^0 u_{it}^{0\prime}
\nonumber\\
&&:= -\Sigma_{u,i}+C_{1Ti}+C_{2Ti}+C_{3Ti}.
\end{eqnarray*}
Denote $\Omega_u$, $C_{NT, 1}$, $C_{NT, 2}$ and $C_{NT, 3}$ to be $N(d_\beta+d_f)\times N(d_\beta+d_f)$  block-diagonal matrices with $N$ diagonal blocks in each of them, and
the $i$-th  blocks of $\Omega_u$, $C_{NT, 1}$, $C_{NT, 2}$ and $C_{NT, 3}$ are $\Sigma_{u,i}$, $C_{1Ti}$, $C_{2Ti}$ and $C_{3Ti}$, respectively. With this notation, we have

\begin{eqnarray}\label{rev5}
\frac{1}{T}\frac{\partial^2 \log L(\Theta_0, F_0)}{\partial \Theta_v\partial\Theta_v'}=-\Omega_u+C_{NT, 1}+C_{NT, 2}+C_{NT, 3}.
\end{eqnarray}

For $\frac{\partial^2 \log L(\Theta_0, F_0)}{\partial \Theta_v\partial F_v'}$, we denote $\Omega_{u\gamma}$ to be a $N(d_\beta+d_f)\times Td_f$ matrix with a $N \times T$ block structure and its $(i,t)$-th block is $
\Omega_{u\gamma, i t} = E\left[\frac{[g_\varepsilon(z^0_{it})]^2 }{[1-G_\varepsilon(z^0_{it})]G_\varepsilon(z^0_{it})}  u^0_{it}\gamma_{0i}'\right]$. With this notation,

\begin{eqnarray*}
\frac{\partial^2 \log L(\Theta, F)}{\partial \theta_i\partial f_t'}&=&-\Omega_{u\gamma, i t}-\left(\frac{[g_\varepsilon(z^0_{it})]^2 }{[1-G_\varepsilon(z^0_{it})]G_\varepsilon(z^0_{it})}  u^0_{it}\gamma_{0i}' - \Omega_{u\gamma, i t}\right)
\nonumber\\
&&+ \frac{[y_{it}-G_\varepsilon(z_{it}^0)]g_\varepsilon^{(1)}(z_{it}^0) }{[1-G_\varepsilon(z_{it}^0)]^2[G_\varepsilon(z_{it}^0)]^2}  \cdot u^0_{it}\gamma_{0i}'^{0\prime}
+  \frac{[y_{it}-G_\varepsilon(z_{it}^0)][g_\varepsilon(z_{it}^0)]^2 [1-2G_\varepsilon(z_{it}^0)]}{[1-G_\varepsilon(z_{it}^0)]^2[G_\varepsilon(z_{it}^0)]^2}  \cdot u^0_{it}\gamma_{0i}'
\nonumber\\
&&:= -\Omega_{u\gamma, i t}+C_{4it}+C_{5it}+C_{6it}.
\end{eqnarray*}
Denote  $C_{NT, 4}$, $C_{NT, 5}$ and $C_{NT, 6}$  to be a $N(d_\beta+d_f)\times Td_f$ matrices with  $N \times T$ block structures and their $(i,t)$-th blocks  are $C_{4it}$, $C_{5it}$ and $C_{6it}$, respectively. We observe that

\begin{eqnarray}\label{rev17}
\frac{\partial^2 \log L(\Theta_0, F_0)}{\partial \Theta_v\partial F_v'}= -\Omega_{u\gamma}+C_{NT, 4}+C_{NT, 5}+C_{NT, 6}.
\end{eqnarray}
By \eqref{rev5} and \eqref{rev17},

\begin{eqnarray}\label{rev18}
\Omega_u\cdot (\widehat{\Theta}_{v}-\Theta_{0v})+ \frac{1}{T} \Omega_{u\gamma} \cdot (\widehat{F}_{v}-F_{0v}) &=&\frac{1}{T}\frac{\partial \log L(\Theta_0, F_0)}{\partial \Theta_v}+(C_{NT, 1}+C_{NT, 2}+C_{NT, 3})(\widehat{\Theta}_{v}-\Theta_{0v})
\nonumber\\
&&+\frac{1}{T}(C_{NT, 4}+C_{NT, 5}+C_{NT, 6})(\widehat{F}_{v}-F_{0v}) +\mathcal{Q}_{NT}
\nonumber\\
&:=&\frac{1}{T}\frac{\partial \log L(\Theta_0, F_0)}{\partial \Theta_v}+\widetilde{\mathcal{Q}}_{NT},
\end{eqnarray} 
where $\widetilde{\mathcal{Q}}_{NT}=\mathcal{Q}_{NT}+(C_{NT, 1}+C_{NT, 2}+C_{NT, 3})(\widehat{\Theta}_{v}-\Theta_{0v})+\frac{1}{T}(C_{NT, 4}+C_{NT, 5}+C_{NT, 6})(\widehat{F}_{v}-F_{0v})$.

From \eqref{rev18}, we know that $\widehat{F}_{v}-F_{0v}$ constitutes the bias terms in $\widehat{\Theta}_{v}-\Theta_{0v}$.  Now we proceed with FOCs for $\widehat{F}$.  We follow analogous arguments to the derivation of \eqref{rev18} to find an expression for $\widehat{F}_{v}-F_{0v}$. By the Taylor expansion at $(\Theta_0, F_0)$, we have

\begin{eqnarray}\label{rev11}
0&=&\frac{1}{N}\frac{\partial \log L(\widehat{\Theta}, \widehat{F})}{\partial F_v}
\nonumber\\
&=&\frac{1}{N}\frac{\partial \log L(\Theta_0, F_0)}{\partial F_v}+\frac{1}{N}\frac{\partial^2 \log L(\Theta_0, F_0)}{\partial F_v\partial F_v'}(\widehat{F}_{v}-F_{0v})+\frac{1}{N}\frac{\partial^2 \log L(\Theta_0, F_0)}{\partial F_v\partial \Theta_v'}(\widehat{\Theta}_{v}-\Theta_{0v}) +\mathcal{J}_{NT},
\nonumber\\
\end{eqnarray}
where $\mathcal{J}_{NT}$ contains the Taylor expansion residuals. Here we have $\mathcal{J}_{NT}=(\mathcal{J}_{N,1}',\cdots, \mathcal{J}_{N, T}')$ with

\begin{eqnarray*}
\mathcal{J}_{Nt}&=&\frac{1}{N}\sum_{r=1}^{d_f}\frac{\partial^3 \log L(\ddot{\Theta}, \ddot{F})}{\partial f_t\partial f_t'\partial f_{tr}}(\widehat{f}_{t}-f_{0t})(\widehat{f}_{tr}-f_{0tr})+\frac{2}{N}\sum_{i=1}^N\sum_{l=1}^{d_\beta+d_f}\frac{\partial^3 \log L(\ddot{\Theta}, \ddot{F})}{\partial f_t\partial f_t'\partial \theta_{il}}(\widehat{f}_{t}-f_{0t})(\widehat{\theta}_{il}-\theta_{0il})
\nonumber\\
&&+\frac{1}{N}\sum_{i=1}^N\sum_{l=1}^{d_\beta+d_f}\frac{\partial^3 \log L(\ddot{\Theta}, \ddot{F})}{\partial f_t\partial \theta_i'\partial \theta_{il}}(\widehat{\theta}_{i}-\theta_{0i})(\widehat{\theta}_{il}-\theta_{0il}),
\end{eqnarray*}
and $(\ddot{\Theta}, \ddot{F})$ lies between $(\widehat{\Theta}, \widehat{F})$ and $(\Theta_0, F_0)$.

For $\frac{1}{N}\frac{\partial^2 \log L(\Theta_0, F_0)}{\partial F_v\partial F_v'}$, recall that it is a block-diagonal matrix with its $t$-th diagonal block as

\begin{eqnarray*}
&&\frac{1}{N}\frac{\partial^2 \log L(\Theta_0, F_0)}{\partial f_t\partial f_t'}= -\Sigma_{\gamma,t}-\left(\frac{1}{N}\sum_{i=1}^N \left\{\frac{[g_\varepsilon(z_{it}^0)]^2 }{[1-G_\varepsilon(z_{it}^0)]G_\varepsilon(z_{it}^0)}  \right\} \gamma_{0i}\gamma_{0i}^{\prime}-\Sigma_{\gamma,t}\right)
\nonumber\\
&&+\frac{1}{N}\sum_{i=1}^N \frac{[y_{it}-G_\varepsilon(z_{it}^0)]g_\varepsilon^{(1)}(z_{it}^0) }{[1-G_\varepsilon(z_{it}^0)]^2[G_\varepsilon(z_{it}^0)]^2}  \cdot \gamma_{0i}\gamma_{0i}^{\prime}
+\frac{1}{N}\sum_{i=1}^N \frac{[y_{it}-G_\varepsilon(z_{it}^0)][g_\varepsilon(z_{it}^0)]^2 [1-2G_\varepsilon(z_{it}^0)]}{[1-G_\varepsilon(z_{it}^0)]^2[G_\varepsilon(z_{it}^0)]^2}  \cdot \gamma_{0i}\gamma_{0i}^{\prime}
\nonumber\\
&&:= -\Sigma_{\gamma,t}+D_{1Nt}+D_{2Nt}+D_{3Nt}.
\end{eqnarray*}
Denote $\Omega_\gamma$, $D_{NT, 1}$, $D_{NT, 2}$ and $D_{NT, 3}$ to be $Td_f\times Td_f$  block-diagonal matrices with $T$ diagonal blocks in each of them, and
the $t$-th  blocks of $\Omega_\gamma$, $D_{NT, 1}$, $D_{NT, 2}$ and $D_{NT, 3}$ are $\Sigma_{\gamma,t}$, $D_{1Nt}$, $D_{2Nt}$ and $D_{3Nt}$, respectively. With this notation, we have

\begin{eqnarray}\label{rev12}
\frac{1}{N}\frac{\partial^2 \log L(\Theta_0, F_0)}{\partial F_v\partial F_v'}=-\Omega_\gamma+D_{NT, 1}+D_{NT, 2}+D_{NT, 3}.
\end{eqnarray}

For $\frac{\partial^2 \log L(\Theta_0, F_0)}{\partial F_v\partial \Theta_v'}$, we know that $\frac{\partial^2 \log L(\Theta_0, F_0)}{\partial F_v\partial \Theta_v'}=\left(\frac{\partial^2 \log L(\Theta_0, F_0)}{\partial \partial \Theta_v F_v'}\right)'$. By \eqref{rev17},  we have $\frac{\partial^2 \log L(\Theta_0, F_0)}{\partial F_v\partial \Theta_v'}= -\Omega_{u\gamma}'+C_{NT, 4}'+C_{NT, 5}'+C_{NT, 6}'.$ Jointly with \eqref{rev12}, it immediately yields that

\begin{eqnarray}\label{rev19}
\Omega_{\gamma} \cdot (\widehat{F}_{v}-F_{0v}) + \frac{1}{N} \Omega_{u\gamma}'\cdot (\widehat{\Theta}_{v}-\Theta_{0v}) &=&\frac{1}{N}\frac{\partial \log L(\Theta_0, F_0)}{\partial F_v}+(D_{NT, 1}+D_{NT, 2}+D_{NT, 3})(\widehat{F}_{v}-F_{0v})
\nonumber\\
&&+\frac{1}{N}(C_{NT, 4}'+C_{NT, 5}'+C_{NT, 6}')(\widehat{\Theta}_{v}-\Theta_{0v}) +\mathcal{J}_{NT}
\nonumber\\
&:=&\frac{1}{N}\frac{\partial \log L(\Theta_0, F_0)}{\partial F_v}+\widetilde{\mathcal{J}}_{NT},
\end{eqnarray} 
where $\widetilde{\mathcal{J}}_{NT}=\mathcal{J}_{NT}+(D_{NT, 1}+D_{NT, 2}+D_{NT, 3})(\widehat{F}_{v}-F_{0v})+\frac{1}{N}(C_{NT, 4}'+C_{NT, 5}'+C_{NT, 6}')(\widehat{\Theta}_{v}-\Theta_{0v})$.

With the condition that  $\Sigma_{u,i}$ and $\Sigma_{\gamma,t}$ are invertible for each $i$ and $t$ in Assumption \ref{Ass2}, we know that $\Omega_u$ and $\Omega_\gamma$ are invertible and their inverse matrices are block-diagonal. The $i$-th diagonal block in $\Omega_u^{-1}$ is $\Sigma^{-1}_{u,i}$ and the $t$-th diagonal block in $\Omega_\gamma^{-1}$ is $\Sigma^{-1}_{\gamma,t}$. Furthermore, 
with \eqref{rev18} and \eqref{rev19}, we can use the inverse formula for the block matrix to find the following expression for $\widehat{\Theta}_{v}-\Theta_{0v}$ :

\begin{eqnarray}\label{rev41}
\widehat{\Theta}_{v}-\Theta_{0v}
&=&\Omega_u^{-1}\cdot \frac{1}{T}\frac{\partial \log L(\Theta_0, F_0)}{\partial \Theta_v}
+\frac{1}{NT}\Omega_u^{-1} \Omega_{u\gamma}\Omega_\gamma^{\ast-1}\Omega_{u\gamma}'\Omega_u^{-1} \cdot \frac{1}{T}\frac{\partial \log L(\Theta_0, F_0)}{\partial \Theta_v}
\nonumber\\
&&-\frac{1}{T}\Omega_u^{\ast-1}\Omega_{u\gamma}\Omega_{\gamma}^{-1}\cdot \frac{1}{N}\frac{\partial \log L(\Theta_0, F_0)}{\partial F_v}
 +\frac{1}{NT}\Omega_u^{-1} \Omega_{u\gamma}\Omega_\gamma^{\ast-1}\Omega_{u\gamma}'\Omega_u^{-1}\widetilde{\mathcal{Q}}_{NT}
\nonumber\\
&&-\frac{1}{T}\Omega_u^{\ast-1}\Omega_{u\gamma}\Omega_{\gamma}^{-1}\widetilde{\mathcal{J}}_{NT}+\Omega_u^{-1}\widetilde{\mathcal{Q}}_{NT}
\nonumber\\
&=&\mathcal{R}_{NT, 1}+\cdots+\mathcal{R}_{NT, 6}, 
\end{eqnarray}
where $\Omega_\gamma^\ast= \Omega_\gamma - \frac{1}{NT}\Omega_{u\gamma}'\Omega_u^{-1}\Omega_{u\gamma}$ and $\Omega_u^\ast = \Omega_u-\frac{1}{NT}\Omega_{u\gamma} \Omega_{\gamma}^{-1}\Omega_{u\gamma}'$.

We consider these six terms one by one to show the convergence of $\widehat{\Theta}_{v}-\Theta_{0v}$. For $\mathcal{R}_{NT, 1}$, since $\Omega_u^{-1}$ is block diagonal, 

\begin{eqnarray}\label{rev20}
E \|\mathcal{R}_{NT, 1}\|^2&=&\sum_{i=1}^N E \left\|\Sigma_{u,i}^{-1}\cdot \frac{1}{T}\frac{\partial \log L(\Theta_0, F_0)}{\partial \theta_i}\right\|^2 
\nonumber\\
&=&\sum_{i=1}^N \left\|\Sigma_{u,i}^{-1}\cdot \frac{1}{T}\sum_{t=1}^T \frac{[y_{it}-G_\varepsilon(z_{it}^0)]g_\varepsilon(z_{it}^0) }{[1-G_\varepsilon(z_{it}^0)]G_\varepsilon(z_{it}^0)}   u_{it}^0\right\|^2. 
\end{eqnarray}
Recall that $e_{it} =-\frac{y_{it}-G_\varepsilon(z_{it}^0)}{(1- G_\varepsilon(z_{it}^0))G_\varepsilon(z_{it}^0)}$. With this notation, \eqref{rev20} becomes

\begin{eqnarray}\label{rev21}
&&E \|\mathcal{R}_{NT, 1}\|^2 = \sum_{i=1}^N E \left\|\Sigma_{u,i}^{-1}\cdot \frac{1}{T}\sum_{t=1}^T   g_\varepsilon(z_{it}^0)u_{it}^0e_{it}\right\|^2 
\nonumber\\
&\leq&O(1)\frac{1}{T^2} \sum_{i=1}^N\sum_{t=1}^T\sum_{s=1}^T E[|g_\varepsilon(z_{it}^0)|\cdot |g_\varepsilon(z_{is}^0)|\cdot \|u_{is}^0\|\cdot \|u_{it}^0\| \cdot |E[e_{it}e_{is}|\mathcal{W}]|]
\nonumber\\
&\leq&O(1)\frac{c_{\delta}}{T^2}\sum_{i=1}^N \sum_{t=1}^T\sum_{s=1}^T\alpha_{ii}(|t-s|)^{\delta/(4+\delta)}E\left[\|u_{it}^{0}\| \cdot \|u_{is}^{0}\|\cdot  E\left[|e_{it}|^{2+\delta/2} |\, \mathcal{W}\right]^{2/(4+\delta)} \cdot E\left[|e_{is}|^{2+\delta/2} |\, \mathcal{W}\right]^{2/(4+\delta)}\right]
\nonumber\\
&=&O\left(\frac{N}{T}\right),
\end{eqnarray}
where $c_{\delta}=(4+\delta)/\delta\cdot2^{(4+2\delta)/(4+\delta)}$, the second inequality  holds by the fact that conditional on $\mathcal{W}$, 
$e_{it}$ is $\alpha$-mixing,  as we discussed in the previous proofs, and  Davydov's inequality for $\alpha$-mixing process. The last equality holds by the $\alpha$-mixing and moment conditions in  Assumption \ref{Ass2}. From \eqref{rev21} and the fact that $E[\mathcal{R}_{NT, 1}]=0$, we know that

\begin{equation}\label{rev55}
\|\mathcal{R}_{NT, 1}\| = O_P\left(\sqrt{\frac{N}{T}}\right).
\end{equation}
For $\mathcal{R}_{NT, 2}$, we can observe that  
\begin{eqnarray*}
\Omega_{u\gamma}'\Omega_u^{-1} \cdot \left(\frac{1}{T}\frac{\partial \log L(\Theta_0, F_0)}{\partial \Theta_v}\right)=\frac{1}{T} \left(\sum_{i=1}^N\Omega_{u\gamma,i1}'\Sigma_{u,i}^{-1} \frac{\partial \log L(\Theta_0, F_0)}{\partial \theta_i}, \cdots, \sum_{i=1}^N\Omega_{u\gamma,iT}'\Sigma_{u,i}^{-1} \frac{\partial \log L(\Theta_0, F_0)}{\partial \theta_i}\right)'.
\end{eqnarray*}
Therefore, we have

\begin{eqnarray}\label{rev22}
&&E\left\|\Omega_{u\gamma}'\Omega_u^{-1} \cdot \left(\frac{1}{T}\frac{\partial \log L(\Theta_0, F_0)}{\partial \Theta_v}\right)\right\|^2 =\frac{1}{T^2}\sum_{s=1}^T E\left\|\sum_{i=1}^N\Omega_{u\gamma,is}'\Sigma_{u,i}^{-1} \frac{\partial \log L(\Theta_0, F_0)}{\partial \theta_i}\right\|^2
\nonumber\\
&\leq& \frac{1}{T^2} \sum_{i=1}^N \sum_{j=1}^N \sum_{s=1}^T \sum_{t_1=1}^T \sum_{t_2=1}^T \|\Omega_{u\gamma,is}\|\cdot\|\Sigma_{u,i}^{-1}\|\cdot |E[ g_\varepsilon(z_{it_1}^0)g_\varepsilon(z_{jt_2}^0)u_{it_1}^0u_{jt_2}^0e_{it_1}e_{jt_2}] |
\nonumber\\
&\leq & O(1)\frac{1}{T}  \sum_{i=1}^N \sum_{j=1}^N \sum_{t_1=1}^T \sum_{t_2=1}^T \alpha_{ij}(|t_1-t_2|)^{\delta/(4+\delta)}E\left[\|u_{it_1}^{0}\| \cdot \|u_{jt_2}^{0}\|\cdot  E\left[|e_{it_1}|^{2+\delta/2} |\, \mathcal{W}\right]^{2/(4+\delta)} 
\right.
\nonumber\\
&&\left.
\cdot E\left[|e_{jt_2}|^{2+\delta/2} |\, \mathcal{W}\right]^{2/(4+\delta)}\right]
\nonumber\\
&=&O\left(N\right),
\end{eqnarray}
where the last equality holds by the $\alpha$-mixing and moment conditions in  Assumption \ref{Ass2}.

By \eqref{rev22} and the fact that $E\left[\Omega_{u\gamma}'\Omega_u^{-1} \cdot \left(\frac{1}{T}\frac{\partial \log L(\Theta_0, F_0)}{\partial \Theta_v}\right)\right]=0$,

\begin{eqnarray}\label{rev23}
\left\|\Omega_{u\gamma}'\Omega_u^{-1} \cdot \left(\frac{1}{T}\frac{\partial \log L(\Theta_0, F_0)}{\partial \Theta_v}\right)\right\|=O_P(\sqrt{N}).
\end{eqnarray}
Moreover, by the fact  $\|\left(\Omega_\gamma - \frac{1}{NT}\Omega_{u\gamma}'\Omega_u^{-1}\Omega_{u\gamma}\right)^{-1}\|=O(\sqrt{T})$ and $\|\Omega_u^{-1} \Omega_{u\gamma}\|=O(\sqrt{NT})$, we have

\begin{eqnarray}\label{rev46}
\|\mathcal{R}_{NT, 2}\| = O_P\left(1\right).
\end{eqnarray}
For $\mathcal{R}_{NT, 3}$, since $\Omega_{u\gamma}\Omega_{\gamma}^{-1}\cdot \left(\frac{1}{N}\frac{\partial \log L(\Theta_0, F_0)}{\partial F_v}\right)$ has a symmetric structure with $\Omega_{u\gamma}'\Omega_u^{-1} \cdot \left(\frac{1}{T}\frac{\partial \log L(\Theta_0, F_0)}{\partial \Theta_v}\right)$, we can use arguments analogous to those in the proof of \eqref{rev23} to show that

\begin{eqnarray*}
\left\|\Omega_{u\gamma}\Omega_\gamma^{-1} \cdot \left(\frac{1}{N}\frac{\partial \log L(\Theta_0, F_0)}{\partial F_v}\right)\right\|=O_P(\sqrt{T}),
\end{eqnarray*}
which yields that 

\begin{eqnarray}\label{rev47}
\|\mathcal{R}_{NT, 3}\| = O_P\left(\sqrt{\frac{N}{T}}\right).
\end{eqnarray}
For $\mathcal{R}_{NT, 4}$,

\begin{eqnarray}\label{rev29}
\mathcal{R}_{NT, 4} &=& \frac{1}{NT}\Omega_u^{-1} \Omega_{u\gamma}\Omega_\gamma^{\ast-1}\Omega_{u\gamma}'\Omega_u^{-1}\widetilde{\mathcal{Q}}_{NT}
\nonumber\\
&=&\frac{1}{NT}\Omega_u^{-1} \Omega_{u\gamma}\Omega_\gamma^{\ast-1}\Omega_{u\gamma}'\Omega_u^{-1} \mathcal{Q}_{NT}
+\frac{1}{NT}\Omega_u^{-1} \Omega_{u\gamma}\Omega_\gamma^{\ast-1}\Omega_{u\gamma}'
\Omega_u^{-1}(C_{NT, 1}+C_{NT, 2}+C_{NT, 3})(\widehat{\Theta}_{v}-\Theta_{0v})
\nonumber\\
&&+\frac{1}{NT^2}\Omega_u^{-1} \Omega_{u\gamma}\Omega_\gamma^{\ast-1}\Omega_{u\gamma}'\Omega_u^{-1}(C_{NT, 4}+C_{NT, 5}+C_{NT, 6})(\widehat{F}_{v}-F_{0v}).
\end{eqnarray}

We now consider these terms one by one. For the first term on the right-hand side of \eqref{rev29}. Recall that $\mathcal{Q}_{NT}=(\mathcal{Q}_{T,1}',\cdots, \mathcal{Q}_{T, N}')$ where $\mathcal{Q}_{Ti}$ is defined in \eqref{rev25}.  Denote $\mathcal{Q}_{Ti, 1}=\frac{1}{T}\sum_{l=1}^{d_{\beta}+d_f}\frac{\partial^3 \log L(\dot{\Theta}, \dot{F})}{\partial \theta_i\partial\theta_i'\partial \theta_{il}}(\widehat{\theta}_{i}-\theta_{0i})(\widehat{\theta}_{il}-\theta_{0il})$, $\mathcal{Q}_{Ti, 2}=\frac{2}{T}\sum_{t=1}^T\sum_{r=1}^{d_f}\frac{\partial^3 \log L(\dot{\Theta}, \dot{F})}{\partial \theta_i\partial\theta_i'\partial f_{tr}}(\widehat{\theta}_{i}-\theta_{0i})(\widehat{f}_{tr}-f_{0tr})$, 
$\mathcal{Q}_{Ti, 3}=\frac{1}{T}\sum_{t=1}^T\sum_{r=1}^{d_f}\frac{\partial^3 \log L(\dot{\Theta}, \dot{F})}{\partial \theta_i\partial f_t'\partial f_{tr}}(\widehat{f}_{t}-f_{0t})(\widehat{f}_{tr}-f_{0tr})$. With this notation, we have  $\mathcal{Q}_{Ti}=\mathcal{Q}_{Ti,1}+\mathcal{Q}_{Ti,2}+\mathcal{Q}_{Ti,3}$. For $\mathcal{Q}_{Ti,1}$, we have

\begin{eqnarray}\label{rev26}
\sum_{i=1}^N \|\Sigma_{u,i}^{-1}\mathcal{Q}_{Ti,1}\|^2 \leq O_P(1) \sum_{i=1}^N \|\widehat{\theta}_i-\theta_{0i}\|^4=o_P(1) \sum_{i=1}^N \|\widehat{\theta}_i-\theta_{0i}\|^2=o_P\left(\|\widehat{\Theta}_v-\Theta_{0v}\|^2\right),
\end{eqnarray}
where the inequality holds by  Assumption \ref{Ass2} and the equality holds by the uniform consistency in Lemma \ref{Lemmauni}.(1). 

For $\mathcal{Q}_{Ti,2}$, 

\begin{eqnarray}\label{rev27}
\sum_{i=1}^N \|\Sigma_{u,i}^{-1}\mathcal{Q}_{Ti,2}\|^2 &\leq& O_P(1)\left(\sum_{i=1}^N \|\widehat{\theta}_i-\theta_{0i}\|^2\right)\left(\frac{1}{T}\sum_{t=1}^T \|\widehat{f}_t-f_{0t}\|^2 \right)
\nonumber\\
&=&O_P\left(\|\widehat{\Theta}_v-\Theta_{0v} \|^2\cdot \left(\frac{1}{T}\|\widehat{F}_v-F_{0v} \|^2\right)\right),
\end{eqnarray}
where  inequality holds by Cauchy-Schwarz inequality and  Assumption \ref{Ass2}.  

For $\mathcal{Q}_{Ti,3}$, 
\begin{eqnarray}\label{rev28}
\sum_{i=1}^N \|\Sigma_{u,i}^{-1}\mathcal{Q}_{Ti,3}\|^2 &\leq& O_P(N)\left(\frac{1}{T}\sum_{t=1}^T \|\widehat{f}_t-f_{0t}\|^2 \right)^2
=O_P\left( N\cdot \left( \frac{1}{T}\|\widehat{F}_v-F_{0v}\|^2\right)^2\right).
\end{eqnarray}

By \eqref{rev26}, \eqref{rev27} and \eqref{rev28},  the first term on the right-hand side of \eqref{rev29} satisfies that

\begin{eqnarray}\label{rev33}
\|\Omega_u^{-1} \mathcal{Q}_{NT}\|=o_P\left(\|\widehat{\Theta}_v-\Theta_{0v}\|\right)+O_P\left(\|\widehat{\Theta}_v-\Theta_{0v}\|\cdot \frac{1}{\sqrt{T}}\|\widehat{F}_v-F_{0v} \|\right)+O_P\left(\sqrt{N}\cdot \frac{1}{T}\|\widehat{F}_v-F_{0v} \|^2\right).
\nonumber\\
\end{eqnarray}
In addition, since we have $\|\left(\Omega_\gamma - \frac{1}{NT}\Omega_{u\gamma}'\Omega_u^{-1}\Omega_{u\gamma}\right)^{-1}\|=O(\sqrt{T})$ and $\|\left(\Omega_u - \frac{1}{NT}\Omega_{u\gamma}\Omega_\gamma^{-1}\Omega_{u\gamma}'\right)^{-1}\|=O(\sqrt{N})$ by Assumption \ref{Ass2}, we know that $\|\Omega_u^{-1} \Omega_{u\gamma}\Omega_\gamma^{\ast-1}\Omega_{u\gamma}'\|=O(NT)$. Jointly with \eqref{rev33}, it yields that 

\begin{eqnarray}\label{rev34}
\frac{1}{NT}\|\Omega_u^{-1} \Omega_{u\gamma}\Omega_\gamma^{\ast-1}\Omega_{u\gamma}'\Omega_u^{-1} \mathcal{Q}_{NT}\|=o_P\left(\|\widehat{\Theta}-\Theta_0\|\right)+O_P\left(\sqrt{N}\cdot \frac{1}{T}\|\widehat{F}_v-F_{0v} \|^2\right).
\end{eqnarray}

We now proceed with  the rate of convergence for the second term on the right-hand side of \eqref{rev29}. Note that $\Omega_u^{-1} C_{NT, 1}$, $\Omega_u^{-1} C_{NT, 2}$ and $ \Omega_u^{-1} C_{NT, 3}$  are all block-diagonal matrices. For the term with $C_{NT, 1}$, we have

\begin{eqnarray*}
 &&\left\|\Omega_u^{-1} C_{NT, 1}(\widehat{\Theta}_{v}-\Theta_{0v})\right\|^2=\sum_{i=1}^N \left\| \Sigma^{-1}_{u,i} C_{1Ti} (\widehat{\theta}_i-\theta_{0i})\right\|^2
\nonumber\\
&\leq&\sum_{i=1}^N \left\|\Sigma^{-1}_{u,i} \right\|^2 \cdot  \left\|\frac{1}{T}\sum_{t=1}^T \left\{\frac{[g_\varepsilon(z_{it}^0)]^2 }{[1-G_\varepsilon(z_{it}^0)]G_\varepsilon(z_{it}^0)}  \right\} u_{it}^0u_{it}^{0\prime}-\Sigma_{u,i}\right\|^2\cdot \|\widehat{\theta}_i-\theta_{0i}\|^2
\nonumber\\
&\leq & O\left(1\right)\cdot \max_{1\leq i\leq N}\left\|\frac{1}{T}\sum_{t=1}^T \left\{\frac{[g_\varepsilon(z_{it}^0)]^2 }{[1-G_\varepsilon(z_{it}^0)]G_\varepsilon(z_{it}^0)}  \right\} u_{it}^0u_{it}^{0\prime}-\Sigma_{u,i}\right\|^2\cdot  \|\widehat{\Theta}_{v}-\Theta_{0v}\|^2
\nonumber\\
&=&o_P\left(\|\widehat{\Theta}_{v}-\Theta_{0v}\|^2\right) ,
\end{eqnarray*}
where the second inequality holds by Assumption \ref{Ass2} and the last equality holds by the fact that  $\max_{1\leq i\leq N}\left\|\frac{1}{T}\sum_{t=1}^T \left\{\frac{[g_\varepsilon(z_{it}^0)]^2 }{[1-G_\varepsilon(z_{it}^0)]G_\varepsilon(z_{it}^0)}  \right\} u_{it}^0u_{it}^{0\prime}-\Sigma_{u,i}\right\|^2=o_P(1)$ under the condition $N/T^{1+\delta^\ast/4}\rightarrow0$. By Lemma \ref{LemmaT2}, we can  prove it after checking the moment conditions required by the lemma, which can hold immediately by the  fact that $\max_{z_{it}}\{[1-G_\varepsilon(z_{it}^0)]G_\varepsilon(z_{it}^0)\}^{-1}=O(1)$,  $g_\varepsilon(z_{it}^0)$ is uniformly bounded  by Assumption \ref{Ass0} and $E\|u_{it}^0\|^{4+\delta}<\infty$ by  Assumption \ref{Ass2}.

It yields that

\begin{eqnarray}\label{rev2}
 \left\|\Omega_u^{-1} C_{NT, 1}(\widehat{\Theta}_{v}-\Theta_{0v})\right\|=o_P\left( \|\widehat{\Theta}_{v}-\Theta_{0v}\|\right).
\end{eqnarray}

For $\Omega_u^{-1} C_{NT, 2}(\widehat{\Theta}_{v}-\Theta_{0v})$, recall that $e_{it} =-\frac{y_{it}-G_\varepsilon(z_{it}^0)}{(1- G_\varepsilon(z_{it}^0))G_\varepsilon(z_{it}^0)}$. Then we know that $C_{2Ti}=-\frac{1}{T}\sum_{t=1}^T \frac{g_\varepsilon^{(1)}(z_{it}^0) e_{it}}{[1-G_\varepsilon(z_{it}^0)][G_\varepsilon(z_{it}^0)]}  \cdot u_{it}^0 u_{it}^{0\prime}$
and

\begin{eqnarray}\label{rev1}
&&\left\|\Omega_u^{-1} C_{NT, 1}(\widehat{\Theta}_{v}-\Theta_{0v})\right\|^2
\nonumber\\
&\leq& \sum_{i=1}^N  \left\|\Sigma^{-1}_{u,i} \right\|^2\cdot  \left\|\frac{1}{T}\sum_{t=1}^T \frac{g_\varepsilon^{(1)}(z_{it}^0) e_{it}}{[1-G_\varepsilon(z_{it}^0)][G_\varepsilon(z_{it}^0)]}  \cdot u_{it}^0 u_{it}^{0\prime}\right\|^2\cdot \|\widehat{\theta}_i-\theta_{0i}\|^2
\nonumber\\
&\leq & O\left(1\right)\cdot \max_{1\leq i\leq N}\left\|\frac{1}{T}\sum_{t=1}^T \frac{g_\varepsilon^{(1)}(z_{it}^0) e_{it}}{[1-G_\varepsilon(z_{it}^0)][G_\varepsilon(z_{it}^0)]}  \cdot u_{it}^0 u_{it}^{0\prime}\right\|^2\cdot  \|\widehat{\Theta}_{v}-\Theta_{0v}\|^2
\nonumber\\
&=&o_P\left( \|\widehat{\Theta}_{v}-\Theta_{0v}\|^2 \right)
,
\end{eqnarray}
where the last equality holds by Lemma \ref{LemmaT2} and the fact that $E[e_{it}|\mathcal{W}]=0$, $E[|e_{it}|^{4+\delta}|\mathcal{W}]<\infty$, $E[\|u_{it}^0\|^{4+\delta}]<\infty$ by Assumption \ref{Ass2}. Therefore, we have

\begin{eqnarray}\label{rev3}
 \left\|\Omega_u^{-1} C_{NT, 2}(\widehat{\Theta}_{v}-\Theta_{0v})\right\|=o_P\left( \|\widehat{\Theta}_{v}-\Theta_{0v}\|\right).
\end{eqnarray}
Analogously to \eqref{rev3}, we can show that

\begin{eqnarray}\label{rev4}
 \left\|\Omega_u^{-1} C_{NT, 3}(\widehat{\Theta}_{v}-\Theta_{0v})\right\|=o_P\left( \|\widehat{\Theta}_{v}-\Theta_{0v}\|\right).
\end{eqnarray}
By  \eqref{rev2}, \eqref{rev3} and \eqref{rev4},

\begin{eqnarray}\label{rev8}
\left\|\Omega_{u}^{-1}\cdot(C_{NT, 1}+C_{NT, 2}+C_{NT, 3})(\widehat{\Theta}_{v}-\Theta_{0v})\right\| = o_P\left( \|\widehat{\Theta}_{v}-\Theta_{0v}\|\right).
\end{eqnarray}
which immediately yields that

\begin{eqnarray}\label{rev35}
\frac{1}{NT}\left\|\Omega_u^{-1} \Omega_{u\gamma}\Omega_\gamma^{\ast-1}\Omega_{u\gamma}'
\Omega_u^{-1}(C_{NT, 1}+C_{NT, 2}+C_{NT, 3})(\widehat{\Theta}_{v}-\Theta_{0v})\right\| = o_P\left( \|\widehat{\Theta}_{v}-\Theta_{0v}\|\right).
\end{eqnarray}

For the third term on the right-hand side of \eqref{rev29}, recall that $C_{NT, 4}$, $C_{NT, 5}$ and $C_{NT, 6}$ are defined in  \eqref{rev17}. For $C_{NT, 4}$, recall that $C_{4it} = -\zeta_{it} = -\left(\frac{[g_\varepsilon(z^0_{it})]^2 }{[1-G_\varepsilon(z^0_{it})]G_\varepsilon(z^0_{it})}  u^0_{it}\gamma_{0i}' - \Omega_{u\gamma, i t}\right)$. By Cauchy-Schwarz inequality,

\begin{eqnarray}\label{rev36}
\|\Omega_{u\gamma}'\Omega_u^{-1}C_{NT, 4}(\widehat{F}_{v}-F_{0v})\|^2 &=&\sum_{s=1}^T \left\|\sum_{i=1}^N\sum_{t=1}^T \Omega_{u\gamma,is}' \Sigma_{u,i}^{-1} \zeta_{it}(\widehat{f}_t-f_{0t})\right\|^2 
\nonumber\\
&\leq &  \sum_{t=1}^T \sum_{s=1}^T \left\|\sum_{i=1}^N \Omega_{u\gamma,is}' \Sigma_{u,i}^{-1} \zeta_{it} \right\|^2 \cdot \sum_{t=1}^T \| \widehat{f}_t-f_{0t}\|^2.  
\end{eqnarray}

For the first term in the product on the right-hand side of \eqref{rev36},

\begin{eqnarray*}
\sum_{t=1}^T \sum_{s=1}^T E\left\|\sum_{i=1}^N \Omega_{u\gamma,is}' \Sigma_{u,i}^{-1} \zeta_{it} \right\|^2 &\leq& O(T)  \sum_{i=1}^N\sum_{j=1}^N\sum_{t=1}^T \| E[\zeta_{it}\zeta_{jt}']\|
\nonumber\\
&\leq& O(T)  \sum_{i=1}^N\sum_{j=1}^N\sum_{t=1}^T \alpha_{ij}(0)^{\delta/(4+\delta)}E\left[  \|\zeta_{it}\|^{2+\delta/2} \right]^{2/(4+\delta)} \cdot E\left[  \|\zeta_{jt}\|^{2+\delta/2} \right]^{2/(4+\delta)} 
\nonumber\\
&=&O(NT^2),
\end{eqnarray*}
where the second inequality holds by the fact that $\zeta_{it}$ is $\alpha$-mixing under the conditions in Assumption \ref{Ass2}, the Davydov's inequality and $\alpha$-mixing conditions in Assumption \ref{Ass2}. Jointly with \eqref{rev36}, it yields that

\begin{eqnarray}\label{rev362}
\|\Omega_{u\gamma}'\Omega_u^{-1}C_{NT, 4}(\widehat{F}_{v}-F_{0v})\|&=& O_P\left(\sqrt{NT^2}\right)\cdot \|\widehat{F}_v-F_{0v}\|.
\end{eqnarray}

Analogously, we can show that 

\begin{eqnarray}\label{rev37}
\|\Omega_{u\gamma}'\Omega_u^{-1}C_{NT, 5}(\widehat{F}_{v}-F_{0v})\|&=& O_P\left(\sqrt{NT^2}\right)\cdot \|\widehat{F}_v-F_{0v}\|,
\end{eqnarray}
and

\begin{eqnarray}\label{rev38}
\|\Omega_{u\gamma}'\Omega_u^{-1}C_{NT, 6}(\widehat{F}_{v}-F_{0v})\|&=& O_P\left(\sqrt{NT^2}\right)\cdot \|\widehat{F}_v-F_{0v}\|.
\end{eqnarray}
By \eqref{rev362}, \eqref{rev37} and \eqref{rev38},

\begin{eqnarray}\label{rev39}
\frac{1}{NT^2}\|\Omega_u^{-1} \Omega_{u\gamma}\Omega_\gamma^{\ast-1}\Omega_{u\gamma}'\Omega_u^{-1}(C_{NT, 4}+C_{NT, 5}+C_{NT, 6})(\widehat{F}_{v}-F_{0v})\| = O_P\left(\frac{1}{\sqrt{T}} \|\widehat{F}_v-F_{0v}\|\right).
\end{eqnarray}
By \eqref{rev34}, \eqref{rev35} and \eqref{rev39}, we have

\begin{eqnarray}\label{rev48}
\|\mathcal{R}_{NT, 4}\|=o_P\left(\|\widehat{\Theta}_v-\Theta_{0v}\|\right)+O_P\left(\frac{1}{\sqrt{T}} \|\widehat{F}_v-F_{0v}\|\right)+O_P\left(\sqrt{N}\cdot \frac{1}{T}\|\widehat{F}_v-F_{0v} \|^2\right).
\end{eqnarray}

For $\mathcal{R}_{NT, 5}$, recall that

\begin{eqnarray}\label{rev40}
\mathcal{R}_{NT, 5}&=&-\frac{1}{T}\Omega_u^{\ast-1}\Omega_{u\gamma}\Omega_{\gamma}^{-1}\widetilde{\mathcal{J}}_{NT}
\nonumber\\
&=&-\frac{1}{T}\Omega_u^{\ast-1}\Omega_{u\gamma}\Omega_{\gamma}^{-1}\mathcal{J}_{NT}  
-\frac{1}{T}\Omega_u^{\ast-1}\Omega_{u\gamma}\Omega_{\gamma}^{-1} (D_{NT, 1}+D_{NT, 2}+D_{NT, 3})(\widehat{F}_{v}-F_{0v})
\nonumber\\
&&-\frac{1}{NT}\Omega_u^{\ast-1}\Omega_{u\gamma}\Omega_{\gamma}^{-1}(C_{NT, 4}'+C_{NT, 5}'+C_{NT, 6}')(\widehat{\Theta}_{v}-\Theta_{0v}). 
\end{eqnarray}

We can observe that $\mathcal{J}_{NT}$ and $D_{NT, k}$ have similar structures with $\mathcal{Q}_{NT}$ and $C_{NT, k}$, respectively,  for $k=1,2,3$. Therefore, we can follow analogous arguments in the proof of  \eqref{rev33}, \eqref{rev8} and \eqref{rev39} to show the following results for $\mathcal{R}_{NT, 5}$:

\begin{eqnarray*}
\|\Omega_{\gamma}^{-1}\mathcal{J}_{NT}\|=o_P\left(\|\widehat{F}_v-F_{0v}\|\right)+O_P\left(\|\widehat{F}_v-F_{0v}\|\cdot \frac{1}{\sqrt{N}}\|\widehat{\Theta}_v-\Theta_{0v} \|\right)+O_P\left(\sqrt{T}\cdot \frac{1}{N}\|\widehat{\Theta}_v-\Theta_{0v} \|^2\right),
\end{eqnarray*}

\begin{eqnarray*}
\|\Omega_{\gamma}^{-1} (D_{NT, 1}+D_{NT, 2}+D_{NT, 3})(\widehat{F}_{v}-F_{0v})\|=o_P\left( \|\widehat{F}_{v}-F_{0v}\|\right),
\end{eqnarray*}
and

\begin{eqnarray*}
\frac{1}{NT}\|\Omega_u^{\ast-1}\Omega_{u\gamma}\Omega_{\gamma}^{-1}(C_{NT, 4}'+C_{NT, 5}'+C_{NT, 6}')(\widehat{\Theta}_{v}-\Theta_{0v})\|= O_P\left(\frac{1}{\sqrt{N}} \|\widehat{\Theta}_v-\Theta_{0v}\|\right).
\end{eqnarray*}
They jointly yield that 

\begin{eqnarray}\label{rev49}
\|\mathcal{R}_{NT, 5}\|&=& o_P\left(\sqrt{\frac{N}{T}}\|\widehat{F}_v-F_{0v}\|\right)+O_P\left(\frac{1}{\sqrt{N}} \|\widehat{\Theta}_v-\Theta_{0v}\|\right)+O_P\left(\|\widehat{\Theta}_v-\Theta_{0v} \|\cdot \frac{1}{\sqrt{T}} \|\widehat{F}_v-F_{0v}\|  \right)
\nonumber\\
&&+O_P\left( \frac{1}{\sqrt{N}}\|\widehat{\Theta}_v-\Theta_{0v} \|^2\right).
\end{eqnarray}

For $\mathcal{R}_{NT, 6}$, by  \eqref{rev33} and \eqref{rev8},

\begin{eqnarray}\label{rev24}
&&\mathcal{R}_{NT, 6} = \Omega_u^{-1} \widetilde{\mathcal{Q}}_{NT}
\nonumber\\
&& =  \Omega_u^{-1} \mathcal{Q}_{NT}+ \Omega_u^{-1}(C_{NT, 1}+C_{NT, 2}+C_{NT, 3})(\widehat{\Theta}_{v}-\Theta_{0v})+\frac{1}{T}\Omega_u^{-1}(C_{NT, 4}+C_{NT, 5}+C_{NT, 6})(\widehat{F}_{v}-F_{0v})
\nonumber\\
&&=\frac{1}{T}\Omega_u^{-1}(C_{NT, 4}+C_{NT, 5}+C_{NT, 6})(\widehat{F}_{v}-F_{0v})
+o_P\left(\|\widehat{\Theta}_v-\Theta_{0v}\|\right)+O_P\left(\sqrt{N}\cdot \frac{1}{T}\|\widehat{F}_v-F_{0v} \|^2\right)
\nonumber\\
&&\quad +O_P\left(\|\widehat{\Theta}_v-\Theta_{0v}\|\cdot \frac{1}{\sqrt{T}}\|\widehat{F}_v-F_{0v} \|\right).
\end{eqnarray}
For the first term in \eqref{rev24}, we need to use the expansions for $\widehat{F}_{v}-F_{0v}$. Analogously to \eqref{rev41}, we have

\begin{eqnarray}\label{rev42}
\widehat{F}_{v}-F_{0v}
&=&\Omega_\gamma^{-1}\cdot \frac{1}{N}\frac{\partial \log L(\Theta_0, F_0)}{\partial F_v}
+\frac{1}{NT}\Omega_\gamma^{-1} \Omega_{u\gamma}'\Omega_u^{\ast-1}\Omega_{u\gamma}\Omega_\gamma^{-1} \cdot \frac{1}{N}\frac{\partial \log L(\Theta_0, F_0)}{\partial F_v}
\nonumber\\
&&-\frac{1}{N}\Omega_\gamma^{\ast-1}\Omega_{u\gamma}'\Omega_{u}^{-1}\cdot \frac{1}{T}\frac{\partial \log L(\Theta_0, F_0)}{\partial \Theta_v}
 +\frac{1}{NT}\Omega_\gamma^{-1} \Omega_{u\gamma}'\Omega_u^{\ast-1}\Omega_{u\gamma}\Omega_\gamma^{-1}\widetilde{\mathcal{J}}_{NT}
\nonumber\\
&&-\frac{1}{N}\Omega_\gamma^{\ast-1}\Omega_{u\gamma}'\Omega_{u}^{-1}\widetilde{\mathcal{Q}}_{NT}+\Omega_\gamma^{-1}\widetilde{\mathcal{J}}_{NT}
\nonumber\\
&=&\mathcal{P}_{NT, 1}+\cdots+\mathcal{P}_{NT, 6}.
\end{eqnarray}

With \eqref{rev42}, we have

\begin{eqnarray}\label{rev43}
\frac{1}{T}\Omega_u^{-1}C_{NT, 4}(\widehat{F}_{v}-F_{0v})&=&\frac{1}{T}\Omega_u^{-1}C_{NT, 4}\left(\mathcal{P}_{NT, 1}+\cdots+\mathcal{P}_{NT, 6}\right).
\end{eqnarray}
We now consider these six terms one by one. Recall that $\frac{\partial \log L(\Theta_0, F_0)}{\partial f_t} = \sum_{i=1}^N \frac{[y_{it}-G_\varepsilon(z_{it}^0)]g_\varepsilon(z_{it}^0) }{[1-G_\varepsilon(z_{it}^0)]G_\varepsilon(z_{it}^0)}  \gamma_{0i}= - \sum_{i=1}^N g_\varepsilon(z_{it}^0) \gamma_{0i}e_{it}$.

For the first term in \eqref{rev43}, 

\begin{eqnarray}\label{rev44}
&&\frac{1}{T^2}E\|\Omega_u^{-1}C_{NT, 4}\mathcal{P}_{NT, 1}\|^2= \frac{1}{N^2T^2}\left\|\Omega_u^{-1}C_{NT, 4}\Omega_\gamma^{-1}\cdot \frac{\partial \log L(\Theta_0, F_0)}{\partial F_v}\right\|
\nonumber\\
&=&\frac{1}{N^2T^2}\sum_{j=1}^N E \left\|\sum_{i=1}^N\sum_{t=1}^T \Sigma_{u,j}^{-1}C_{4jt}\Sigma_{\gamma,t}^{-1} g_\varepsilon(z_{it}^0) \gamma_{0i}e_{it} \right\|^2
\nonumber\\
&\leq & O\left(\frac{1}{NT^2}\right)  \sum_{i=1}^N\sum_{j=1}^N \sum_{t=1}^T\sum_{s=1}^T |E\left[ E[e_{it}e_{js}|\mathcal{W}]\right]
\nonumber\\
&\leq &
O\left(\frac{1}{NT^2}\right)  \sum_{i=1}^N\sum_{j=1}^N \sum_{t=1}^T\sum_{s=1}^T \alpha_{ij}(|t-s|)^{\delta/(4+\delta)}E\left[ E\left[|e_{it}|^{2+\delta/2} |\, \mathcal{W}\right]^{2/(4+\delta)} 
\cdot E\left[|e_{js}|^{2+\delta/2} |\, \mathcal{W}\right]^{2/(4+\delta)}\right]
\nonumber\\
&=&O\left(\frac{1}{T}\right).
\end{eqnarray}
 By \eqref{rev44}, we have 

\begin{eqnarray}\label{rev50}
\frac{1}{T}\|\Omega_u^{-1}C_{NT, 4}\mathcal{P}_{NT, 1}\|=O_P\left(\frac{1}{\sqrt{T}}\right).
\end{eqnarray}
$\mathcal{P}_{NT, 2}$, $\mathcal{P}_{NT, 3}$, $\mathcal{P}_{NT, 4}$ and $\mathcal{P}_{NT, 5}$ have similar structures with $\mathcal{R}_{NT, 2}$, $\mathcal{R}_{NT, 3}$, $\mathcal{R}_{NT, 4}$ and $\mathcal{R}_{NT, 5}$, respectively. Therefore, we can use analogous arguments in the proofs of \eqref{rev46}, \eqref{rev47}, \eqref{rev48} and \eqref{rev49} and obtain the following results: 

\begin{eqnarray*}
\|\mathcal{P}_{NT, 2}\| = O_P\left(1\right),\quad \|\mathcal{P}_{NT, 3}\| = O_P\left(\sqrt{\frac{T}{N}}\right),
\end{eqnarray*}
\begin{eqnarray*}
\|\mathcal{P}_{NT, 4}\|=o_P\left(\|\widehat{F}-F_{0v}\|\right)+O_P\left(\frac{1}{\sqrt{N}} \|\widehat{\Theta}_v-\Theta_{0v}\|\right)+O_P\left(\sqrt{T}\cdot \frac{1}{N}\|\widehat{\Theta}_v-\Theta_{0v} \|^2\right),
\end{eqnarray*}
and
\begin{eqnarray*}
\|\mathcal{P}_{NT, 5}\|&=& o_P\left(\sqrt{\frac{T}{N}}\|\widehat{\Theta}_v-\Theta_{0v}\|\right)+O_P\left(\frac{1}{\sqrt{T}} \|\widehat{F}_v-F_{0v}\|\right)+O_P\left(\|\widehat{F}_v-F_{0v} \|\cdot \frac{1}{\sqrt{N}} \|\widehat{\Theta}_v-\Theta_{0v}\|  \right)
\nonumber\\
&&+O_P\left( \frac{1}{\sqrt{T}}\|\widehat{F}_v-F_{0v} \|^2\right).
\end{eqnarray*}
Jointly with \eqref{rev50} and the fact  $\|\Omega_u^{-1}C_{NT, 4}\|=O_P(\sqrt{NT})$, they yield that

\begin{eqnarray}\label{rev53}
&&\frac{1}{T}\Omega_u^{-1}C_{NT, 4}\left(\mathcal{P}_{NT, 2}+\cdots+\mathcal{P}_{NT, 5}\right)=O_P\left(1\right)+O_P\left(\sqrt{\frac{N}{T}}\right)+o_P\left(\sqrt{\frac{N}{T}}\|\widehat{F}-F_{0v}\|\right)
\nonumber\\
&&+O_P\left(\frac{1}{\sqrt{T}} \|\widehat{\Theta}_v-\Theta_{0v}\|\right)
+ o_P\left(\|\widehat{\Theta}_v-\Theta_{0v}\|\right)+O_P\left( \frac{1}{\sqrt{N}}\|\widehat{\Theta}_v-\Theta_{0v} \|^2\right)+O_P\left(\sqrt{N}\cdot  \frac{1}{T}\|\widehat{F}_v-F_{0v} \|^2\right)
\nonumber\\
&&+O_P\left(  \|\widehat{\Theta}_v-\Theta_{0v}\| \cdot \frac{1}{\sqrt{T}} \|\widehat{F}_v-F_{0v} \| \right).
\end{eqnarray}
We now proceed with $\frac{1}{T}\Omega_u^{-1}C_{NT, 4}\mathcal{P}_{NT, 6}$. Following analogous arguments in the proofs of \eqref{rev24}, we can show that

\begin{eqnarray}\label{rev51}
&&\frac{1}{T}\Omega_u^{-1}C_{NT, 4}\mathcal{P}_{NT, 6} = \frac{1}{T}\Omega_u^{-1}C_{NT, 4} \Omega_\gamma^{-1} \widetilde{\mathcal{J}}_{NT}
\nonumber\\
&&=\frac{1}{NT}\Omega_u^{-1}C_{NT, 4} \Omega_\gamma^{-1}(C_{NT, 4}'+C_{NT, 5}'+C_{NT, 6}')(\widehat{\Theta}_{v}-\Theta_{0v})
+o_P\left(\sqrt{\frac{N}{T}} \|\widehat{F}_v-F_{0v}\|\right)
\nonumber\\
&&\quad +O_P\left(\frac{1}{\sqrt{N}}\|\widehat{\Theta}_v-\Theta_{0v} \|^2\right)+O_P\left(  \|\widehat{\Theta}_v-\Theta_{0v}\| \cdot \frac{1}{\sqrt{T}}\|\widehat{F}_v-F_{0v} \|\right).
\end{eqnarray}
We can then have

\begin{eqnarray*}
E\|\Omega_u^{-1}C_{NT, 4}\Omega_\gamma^{-1}C_{NT, 4}'\|^2&=&\sum_{i=1}^N\sum_{j=1}^N E\left\|\sum_{t=1}^T\Sigma_{u,i}^{-1}\zeta_{it}\Sigma_{\gamma,t}^{-1}\zeta_{jt}'\right\|^2
\nonumber\\
&\leq & O(1)\sum_{i=1}^N\sum_{j=1}^N\sum_{t=1}^T\sum_{s=1}^T \left|E[\zeta_{it}'\zeta_{is}\zeta_{jt}'\zeta_{js}]\right|
\nonumber\\
&=& O(NT(N\vee T)), 
\end{eqnarray*}
which can immediately yield that 

\begin{eqnarray*}
\left\|\frac{1}{T}\Omega_u^{-1}C_{NT, 4}\cdot \frac{1}{N}\Omega_\gamma^{-1}C_{NT, 4}'(\widehat{\Theta}_{v}-\Theta_{0v})\right\| =O_P\left(\frac{1}{\sqrt{N}\wedge\sqrt{T}} \|\widehat{\Theta}_{v}-\Theta_{0v}\|\right).
\end{eqnarray*}

Analogously, we can obtain the same rates of convergence for $\frac{1}{T}\Omega_u^{-1}C_{NT, 4}\cdot \frac{1}{N}\Omega_\gamma^{-1}C_{NT, 5}'(\widehat{\Theta}_{v}-\Theta_{0v})$ and $\frac{1}{T}\Omega_u^{-1}C_{NT, 4}\cdot \frac{1}{N}\Omega_\gamma^{-1}C_{NT, 6}'(\widehat{\Theta}_{v}-\Theta_{0v})$. Therefore, we have

\begin{eqnarray}\label{rev52}
\frac{1}{T}\|\Omega_u^{-1}C_{NT, 4}\mathcal{P}_{NT, 6}\| &=& O_P\left(\frac{1}{\sqrt{N}\wedge\sqrt{T}} \|\widehat{\Theta}_{v}-\Theta_{0v}\|\right)+o_P\left(\sqrt{\frac{N}{T}} \|\widehat{F}_v-F_{0v}\|\right)
\nonumber\\
&&+O_P\left(\frac{1}{\sqrt{N}}\|\widehat{\Theta}_v-\Theta_{0v} \|^2\right)+O_P\left(  \|\widehat{\Theta}_v-\Theta_{0v}\| \cdot \frac{1}{\sqrt{T}}\|\widehat{F}_v-F_{0v} \|\right).
\end{eqnarray}

By \eqref{rev43}, \eqref{rev50}, \eqref{rev53} and \eqref{rev52}, 
 
\begin{eqnarray}\label{rev54}
\frac{1}{T}\|\Omega_u^{-1}C_{NT, 4}(\widehat{F}_{v}-F_{0v})\|&=& 
O_P\left(1\right)+O_P\left(\sqrt{\frac{N}{T}}\right)+o_P\left(\sqrt{\frac{N}{T}}\|\widehat{F}-F_{0v}\|\right)+o_P\left(\|\widehat{\Theta}_v-\Theta_{0v}\|\right)
\nonumber\\
&&
+O_P\left( \frac{1}{\sqrt{N}}\|\widehat{\Theta}_v-\Theta_{0v} \|^2\right)+O_P\left(\sqrt{N}\cdot  \frac{1}{T}\|\widehat{F}_v-F_{0v} \|^2\right)
\nonumber\\
&&
+O_P\left(  \|\widehat{\Theta}_v-\Theta_{0v}\| \cdot \frac{1}{\sqrt{T}}\|\widehat{F}_v-F_{0v} \|\right).
\end{eqnarray}
It completes the computation of the rate of convergence for the first term in \eqref{rev24}. Following analogous arguments, we can show that $\frac{1}{T}\|\Omega_u^{-1}C_{NT, 5}(\widehat{F}_{v}-F_{0v})\|$ and $\frac{1}{T}\|\Omega_u^{-1}C_{NT, 6}(\widehat{F}_{v}-F_{0v})\|$ are also bounded  in probability by these seven terms in \eqref{rev54}.  Jointly with \eqref{rev24}, it yields that

\begin{eqnarray}\label{rev56}
\|\mathcal{R}_{NT, 6}\| &=& O_P\left(1\right)+O_P\left(\sqrt{\frac{N}{T}}\right)+o_P\left(\sqrt{\frac{N}{T}}\|\widehat{F}-F_{0v}\|\right)+o_P\left(\|\widehat{\Theta}_v-\Theta_{0v}\|\right)
\nonumber\\
&&
+O_P\left( \frac{1}{\sqrt{N}}\|\widehat{\Theta}_v-\Theta_{0v} \|^2\right)+O_P\left(\sqrt{N}\cdot  \frac{1}{T}\|\widehat{F}_v-F_{0v} \|^2\right)
\nonumber\\
&&
+O_P\left( \|\widehat{\Theta}_v-\Theta_{0v}\| \cdot \frac{1}{\sqrt{T}}\|\widehat{F}_v-F_{0v} \|\right).
\end{eqnarray}
In summary of the results in \eqref{rev55}, \eqref{rev46}, \eqref{rev47}, \eqref{rev48}, \eqref{rev49} and \eqref{rev56}, we have

\begin{eqnarray}\label{rev57}
\frac{1}{\sqrt{N}}\|\widehat{\Theta}_v-\Theta_{0v} \| = O_P \left(\frac{1}{\sqrt{N}\wedge \sqrt{T}}\right)+o_P\left(\frac{1}{\sqrt{N}}\|\widehat{\Theta}_v-\Theta_{0v}\|\right)+o_P\left(\frac{1}{\sqrt{T}}\|\widehat{F}-F_{0v}\|\right).
\end{eqnarray}
Analogously to (\ref{rev57}), we can establish the following result for $\widehat{F}_v-F_{0v}$:

\begin{eqnarray}\label{rev58}
\frac{1}{\sqrt{T}}\|\widehat{F}_v-F_{0v} \| = O_P \left(\frac{1}{\sqrt{N}\wedge \sqrt{T}}\right)+o_P\left(\frac{1}{\sqrt{N}}\|\widehat{\Theta}_v-\Theta_{0v}\|\right)+o_P\left(\frac{1}{\sqrt{T}}\|\widehat{F}-F_{0v}\|\right).
\end{eqnarray}

By \eqref{rev57} and \eqref{rev58}, we can finally have
\begin{eqnarray}\label{rev59}
\frac{1}{\sqrt{N}}\|\widehat{\Theta}_v-\Theta_{0v} \|=O_P \left(\frac{1}{\sqrt{N}\wedge \sqrt{T}}\right),\quad \frac{1}{\sqrt{T}}\|\widehat{F}_v-F_{0v} \| = O_P \left(\frac{1}{\sqrt{N}\wedge \sqrt{T}}\right).
\end{eqnarray} 

The proof of Lemma \ref{Lemmauni} is therefore completed.  \hspace*{\fill}{$\blacksquare$}

\bigskip

\noindent {\bf Proof of Lemma \ref{LemmaAPE}:}

Write

\begin{eqnarray*}
\widehat{\Delta}_{i}-\Delta_{i} &=& \frac{1}{T}\sum_{t=1}^T g(z^0_{it})(\widehat{\beta}_i-\beta_{0i})+\frac{1}{T}\sum_{t=1}^T(g(\widehat{z}_{it})-g(z^0_{it}))\beta_{0i}
\nonumber\\
&&+\frac{1}{T}\sum_{t=1}^T(g(\widehat{z}_{it})-g(z^0_{it})) (\widehat{\beta}_i-\beta_{0i})
\nonumber\\
&:=&\widehat{\Delta}_{1i}+\widehat{\Delta}_{2i}+\widehat{\Delta}_{3i}.
\end{eqnarray*}

By Assumption \ref{Ass0} and Lemma \ref{Lemmauni}, we have 

\begin{eqnarray}\label{APE1}
\max_{1\leq i\leq N}\|\widehat{\Delta}_{1i}\| =o_P(1).
\end{eqnarray}

For $\widehat{\Delta}_{2i}$, by Taylor expansion, 
\begin{eqnarray*}
\widehat{\Delta}_{2i}&=&\frac{1}{T}\sum_{t=1}^T g^{(1)}_\varepsilon(z^0_{it})(\widehat{z}_{it}-z^0_{it})\beta_{0i}+\frac{1}{T}\sum_{t=1}^T g^{(2)}_\varepsilon(\dot{z}_{it})(\widehat{z}_{it}-z^0_{it})^2\beta_{0i},
\end{eqnarray*}
where $\dot{z}_{it}$ lies between $z_{it}$ and $\widehat{z}_{it}$.
By Assumption \ref{Ass2} and Lemma \ref{Lemmauni}, it is straightforward to show that $\max_{1\leq i\leq N}\| \widehat{z}_{it}-z^0_{it}\|=o_P(1)$. Therefore, 

\begin{eqnarray}\label{APE2}
\max_{1\leq i\leq N}\|\widehat{\Delta}_{2i}\| =o_P(1).
\end{eqnarray}

For $\widehat{\Delta}_{3i}$, 

\begin{eqnarray}\label{APE3}
\max_{1\leq i\leq N}\|\widehat{\Delta}_{3i}\|=\max_{1\leq i\leq N}\left\|\frac{1}{T}\sum_{t=1}^T(g(\widehat{z}_{it})-g(z^0_{it})) \right\|\cdot  \max_{1\leq i\leq N}\|\widehat{\beta}_i-\beta_{0i}\|=o_P(1).
\end{eqnarray}

By \eqref{APE1}, \eqref{APE2} and \eqref{APE3}, we have Lemma \ref{LemmaAPE} holds. 
\hspace*{\fill}{$\blacksquare$}

\subsection{Technical Lemmas with Proofs}\label{TechLem}

\begin{lemma}\label{LemmaT1}
If $\xi_{t}$ satisfies that $E\left[\xi_{t}\right]=0$, $E\left[\left\|\xi_{t}\right\|^{2+\delta/2}\right]<\infty$ and $\xi_{t}$  is an $\alpha$-mixing process with the $\alpha$-mixing coefficient such that $\sum_{t=0}^{T}\alpha(t)^{\delta/(4+\delta)}=O(1)$ , then by Theorem 4.1 of \cite{SY1996}, we have
\begin{equation*}
E\left[\left\|\frac{1}{T}\sum_{t=1}^T \xi_{t}\right\|^{2+\delta^\ast/2}\right]\leq \frac{C}{T^{1+\delta^\ast/4}}E\left[\left\|\xi_{t}\right\|^{2+\delta^\ast/2}\right],
\end{equation*} 
where $C$ is a constant and $0<\delta^\ast<\delta$.
\end{lemma}

\begin{lemma}\label{LemmaT2}
If $\xi_{it}$ satisfies that $E\left[\xi_{it}\right]=0$, $E\left[\left\|\xi_{it}\right\|^{2+\delta/2}\right]<\infty$, $\sum_{i=1}^N E\left[\left\|\xi_{it}\right\|^{2+\delta/2}\right]=O(N)$ and $\xi_{it}$  is an $\alpha$-mixing process 
satisfying the $\alpha$-mixing conditions in  Assumption \ref{Ass2},  we have for any given $\varepsilon>0$,
\begin{equation*}
P\left(\max_{1\leq i\leq N}\left\|\frac{1}{T}\sum_{t=1}^T\xi_{it}\right\|\geq \varepsilon \right)=O\left(\frac{N}{T^{1+\delta^\ast/4}}\right),
\end{equation*} 
where $0<\delta^\ast<\delta$.
\end{lemma}

\begin{lemma}\label{LemmaT3}
Under Assumptions \ref{Ass0}-\ref{Ass2}, as $(N,T)\to (\infty,\infty)$,
\begin{enumerate}
\item  For  $\forall i$, \ $\frac{1}{T}\sum_{t=1}^T  \left(\frac{[g_\varepsilon(z_{it}^0)]^2 }{[1-G_\varepsilon(z_{it}^0)]G_\varepsilon(z_{it}^0)}  u_{it}^0u_{it}^{0\prime}-E\left[\frac{[g_\varepsilon(z_{it}^0)]^2 }{[1-G_\varepsilon(z_{it}^0)]G_\varepsilon(z_{it}^0)}  u_{it}^0u_{it}^{0\prime}\right]\right)=O_P\left(\frac{1}{\sqrt{T}}\right)$.
\item  For  $\forall t$, \  $\frac{1}{N}\sum_{i=1}^N \left(\frac{[g_\varepsilon(z_{it}^0)]^2 }{[1-G_\varepsilon(z_{it}^0)]G_\varepsilon(z_{it}^0)}   \gamma_{0i}\gamma_{0i}'-E\left[\frac{[g_\varepsilon(z_{it}^0)]^2 }{[1-G_\varepsilon(z_{it}^0)]G_\varepsilon(z_{it}^0)}   \gamma_{0i}\gamma_{0i}'\right]\right)=O_P\left(\frac{1}{\sqrt{N}}\right)$.
\end{enumerate}
\end{lemma}

\begin{lemma}\label{LemmaT4}
Let Assumptions \ref{Ass0}-\ref{Ass3} hold. As $(N,T)\to (\infty,\infty)$,
\begin{equation*}
\frac{1}{\sqrt{T}}\frac{\partial \log L(B_0, F_0,\Gamma_0)}{\partial \theta_i}\to_D N(0,\Sigma_{\theta,i}),
\end{equation*}
for  $\forall i$, where $\Sigma_{\theta,i}$ is defined in Assumption \ref{Ass3}.
\end{lemma}

Lemma \ref{LemmaT1} holds immediately by Theorem 4.1 of \cite{SY1996}. Therefore, its proof is omitted in this paper. We now provide the proofs for the rest of technical lemmas. 

\bigskip

\noindent \textbf{Proof of Lemma \ref{LemmaT2}:}

By the properties of the probability function, we have  

\begin{eqnarray*}
P\left(\max_{1\leq i\leq N}\left\|\frac{1}{T}\sum_{t=1}^T\xi_{it}\right\|\geq \varepsilon \right)&\leq &\sum_{i=1}^N P\left(\left\|\frac{1}{T}\sum_{t=1}^T\xi_{it}\right\|\geq \varepsilon \right).
\end{eqnarray*}
By Chebyshev's inequality, 

\begin{eqnarray*}
\sum_{i=1}^N P\left(\left\|\frac{1}{T}\sum_{t=1}^T\xi_{it}\right\|\geq \varepsilon \right)&\leq& \sum_{i=1}^N\frac{E\left[\left\|\frac{1}{T}\sum_{t=1}^T \xi_{it}\right\|^{2+\delta^\ast/2}\right]}{\varepsilon^{2+\delta^\ast/2}}
\nonumber\\
&\leq&O\left(\frac{1}{T^{1+\delta^\ast/4}}\right)\cdot \sum_{i=1}^N E\left[\left\|\xi_{it}\right\|^{2+\delta^\ast/2}\right]
\nonumber\\
&=&O\left(\frac{N}{T^{1+\delta^\ast/4}}\right),
\end{eqnarray*}
where the second inequality holds by Lemma \ref{LemmaT1}. The proof of Lemma \ref{LemmaT2} is therefore completed.  \hspace*{\fill}{$\blacksquare$}

\bigskip

\noindent \textbf{Proof of Lemma \ref{LemmaT3}:}

(1). For simplicity of notation, we denote that  $\xi_{it}=\frac{[g_\varepsilon(z_{it}^0)]^2 }{[1-G_\varepsilon(z_{it}^0)]G_\varepsilon(z_{it}^0)}  u_{it}^0u_{it}^{0\prime}-E\left[\frac{[g_\varepsilon(z_{it}^0)]^2 }{[1-G_\varepsilon(z_{it}^0)]G_\varepsilon(z_{it}^0)}  u_{it}^0u_{it}^{0\prime}\right]$. It suffices to show that $E\left[\frac{1}{T}\sum_{t=1}^T \xi_{it}\right]=0$  and $E\left[\left\|\frac{1}{T}\sum_{t=1}^T\xi_{it}\right\|^2\right]=O\left(\frac{1}{T}\right)$. The first moment is obvious and therefore we only consider the second moment here. Note that  $\xi_{it}$ is an $\alpha$-mixing process and satisfies the conditions in Assumption \ref{Ass3}.

\begin{eqnarray*}
E\left[\left\|\frac{1}{T}\sum_{t=1}^T\xi_{it}\right\|^2\right] &=& \frac{1}{T^2}\sum_{t=1}^T\sum_{s=1}^T\sum_{l_1=1}^{d_\beta+d_f}\sum_{l_2=1}^{d_\beta+d_f}|E(\xi^{(l_1)}_{it}\xi^{(l_2)}_{is})|
\nonumber\\
&\leq& c_{\delta} \frac{1}{T^2}\sum_{t=1}^T\sum_{s=1}^T\sum_{l_1=1}^{d_\beta+d_f} \sum_{l_1=1}^{d_\beta+d_f}  \alpha_{ii}(|t-s|)^{\delta/(4+\delta)} E\left[|\xi^{(l_1)}_{it}|^{2+\delta/2} \right]^{2/(4+\delta)}E\left[|\xi^{(l_2)}_{is}|^{2+\delta/2} \right]^{2/(4+\delta)} 
\nonumber\\
&=&O\left(\frac{1}{T}\right),
\end{eqnarray*}
where $c_{\delta}=(4+\delta)/\delta\cdot2^{(4+2\delta)/(4+\delta)}$, the  inequality  holds by Davydov's inequality for $\alpha$-mixing process  and the last equality holds by the $\alpha$-mixing and moment conditions in Assumption \ref{Ass2}. Then Lemma \ref{LemmaT3}.(1) holds by Chebyshev's inequality.

(2). For Lemma \ref{LemmaT3}.(2), we redefine $\xi_{it}= \frac{[g_\varepsilon(z_{it}^0)]^2 }{[1-G_\varepsilon(z_{it}^0)]G_\varepsilon(z_{it}^0)}   \gamma_{0i}\gamma_{0i}'-E\left[\frac{[g_\varepsilon(z_{it}^0)]^2 }{[1-G_\varepsilon(z_{it}^0)]G_\varepsilon(z_{it}^0)}   \gamma_{0i}\gamma_{0i}'\right]$. Then for its first moment, we can see that $E\left[\frac{1}{N}\sum_{i=1}^N \xi_{it}\right]=0$. For the second moment, 

\begin{eqnarray*}
E\left[\left\|\frac{1}{N}\sum_{i=1}^N\xi_{it}\right\|^2\right]&=&\frac{1}{N^2}\sum_{i=1}^N\sum_{j=1}^N\sum_{l_1=1}^{d_\beta+d_f}\sum_{l_2=1}^{d_\beta+d_f}|E(\xi^{(l_1)}_{it}\xi^{(l_2)}_{jt})|
\nonumber\\
&\leq& c_{\delta} \frac{1}{N^2}\sum_{i=1}^N\sum_{j=1}^N\sum_{l_1=1}^{d_\beta+d_f} \sum_{l_1=1}^{d_\beta+d_f}  \alpha_{ij}(0)^{\delta/(4+\delta)} E\left[|\xi^{(l_1)}_{it}|^{2+\delta/2} \right]^{2/(4+\delta)}E\left[|\xi^{(l_2)}_{jt}|^{2+\delta/2} \right]^{2/(4+\delta)} 
\nonumber\\
&=&O\left(\frac{1}{N}\right),
\end{eqnarray*}
where the  inequality  holds by Davydov's inequality and the last equality holds by the $\alpha$-mixing and moment conditions in Assumption \ref{Ass2}. We therefore conclude that Lemma \ref{LemmaT3}.(2) holds. \hspace*{\fill}{$\blacksquare$}

\bigskip

\noindent \textbf{Proof of Lemma \ref{LemmaT4}:}

Recall that 

\begin{eqnarray*}
\frac{\partial \log L(\Theta_0, F_0 )}{\partial \theta_i}
&=&\sum_{t=1}^T \frac{[y_{it}-G_\varepsilon(z^0_{it})]g_\varepsilon(z^0_{it}) }{[1-G_\varepsilon(z^0_{it})]G_\varepsilon(z^0_{it})}   u^0_{it} = -\sum_{t=1}^T g_\varepsilon(z^0_{it}) u^0_{it}e_{it}.
\end{eqnarray*}

As we have discussed in the proof of \eqref{ab2}, conditional on $\mathcal{W}=\{w_{it}^0, i,t\geq 1\}$, $e_{is}$ is $\alpha$-mixing, because $\varepsilon_{it}$ is $\alpha$-mixing and independent of $\mathcal{W}$ under Assumption \ref{Ass2}. Therefore,  we can apply the conventional large-block and small-block technique for $\alpha$-mixing process to obtain its asymptotic distributions. 

By partitioning the set ${1,2,\cdots,T}$ into $2\kappa_T+1$ subsets with large block with size $l_T$, small block with size $s_T$ and the remaining set with size $T-\kappa_T(l_T+s_T)$, we can choose $l_T$, $s_T$ to make the following conditions hold:
\begin{equation*}
s_T\rightarrow\infty,\text{\quad} \frac{s_T}{l_T}\rightarrow 0, \text{\quad} \frac{l_T}{T}\rightarrow0,\text{\quad and \quad} \kappa_T=\left[\frac{T}{l_T+s_T}\right]=O\left(s_T\right),
\end{equation*}
where $[m]$ operator define the largest integer which is bounded by  $m$.

Let $\nu_{it}=\frac{1}{\sqrt{T}} g_\varepsilon(z^0_{it}) u^0_{it}e_{it}$. We can observe that by the law of iterated expectations
\begin{eqnarray*}
E[\nu_{it}]&=&E[E[\nu_{it}|\mathcal{W}]]= \frac{1}{\sqrt{T}} E[g_\varepsilon(z^0_{it}) u^0_{it}E[ e_{it}|\mathcal{W}]]=0.
\end{eqnarray*}

Define
\begin{equation*}
  \widetilde{\nu}_{i\rho}=\sum_{t=(\rho-1)(l_T+s_T)+1}^{\rho l_T+(\rho-1)s_T} \nu_{it},\,\widehat{\nu}_{i\rho}=\sum_{t=\rho l_T+(\rho-1)s_T+1}^{\rho (l_T+s_T)} \nu_{it},\,
  \bar{\nu}_i=\sum_{t=\kappa_T(l_T+s_T)+1}^{T} \nu_{it},
\end{equation*}
for $\rho=1,\ldots,\kappa_T$.
We have
\begin{equation*}
  \sum_{t=1}^T\nu_{it}=\sum_{\rho=1}^{\kappa_T}\widetilde{\nu}_{i\rho}+\sum_{\rho=1}^{\kappa_T}\widehat{\nu}_{i\rho}+\bar{\nu}_i.
\end{equation*}
Following analogous arguments in (A.6) of \cite{Gao2012b} (through computing the second moments), we can have the following results for  $ \widehat{\nu}_{i\rho}$ and $\bar{\nu}_i$:

\begin{equation}\label{tec33}
\left\|\sum_{\rho=1}^{\kappa_T}\widehat{\nu}_{i\rho}\right\|=O_P\left(\sqrt{\frac{\kappa_Ts_T}{T}}\right),\,
\left\|\bar{\nu}_i\right\|=O_P\left(\sqrt{\frac{T-\kappa_T(l_T+s_T)}{T}}\right).
\end{equation}

For $\sum_{\rho=1}^{\kappa_T}\widetilde{\nu}_\rho$, by Proposition 2.6 in \cite{FanYao}, we have the following inequality for the characteristic function of 
$\widetilde{\nu}_{i\rho}$,

\begin{eqnarray*}
&&\left|E\left[\exp\left\{\underline{i}\tau\sum_{\rho=1}^{\kappa_T}\widetilde{\nu}_{i\rho} \right\}\right]-\prod_{\rho=1}^{\kappa_T}E\left[\exp\left\{\underline{i}\tau \widetilde{\nu}_{i\rho}\right\} \right] \right|\leq16(\kappa_T-1)\alpha_{ii}(s_T)
\nonumber\\
&&=o(1),
\end{eqnarray*}
where  $\underline{i}$ is the imaginary unit and $\tau$ is the argument in the characteristic function. 

In addition, the variance of large blocks is given by 

\begin{eqnarray*}
\sum_{\rho=1}^{\kappa_T} \text{Var} \left(\widetilde{\nu}_{i\rho}\right)&=&\sum_{\rho=1}^{\kappa_T} \text{Var} \left(\sum_{t=(\rho-1)(l_T+s_T)+1}^{\rho l_T+(\rho-1)s_T} \nu_{it}\right)
\nonumber\\
&=&\sum_{\rho=1}^{\kappa_T} \sum_{t=(\rho-1)(l_T+s_T)+1}^{\rho l_T+(\rho-1)s_T}  \sum_{s=(\rho-1)(l_T+s_T)+1}^{\rho l_T+(\rho-1)s_T} E[ \nu_{it}\nu_{is}]
\nonumber\\
&=& \frac{1}{T}\sum_{\rho=1}^{\kappa_T} \sum_{t=(\rho-1)(l_T+s_T)+1}^{\rho l_T+(\rho-1)s_T} 
\sum_{s=(\rho-1)(l_T+s_T)+1}^{\rho l_T+(\rho-1)s_T}
 E [g_\varepsilon(z^0_{it}) g_\varepsilon(z^0_{is}) e_{it}e_{is} u^0_{it}u_{is}^{0\prime}]
 \nonumber\\
&=& \Sigma_{\theta,i}(1+O(1)),
\end{eqnarray*}
where $\Sigma_{\theta,i}=\lim_{T\rightarrow\infty} \frac{1}{T}\sum_{t=1}^T\sum_{s=1}^T E [g_\varepsilon(z^0_{it}) g_\varepsilon(z^0_{is}) e_{it}e_{is} u^0_{it}u_{is}^{0\prime}]$.
Therefore,  the Feller condition is satisfied.

In addition, for any $\varepsilon>0$, we have

\begin{eqnarray*}
E\left[\left\|\widetilde{\nu}_{i\rho}\right\|^2 I\left\{\left\|\widetilde{\nu}_{i\rho}\right\|\geq \varepsilon\right\}\right]&\leq& \left(E\left[\left\|\widetilde{\nu}_{i\rho}\right\|^3\right]\right)^{\frac{2}{3}}\left(Pr\left[\left\|\widetilde{\nu}_{i\rho}\right\|\geq\varepsilon\right]\right)^{\frac{1}{3}}
\nonumber\\
 &\leq& \varepsilon^{-\frac{2}{3}}\left(E\left[\left\|\widetilde{\nu}_{i\rho}\right\|^3\right]\right)^{\frac{2}{3}}\left(E\left[\left\|\widetilde{\nu}_{i\rho}\right\|^2\right]\right)^{\frac{1}{3}}.
\end{eqnarray*}
\indent
It is clear to see that

\begin{equation*}
 E\left[\left\|\widetilde{\nu}_{i\rho}\right\|^2\right]=O\left(\frac{l_T}{T}\right).
\end{equation*}
By Lemma \ref{LemmaT1},

\begin{eqnarray*}
E\left[\left\|\widetilde{\nu}_{i\rho}\right\|^3\right]=O\left( \left(\frac{l_T}{T}\right)^{\frac{3}{2}}\right).
\end{eqnarray*}
We have

\begin{equation*}
  E\left[\left\|\widetilde{\nu}_{i\rho}\right\|^2 I\left\{\left\|\widetilde{\nu}_{i\rho}\right\|\geq \varepsilon\right\}\right]=O\left(\left(\frac{l_T}{T}\right)^{\frac{4}{3}}\right),
\end{equation*}
and therefore

\begin{eqnarray*}
&&\sum_{\rho=1}^{\kappa_T} E\left[\left\|\widetilde{\nu}_{i\rho}\right\|^2 I\left\{\left\|\widetilde{\nu}_{i\rho}\right\|\geq \varepsilon\right\}\right]=O\left(\left(\frac{l_T}{T}\right)^{\frac{4}{3}}\kappa_T\right)
=o(1).
\end{eqnarray*}
It yields that the Lindeberg condition is satisfied. Since Feller condition and Lindeberg condition can both be satisfied, we can conclude that Lemma \ref{LemmaT4} holds. \hspace*{\fill}{$\blacksquare$}
 
\end{appendices}

\end{document}